\nonstopmode
\documentclass[twocolumn]{aastex63}
\usepackage[utf8]{inputenc}
\usepackage{hyperref}
\usepackage{float}
\usepackage{verbatim}
 \usepackage{natbib}

\submitjournal{AAS Journals}
\usepackage{xcolor}

\begin{document}

\title{TIC 168789840: A Sextuply-Eclipsing Sextuple Star System}

\correspondingauthor{Brian P. Powell}
\email{brian.p.powell@nasa.gov}

\author[0000-0003-0501-2636]{Brian P. Powell}
\affiliation{NASA Goddard Space Flight Center, 8800 Greenbelt Road, Greenbelt, MD 20771, USA}
%
\author[0000-0001-9786-1031]{Veselin~B.~Kostov}
\affiliation{NASA Goddard Space Flight Center, 8800 Greenbelt Road, Greenbelt, MD 20771, USA}
\affiliation{SETI Institute, 189 Bernardo Ave, Suite 200, Mountain View, CA 94043, USA}
%
\author[0000-0003-3182-5569]{Saul A. Rappaport}
\affiliation{Department of Physics, Kavli Institute for Astrophysics and Space Research, M.I.T., Cambridge, MA 02139, USA}
%
\author[0000-0002-8806-496X]{Tam\'as Borkovits}
\affiliation{Baja Astronomical Observatory of University of Szeged, H-6500 Baja, Szegedi út, Kt. 766, Hungary}
\affiliation{Konkoly Observatory, Research Centre for Astronomy and Earth Sciences, H-1121 Budapest, Konkoly Thege Miklós út 15-17, Hungary}
\affiliation{ELTE Gothard Astrophysical Observatory, H-9700 Szombathely, Szent Imre h. u. 112, Hungary}
%
\author[0000-0001-9383-7704]{Petr Zasche}
\affiliation{Astronomical Institute, Charles University, Faculty of Mathematics and Physics, V Hole\v{s}ovi\v{c}k\'ach 2, CZ-180 00, Praha 8, Czech Republic}
%
\author[0000-0002-2084-0782]{Andrei Tokovinin}
\affiliation{Cerro Tololo Inter-American Observatory | NSF's NOIRab, Casilla 603, La Serena, Chile}
%
\author[0000-0002-0493-1342]{Ethan Kruse}
\affiliation{NASA Goddard Space Flight Center, 8800 Greenbelt Road, Greenbelt, MD 20771, USA}
\affiliation{Universities Space Research Association, 7178 Columbia Gateway Drive, Columbia, MD 21046}
%
\author[0000-0001-9911-7388]{David W. Latham}
\affiliation{Center for Astrophysics $\vert$ Harvard \& Smithsonian, 60 Garden Street, Cambridge, MA 02138, USA}
%
\author[0000-0001-7516-8308]{Benjamin~T.~Montet}
\affiliation{School of Physics, University of New South Wales, Sydney NSW 2052, Australia}
%
\author[0000-0002-4625-7333]{Eric L. N. Jensen} 
\affiliation{Dept.\ of Physics \& Astronomy, Swarthmore College, Swarthmore PA 19081, USA}
%
\author[0000-0002-7778-3117]{Rahul Jayaraman}
\affiliation{Department of Physics, Kavli Institute for Astrophysics and Space Research, M.I.T., Cambridge, MA 02139, USA}
%
\author[0000-0001-6588-9574]{Karen A.\ Collins}
\affiliation{Center for Astrophysics $\vert$ Harvard \& Smithsonian, 60 Garden Street, Cambridge, MA 02138, USA}
%
\author[0000-0002-0967-0006]{Martin Ma\v{s}ek}
\affiliation{FZU - Institute of Physics of the Czech Academy of Sciences, Na Slovance 1999/2, CZ-182~21, Praha, Czech Republic}
%
\author[0000-0002-3439-1439]{Coel Hellier}
\affiliation{Astrophysics Group, Keele University, Staffordshire, ST5 5BG, United Kingdom}
%
\author[0000-0002-5674-2404]{Phil Evans} 
\affiliation{El Sauce Observatory, Coquimbo Province, Chile}
%
\author[0000-0001-5603-6895]{Thiam-Guan Tan}
\affiliation{Perth Exoplanet Survey Telescope, Australia}
%
\author[0000-0001-5347-7062]{Joshua E. Schlieder}
\affiliation{NASA Goddard Space Flight Center, 8800 Greenbelt Road, Greenbelt, MD 20771, USA}
%
\author[0000-0002-5286-0251]{Guillermo Torres}
\affiliation{Center for Astrophysics $\vert$ Harvard \& Smithsonian, 60 Garden Street, Cambridge, MA 02138, USA}
%
\author{Alan P. Smale}
\affiliation{NASA Goddard Space Flight Center, 8800 Greenbelt Road, Greenbelt, MD 20771, USA}
%
\author{Adam H. Friedman}
\affiliation{University of Michigan, 500 S State St, Ann Arbor, MI 48109}
\affiliation{NASA Goddard Space Flight Center, 8800 Greenbelt Road, Greenbelt, MD 20771, USA}
%
\author[0000-0001-7139-2724]{Thomas~Barclay}
\affiliation{NASA Goddard Space Flight Center, 8800 Greenbelt Road, Greenbelt, MD 20771, USA}
\affiliation{University of Maryland, Baltimore County, 1000 Hilltop Circle,
Baltimore, MD 21250, USA}
%
\author[0000-0002-5665-1879]{Robert Gagliano}
\affiliation{Amateur Astronomer, Glendale, AZ 85308}
\author[0000-0003-1309-2904]{Elisa V. Quintana}
\affiliation{NASA Goddard Space Flight Center, 8800 Greenbelt Road, Greenbelt, MD 20771, USA}
%
\author[0000-0003-3988-3245]{Thomas L. Jacobs}
\affiliation{Amateur Astronomer, 12812 SE 69th Place, Bellevue, WA 98006}
\author[0000-0002-0388-8004]{Emily A. Gilbert}
\affiliation{Department of Astronomy and Astrophysics, University of
Chicago, 5640 S. Ellis Ave, Chicago, IL 60637, USA}
\affiliation{University of Maryland, Baltimore County, 1000 Hilltop Circle, Baltimore, MD 21250, USA}
\affiliation{The Adler Planetarium, 1300 South Lakeshore Drive, Chicago, IL 60605, USA}
\affiliation{NASA Goddard Space Flight Center, 8800 Greenbelt Road, Greenbelt, MD 20771, USA}
\affiliation{GSFC Sellers Exoplanet Environments Collaboration}
%
\author[0000-0002-2607-138X]{Martti~H.~Kristiansen}
\affil{Brorfelde Observatory, Observator Gyldenkernes Vej 7, DK-4340 T\o{}ll\o{}se, Denmark}
\affil{DTU Space, National Space Institute, Technical University of Denmark, Elektrovej 327, DK-2800 Lyngby, Denmark}
\author[0000-0001-8020-7121]{Knicole D. Col\'{o}n}
\affiliation{NASA Goddard Space Flight Center, 8800 Greenbelt Road, Greenbelt, MD 20771, USA}
%
\author[0000-0002-8527-2114]{Daryll M. LaCourse}
\affiliation{Amateur Astronomer, 7507 52nd Place NE Marysville, WA 98270}
\author[0000-0001-8472-2219]{Greg Olmschenk}
\affiliation{NASA Goddard Space Flight Center, 8800 Greenbelt Road, Greenbelt, MD 20771, USA}
\affiliation{Universities Space Research Association, 7178 Columbia Gateway Drive, Columbia, MD 21046}
%
\author{Mark Omohundro}
\affiliation{Citizen Scientist, c/o Zooniverse, Department of Physics, University of Oxford, Denys Wilkinson Building, Keble Road, Oxford, OX13RH, UK}
\author[0000-0002-2942-8399]{Jeremy D. Schnittman}
\affiliation{NASA Goddard Space Flight Center, 8800 Greenbelt Road, Greenbelt, MD 20771, USA}
%
\author{Hans M. Schwengeler}
\affiliation{Citizen Scientist, Planet Hunter, Bottmingen, Switzerland}
\author{Richard K. Barry}
\affiliation{NASA Goddard Space Flight Center, 8800 Greenbelt Road, Greenbelt, MD 20771, USA}
%
\author{Ivan A. Terentev}
\affiliation{Citizen Scientist, Planet Hunter, Petrozavodsk, Russia}
\author{Patricia Boyd}
\affiliation{NASA Goddard Space Flight Center, 8800 Greenbelt Road, Greenbelt, MD 20771, USA}
%
\author{Allan R. Schmitt}
\affiliation{Citizen Scientist, 616 W. 53rd. St., Apt. 101, Minneapolis, MN 55419, USA}
\author[0000-0002-8964-8377]{Samuel N. Quinn}
\affil{Center for Astrophysics $\vert$ Harvard \& Smithsonian, 60 Garden Street, Cambridge, MA 02138, USA}
%
\author[0000-0001-7246-5438]{Andrew Vanderburg}
\affiliation{Department of Astronomy, University of Wisconsin-Madison, Madison, WI 53706, USA}
%
\author[0000-0003-0987-1593]{Enric Palle} 
\affiliation{Instituto de Astrof\'\i sica de Canarias (IAC), 38205 La Laguna, Tenerife, Spain}
\affiliation{Departamento de Astrof\'\i sica, Universidad de La Laguna (ULL), 38206, La Laguna, Tenerife, Spain}
%
\author{James Armstrong} 
\affiliation{Institute for Astronomy, University of Hawaii, Maui, HI 96768, USA}
%
\author[0000-0003-2058-6662]{George R. Ricker}
\affiliation{Department of Physics, Kavli Institute for Astrophysics and Space Research, M.I.T., Cambridge, MA 02139, USA}
%
\author[0000-0001-6763-6562]{Roland Vanderspek}
\affiliation{Department of Physics, Kavli Institute for Astrophysics and Space Research, M.I.T., Cambridge, MA 02139, USA}
%
\author[0000-0002-6892-6948]{S.~Seager}
\affiliation{Department of Physics, Kavli Institute for Astrophysics and Space Research, M.I.T., Cambridge, MA 02139, USA}
\affiliation{Department of Earth, Atmospheric and Planetary Sciences, Massachusetts Institute of Technology, Cambridge, MA 02139, USA}
\affiliation{Department of Aeronautics and Astronautics, MIT, 77 Massachusetts Avenue, Cambridge, MA 02139, USA}
%
\author{Joshua N. Winn}
\affiliation{Department of Astrophysical Sciences, Princeton University, Princeton, NJ 08544, USA}
%
\author[0000-0002-4715-9460]{Jon M. Jenkins}
\affiliation{NASA Ames Research Center, Moffett Field, CA 94035, USA}
%
\author[0000-0003-1963-9616]{Douglas A. Caldwell}
\affiliation{NASA Ames Research Center, Moffett Field, CA 94035, USA}
\affiliation{SETI Institute, 189 Bernardo Ave, Suite 200, Mountain View, CA 94043, USA}
%
\author[0000-0002-5402-9613]{Bill Wohler}
\affiliation{NASA Ames Research Center, Moffett Field, CA 94035, USA}
\affiliation{SETI Institute, 189 Bernardo Ave, Suite 200, Mountain View, CA 94043, USA}
%
\author{Bernie Shiao}
\affiliation{Mikulski Archive for Space Telescopes, 3700 San Martin Drive, Baltimore, MD 21218 USA}
%
\author[0000-0002-7754-9486]{Christopher~J.~Burke}
\affiliation{Department of Physics, Kavli Institute for Astrophysics and Space Research, M.I.T., Cambridge, MA 02139, USA}
\author[0000-0002-6939-9211]{Tansu~Daylan}
\affiliation{Department of Physics, Kavli Institute for Astrophysics and Space Research, M.I.T., Cambridge, MA 02139, USA}
\affiliation{Kavli Fellow}
%
\author{Joel Villase{\~n}or}
\affiliation{Department of Physics, Kavli Institute for Astrophysics and Space Research, M.I.T., Cambridge, MA 02139, USA}

\received{December 11, 2020}
\revised{January 9, 2021}
\accepted{MMM DD, YYYY}

\begin{abstract}
We report the discovery of a sextuply-eclipsing sextuple star system from {\em TESS} data, TIC 168789840, also known as TYC 7037-89-1, the first known sextuple system consisting of three eclipsing binaries. The target was observed in Sectors 4 and 5 during Cycle 1, with lightcurves extracted from {\em TESS} Full Frame Image data. It was also previously observed by the WASP survey and ASAS-SN. The system consists of three gravitationally-bound eclipsing binaries in a hierarchical structure of an inner quadruple system with an outer binary subsystem.  Follow-up observations from several different observatories were conducted as a means of determining additional parameters. The system was resolved by speckle interferometry with a 0\farcs42 separation between the inner quadruple and outer binary, inferring an estimated outer period of $\sim$2 kyr. It was determined that the fainter of the two resolved components is an 8.217 day eclipsing binary, which orbits the inner quadruple that contains two eclipsing binaries with periods of 1.570 days and 1.306 days.   MCMC analysis of the stellar parameters has shown that the three binaries of TIC 168789840 are ``triplets'', as each binary is quite similar to the others in terms of mass, radius, and $T_{\rm eff}$.  As a consequence of its rare composition, structure, and orientation, this object can provide important new insight into the formation, dynamics, and evolution of multiple star systems.  Future observations could reveal if the intermediate and outer orbital planes are all aligned with the planes of the three inner eclipsing binaries.
\end{abstract}



\section{Introduction}\label{sec:intro}

The Transiting Exoplanet Survey Satellite ({\em TESS}) mission \citep{Ricker14} has dramatically improved our ability to discover multiple star systems. Though it is more prone to systematics than the {\em Kepler} telescope and has a poorer angular resolution (21\arcsec \ per pixel for {\em TESS} vs.~3.98\arcsec \ per pixel for {\em Kepler}), the breadth of observation of {\em TESS}, encompassing nearly the entire sky, has allowed for the identification of many candidate multiple star systems through the analysis of eclipses in the lightcurves. In fact, a collaboration between the NASA Goddard Space Flight Center (GSFC) Astrophysics Science Division and the MIT Kavli Institute, in conjunction with expert visual surveyors, has found well over 100 triple and quadruple star system candidates. This number will continue to increase as {\em TESS} proceeds with the extended mission at faster observation cadence (10 minutes for cycles 3 and 4 vs.~30 minutes for cycles 1 and 2), enabling researchers to capture shorter-duration eclipse events.  We also note that lightcurves from cycles 1 and 2 have yet to be fully exploited.

The large majority of our discovered candidate triple and quadruple star systems are quadruples, followed by triples. Though quadruple systems are much more rare than triple systems, the large outer orbit of the third star in a hierarchical triple, necessary for stability, substantially reduces the probability that the eclipse or occultation of the third star will be visually noticed in a {\em TESS} lightcurve.  Beyond quadruple stars, the probability of systems with more components being identified via photometry alone is remote, as the formation of sextuple systems is likely quite rare. This low probability is compounded by the requirement that each binary must be oriented in such a manner that they are all eclipsing.  Though simulations of stellar system formation have found that a sextuple system consisting of two inner triples is nearly ten times more likely to form than a system of three close binaries \citep{mult}, the visual detection of all the eclipses in a sextuple consisting of two triples is far less likely, again, due to the large outer orbit of the third star in each triple.

\begin{figure*}
    \centering
    \includegraphics[width=0.7\textwidth]{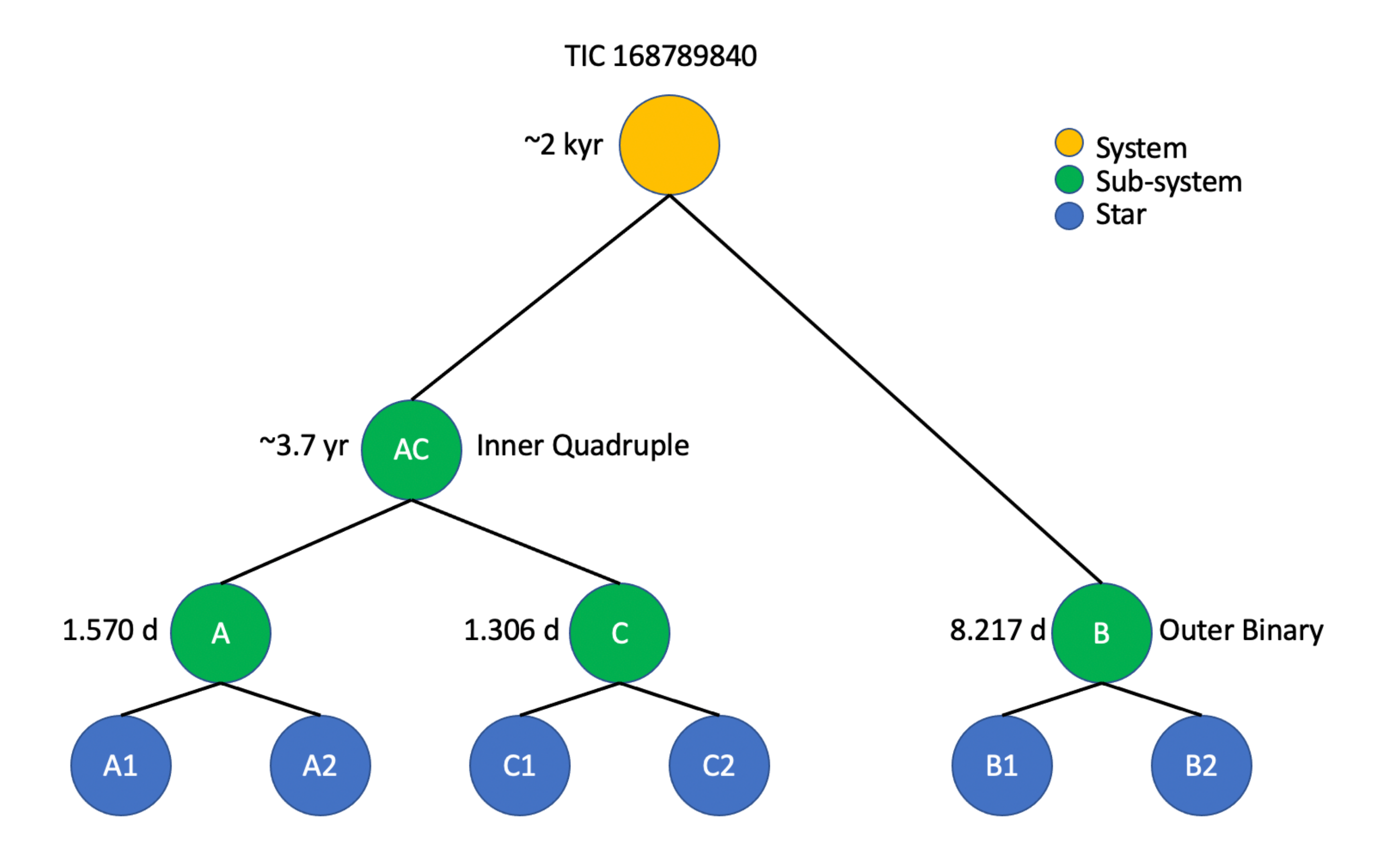}
    \caption{Structure of TIC 168789840, a sextuple system of three eclipsing binaries arranged as inner quadruple AC and outer binary B. In this work we will discuss how we arrived at this configuration.}
   \label{fig:structure}
\end{figure*}

In this work we present a sextuple system which exhibits all six eclipses (three primary and three secondary) discovered with {\em TESS}. We show that TIC 168789840 consists of three close binaries. The inner quadruple system with a period of $\sim$3.7 yrs is comprised of two eclipsing binaries (which we provide the names ``A'' and ``C''), at periods of 1.570 and 1.306 days, respectively; the inner quadruple is orbited by another eclipsing binary (which we call ``B''), with a period of 8.217 days, at a period of $\sim$2 kyr.  The structure of the system, shown in Figure~\ref{fig:structure}, will be the nomenclature that will be used for the rest of this paper. Prior to the discovery of TIC 168789840, there were 17 known sextuple star systems according to the June 2020 update of the Multiple Star Catalog \citep{msc}.  TIC 168789840 is the first that is sextuply eclipsing, with the caveat that Jayaraman, Rappaport, Borkovits, Zasche et al.~are currently analyzing another such system that will be published in the near future.

There are several known sextuple systems with a similar structure to that of TIC 168789840.  One of these systems, ADS 9731, is a resolved visual quadruple system known for more than a century;
two of its components were further determined to be close spectroscopic binaries by \cite{1998AstL...24..795T}.  88 Tauri, suspected to be a spectroscopic quintuple by \cite{1988A&A...200..175B}, was later determined to be a sextuple system of three binaries \citep{1997A&AS..124...75T}, with the two binaries comprising the inner quadruple having an 18 yr period \citep{2007ApJ...669.1209L}.  Interestingly, of the known sextuple systems, TIC 168789840 is most similar to the famous Castor system, which also contains three close binaries.  

Castor, among the brightest star systems in the sky, was originally identified as a visual binary system in 1719 by Bradley and Pound.  \cite{1897ApJ.....5....1B} found that one of its components was a spectroscopic binary, and \cite{1905PASP...17..173C} discovered that another component was also a binary.  \cite{1920PASP...32..276A} found that there was a third component which was also a binary, completing the discovery of the Castor system as the first known sextuple star system.  The mass and radius ratios of the binaries of TIC 168789840, in addition to the close orbits of the binaries, are found in this work to be quite similar to those determined by the extensive analysis of Castor.

The rest of the paper is organized as follows. Section \ref{sec:detection} outlines the initial detection and analysis of the {\em TESS} data; the disentaglement of the individual EB lightcurves is presented in Section \ref{sec:disentanglement}. In Sections \ref{sec:archival} and \ref{sec:follow-up} we present the analysis of archival data and our follow-up observations, respectively. The comprehensive analysis of the parameters of the system and the corresponding discussion of the results is presented in Section \ref{sec:sysparms}. Finally, we draw our conclusions in Section \ref{sec:summary}.

\section{Detection}
\label{sec:detection}

Using the 129,000-core {\it Discover} supercomputer at the NASA Center for Climate Simulation (NCCS) at NASA GSFC, we are building Full-Frame-Image (FFI) lightcurves for all stars observed by {\em TESS} up to 15th magnitude.  All original and calibrated FFIs are produced by the {\em TESS} Science Processing Operations Center (SPOC, \citealt{jenkinsSPOC2016}).  Target lists were created through a parallelized implementation of {\tt tess-point} \citep{tess-point} on the {\em TESS} Input Catalog (TIC) provided by the Mikulski Archive for Space Telescopes (MAST).  The lightcurves for each sector were constructed in 1-4 days of wall clock time (for a total of over 100 CPU-years to date), depending on the density of targets in the sector, through a parallelized implementation of the \texttt{eleanor} Python module \citep{eleanor}. From these lightcurves, we are performing a search for multiple stellar systems using targets from the GSFC {\em TESS} Eclipsing Binary (EB) Catalog (Kruse et al. 2021, in prep). 

\begin{figure*}
    \centering
    \includegraphics[width=0.8\textwidth]{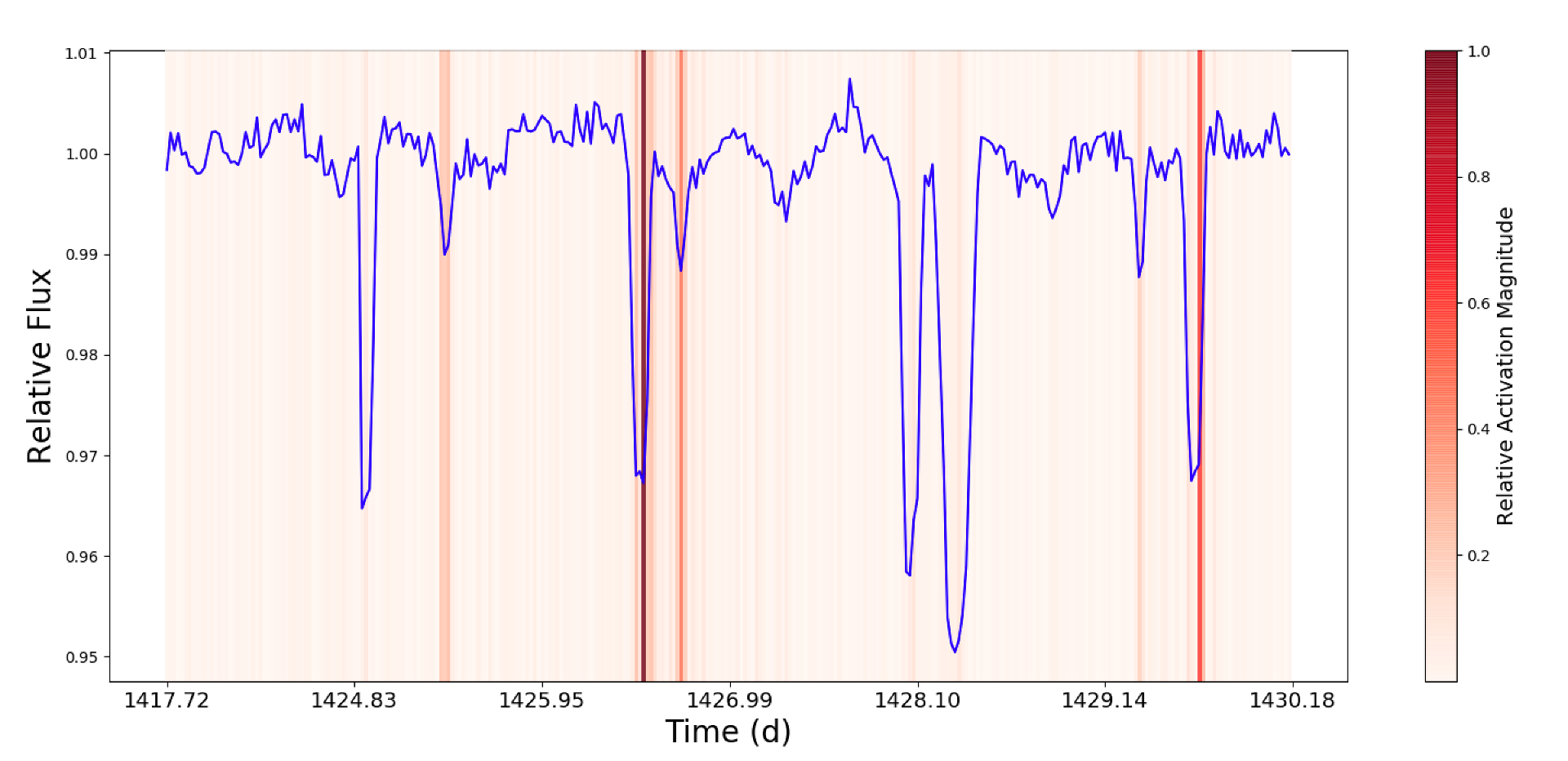}
    \caption{Saliency map -- indicating features of importance to the classification -- of a segment of the TIC 168789840 {\tt eleanor} raw flux lightcurve.  The neural network activates on the feature of the eclipse.  Some eclipses activate more strongly than others, which is a function of the response of the neural network to the lightcurve as well as idiosyncrasies of the training data. The map was created using \texttt{keras-vis} \citep{kerasvis} on the penultimate layer of the EB classifier neural network.}
    \label{fig:activation_map}
\end{figure*}

This catalog of eclipsing binaries was generated by a neural network classifier.  This neural network was trained on the NCCS Advanced Data Analytics PlaTform (ADAPT) GPU cluster to classify a lightcurve (as either an EB or not an EB) based only on the feature of the eclipse, neglecting any periodicity or time-dependency. The neural network is a one-dimensional adaptation of the ResNet \citep{2015arXiv151203385H} structure to accommodate the data shape of a lightcurve, built in Python using {\tt keras} \citep{keras}/{\tt tensorflow}\citep{tensorflow}.  A strength of this approach is that it allows for the identification of single-eclipse EBs.  As such, a lightcurve with an eclipse recognizable by the neural network, no matter the number of eclipses occurring in a single lightcurve, will be properly classified as an EB by the neural network.  Figure \ref{fig:activation_map} shows the activation of the neural network on the feature of the eclipse in a segment of the TIC 168789840 lightcurve, which demonstrates that each eclipse does not need to be individually identified by the neural network in order for the lightcurve to be classified as an EB.  The lack of a periodicity or similarity constraint allows for a lightcurve with multiple irregular eclipses, such as TIC 168789840, to be classified as an EB.  

\begin{figure}
    \centering
    \includegraphics[width=0.3\textwidth]{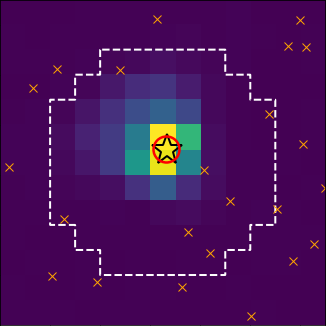}
    \includegraphics[width=0.3\textwidth]{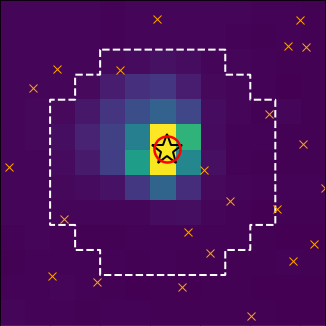}
    \includegraphics[width=0.3\textwidth]{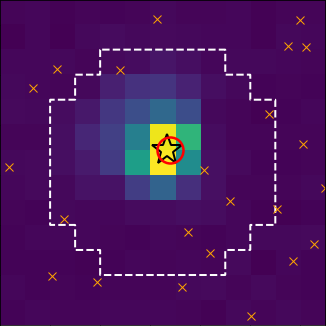}

    \caption{Photocenter analysis for the three primary eclipses of TIC 168789840 for Sector 4. The panels show the mean difference image for pair A (top), B (middle) and C (bottom). The large open circle represents the average difference image photocenter and the large star symbol represents the catalog position of the target. The small x symbols represent nearby stars from the TIC within 10 magnitude difference. The difference image photocenters for all three sets of eclipses are on-target.}
    \label{fig:centroids_small}
\end{figure}

Lightcurves with multiple sets of eclipses are manually flagged as meriting further investigation.  While the overwhelming majority of these lightcurves are determined to be false positives caused by close proximity of two or more EBs blending into a single lightcurve, there remains a fraction which cannot be explained by such contamination.  This is determined through photocenter analysis, the output of which for TIC 168789840 is shown in Figure~\ref{fig:centroids_small}. The analysis follows the difference imaging procedure of \cite{diff} as adapted into the DAVE vetting pipeline \citep{Kostov19}. 

\begin{figure*}
    \centering
    \includegraphics[width=0.75\textwidth]{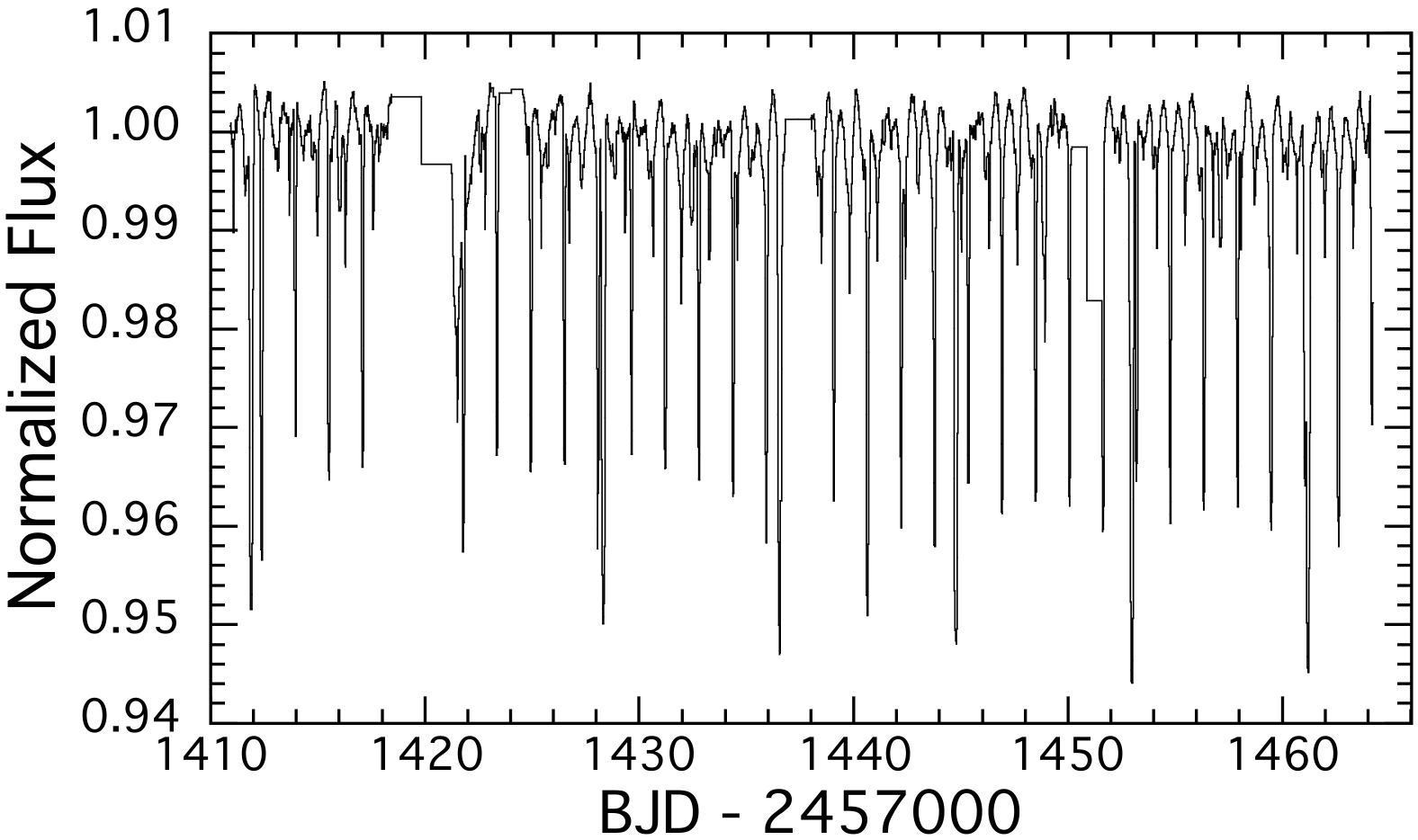} \hglue0.5cm
    \includegraphics[width=0.85\textwidth]{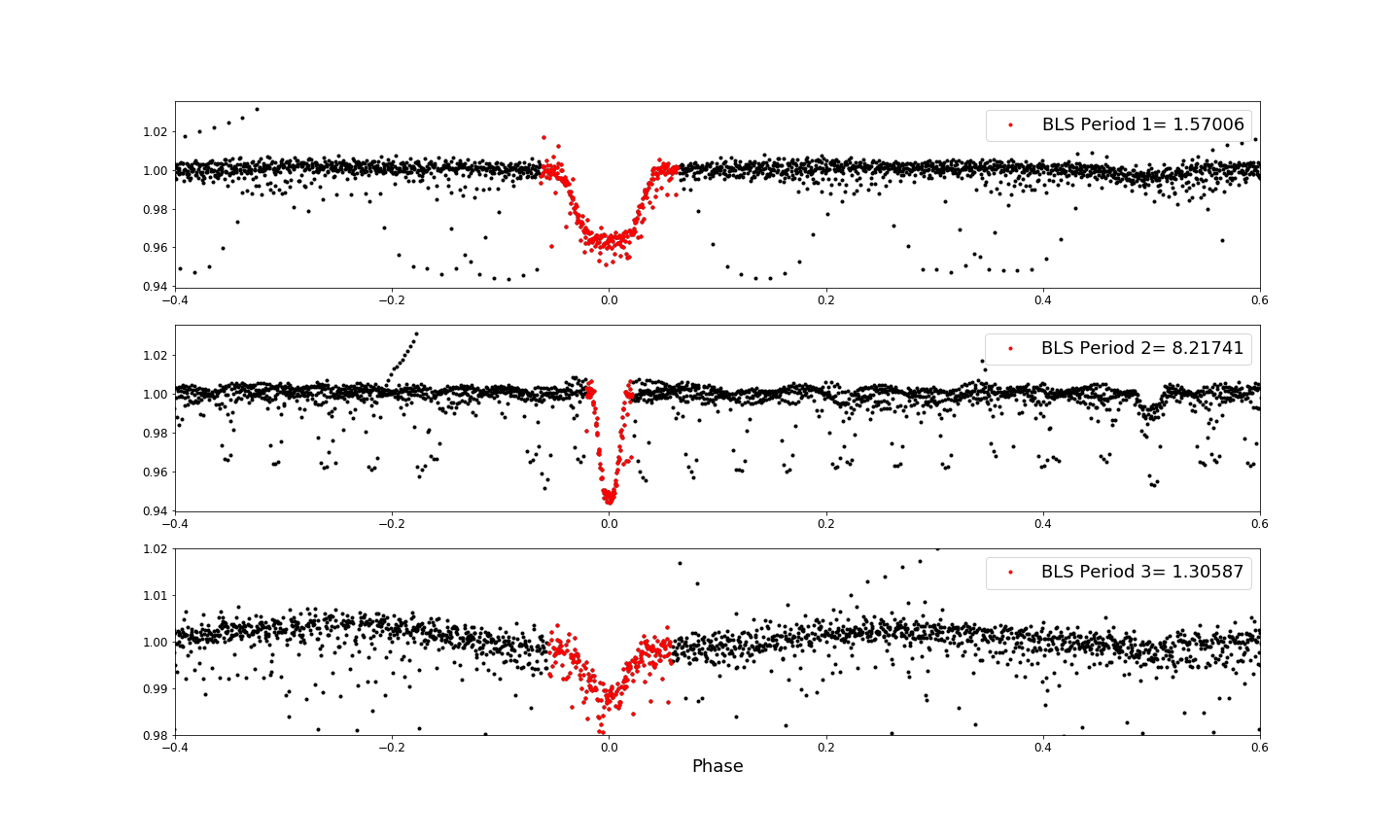}
    \caption{Upper panel: The {\em TESS} lightcurve of TIC 168789840 for sectors 4 and 5. Several eclipses are blended, most notably around time 1461. The rest of the panels show the lightcurve phase-folded on the three distinct periods of the three EBs as measured by BLS.
     }
    \label{fig:bls}
\end{figure*}

Briefly, we first perform a Box Least Squares (BLS) analysis \citep{bls} of the {\em TESS} lightcurve to measure the ephemerides for all sets of eclipses. As demonstrated in Figure~\ref{fig:bls}, the {\em TESS} lightcurve of TIC 168789840 shows three distinct periods with primary and secondary eclipses. Next, for each eclipse of each set we create the out-of-eclipse images and difference images, measure the corresponding center-of-light (photocenter) by calculating the respective x- and y-moments and, finally, compare the average out-of-eclipse photocenter to the average difference image photocenter for each set. A significant shift between these two photocenters would indicate a potential false positive due to a nearby field star. For TIC 168789840, there are no significant differences between the measured out-of-eclipse and difference-image photocenters for all sets of eclipses, indicating that the target is their source. 

\section{Disentanglement of the Lightcurves}
\label{sec:disentanglement}

\subsection{Fourier Method}
\label{sec:FT}

After identifying the sources of all the eclipses to be on target (i.e.~belonging to TIC 168789840), we needed to disentangle the combined photometry to create lightcurves for each of the three eclipsing binaries.  Here we introduce one of our two methods for disentangling the photometric lightcurves of the three eclipsing binaries,  superposed in the {\em TESS} data.  This approach, which amounts to a Fourier decomposition, requires prior knowledge of the orbital periods.  In this particular case, the periods were determined from Lomb-Scargle and BLS transforms of the {\em TESS} data as well as archival ASAS-SN data (see Section \ref{sec:asassn}). 

To represent the three binaries in the lightcurve, we fit a harmonic series of the following form to the entire 27-day {\em TESS} data train:
\begin{equation}
F(t) = \sum_{m=1}^3 \left(\sum_{n=1}^{50} \alpha^{(m)}_n \sin(\omega_n t) + \beta^{(m)}_n \cos(\omega_n t)\right) + \gamma ,
\end{equation} 
where $\omega_n$ is the $n$th orbital frequency in the series representing the $m$th binary, and is given by $2\pi n/P_m$, where $P_m$ is the orbital period of the $m$th binary.  In all there are $3 \times 50 \times 2+1 = 301$ linear coefficients to be fit, i.e., all the $\alpha_n, \beta_n$, and $\gamma$ (the latter being the constant background level, which is $\simeq 1$ if the lightcurve is normalized). We note that the values of $\omega_n$ are unrelated to the usual orthogonal frequencies used in an FFT which are given by integer multiples of $2 \pi/T$, where T is the duration of the observation interval.

These coefficients can all be fitted simultaneously with the inversion of a single $301 \times 301$  $\chi^2$ matrix, which takes much less than a minute on a standard laptop.  While we used 50 harmonics in this case, we have found that 30 harmonic terms are sufficient to effectively reconstruct most binary lightcurves, except those with very deep and/or sharp eclipses. 

The next and final step in the procedure is to reconstruct the lightcurve for the $m$th binary via the following sum:
\begin{equation}
F_m(t_j) =  \sum_{n=1}^{50} \alpha^{(m)}_n \sin(\omega_n t_j) + \beta^{(m)}_n \cos(\omega_n t_j) + \gamma
\end{equation} 
where $j$ is the $j$th data point.

\begin{figure*}
  \centering
  \includegraphics[width=0.3\linewidth]{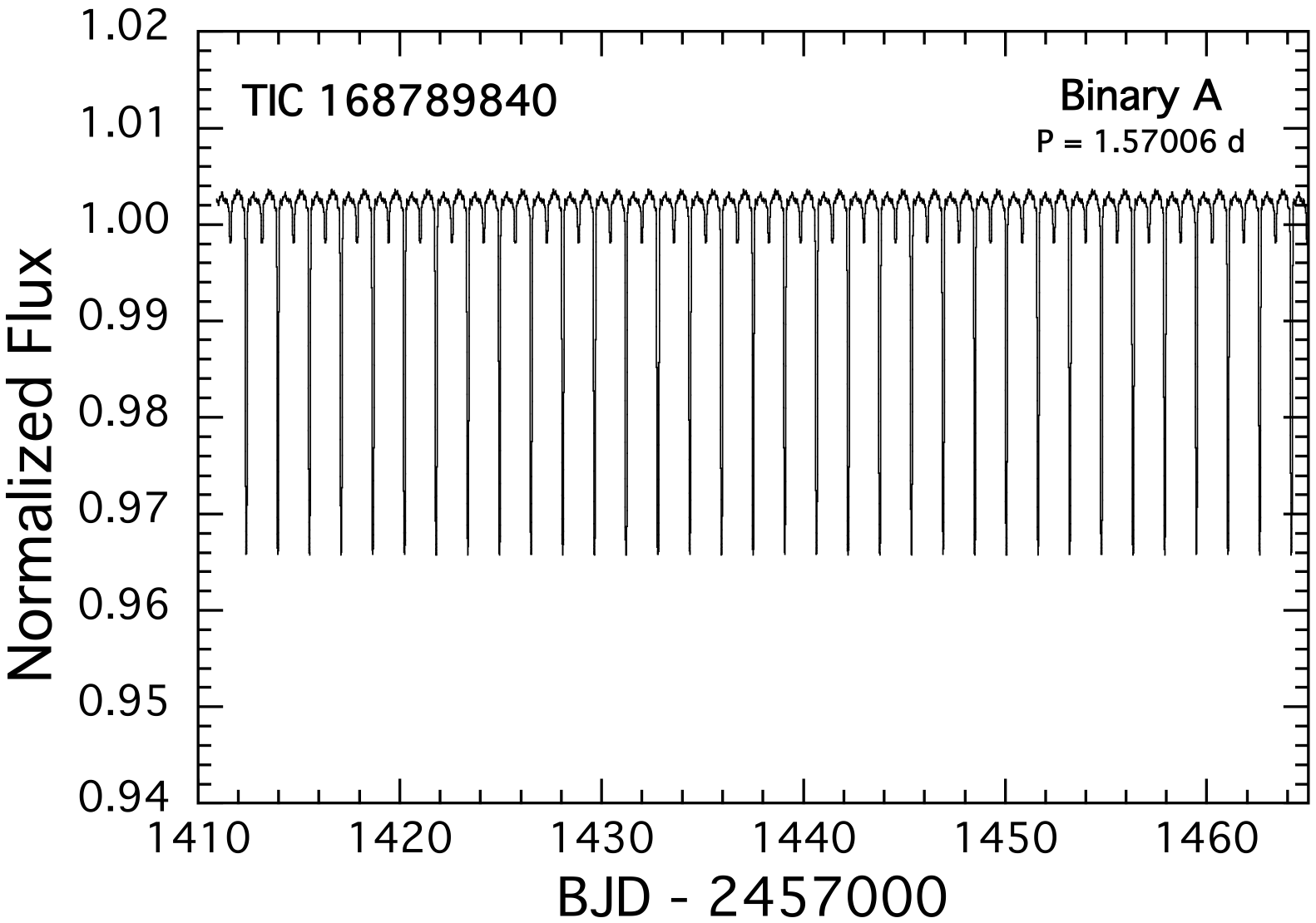} \hglue0.1cm
  \includegraphics[width=0.3\linewidth]{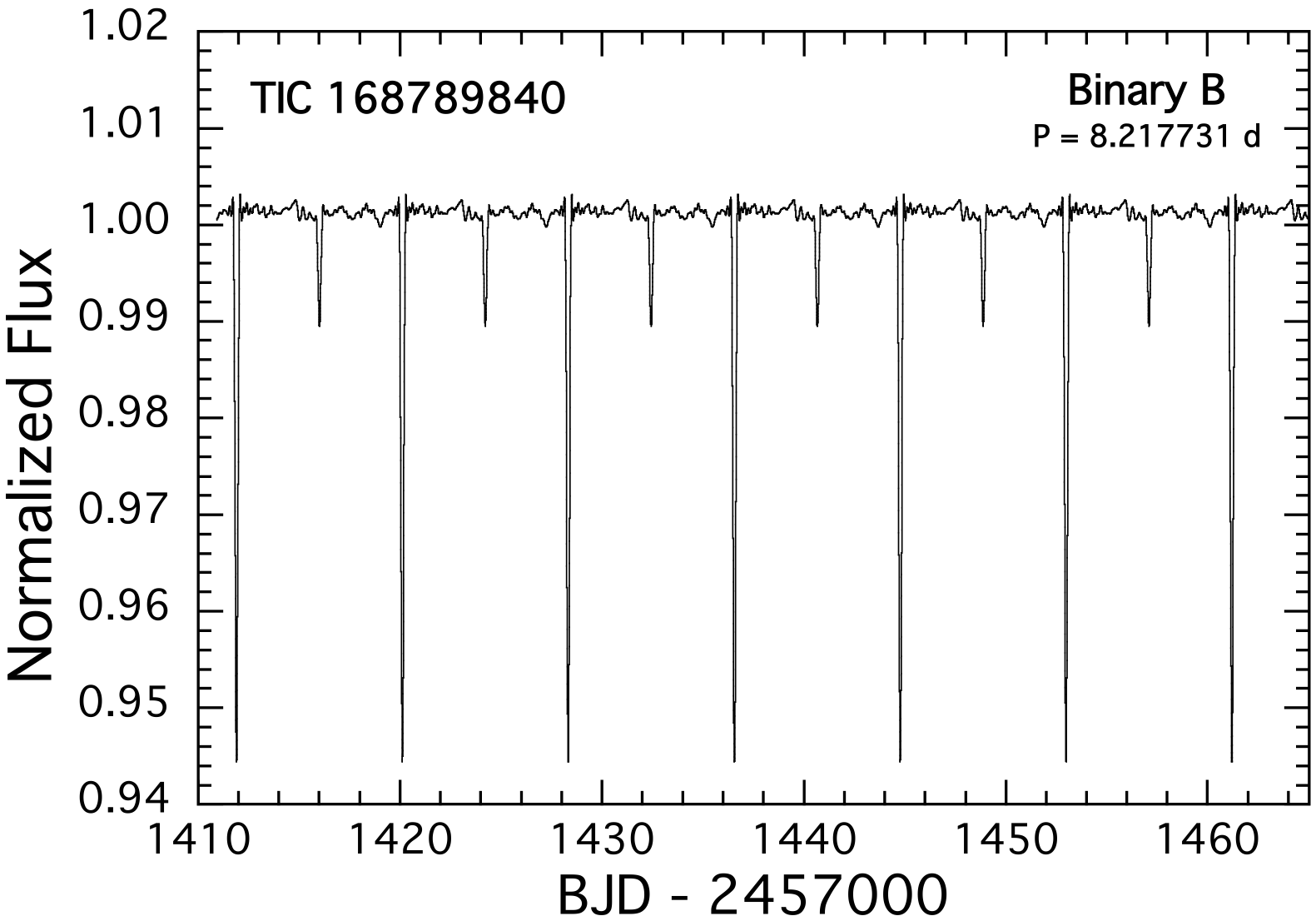} \hglue0.1cm
  \includegraphics[width=0.3\linewidth]{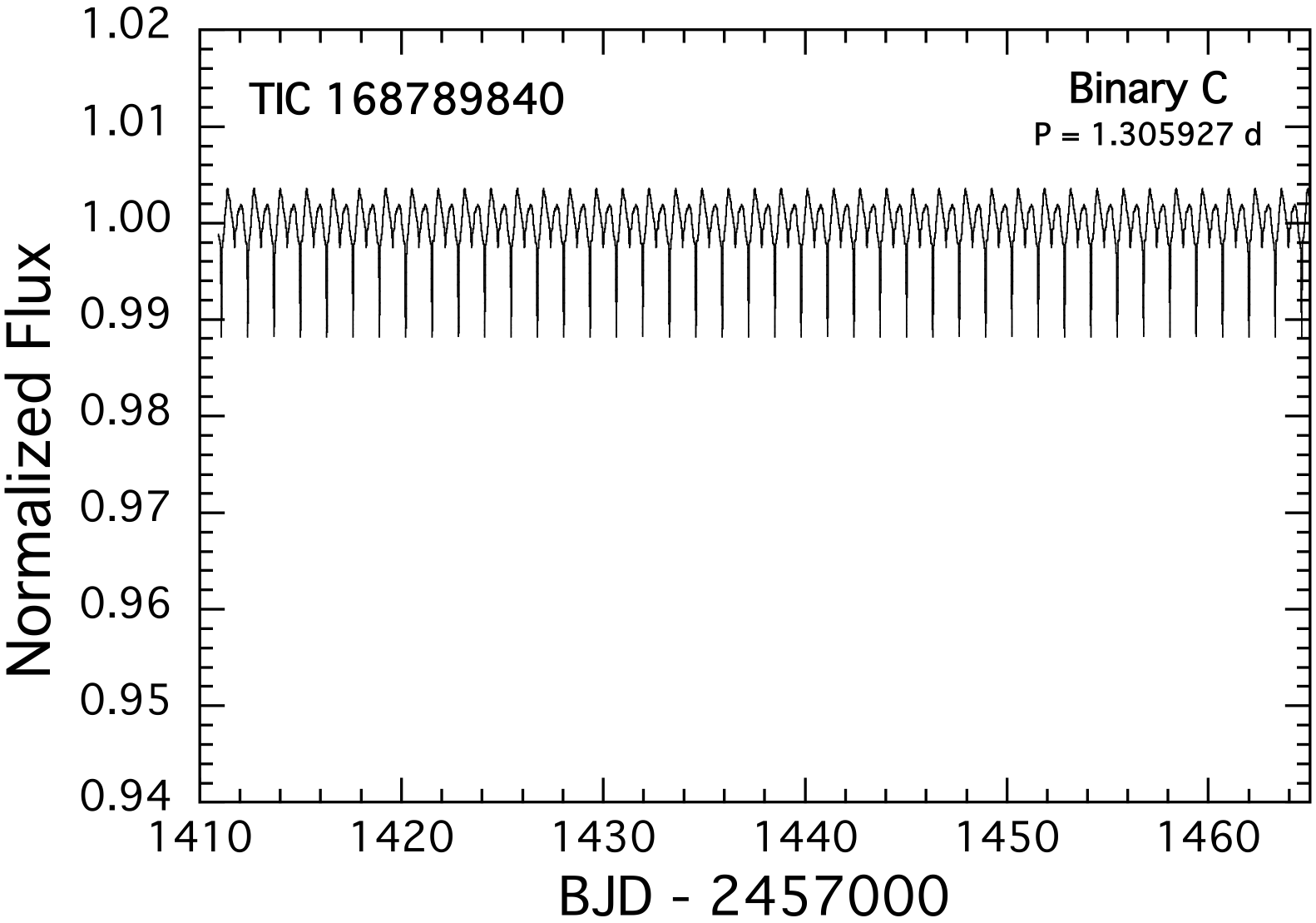}
\caption{Reconstructed {\em TESS} lightcurves for the A, B, and C binaries.  These are presented on the same y-axis in order to visualize the relative contributions from each binary.}
\label{fig:recons}
\end{figure*}

The results of the Fourier disentanglement for the {\em TESS} lightcurve of TIC 168789840 are shown in Figure~\ref{fig:recons}.  The three panels show the reconstructed lightcurves for the A, B, and C binaries with periods of 1.570 days, 8.217 days, and 1.306 days, respectively.  These are perfectly conventional EBs with two eclipses per period, and, at first glance, the lightcurves seem to indicate circular orbits. 

Finally, we discuss an important caveat to this Fourier-based method for disentangling multiple superposed lightcurves. This technique works best if none of the harmonics of one eclipsing binary overlaps, within a resolution element ($2 \pi/T$), of any of the harmonics from the other binaries. If there is significant overlap among any of the lower harmonics (e.g., for $n \lesssim 5$) then this technique {\em may} have problems with the degenerate frequencies. If the overlap is among the higher harmonics (e.g., for $n \gtrsim 15$), then this effect is probably negligible. In the case of TIC 168789840 the 4th harmonic of the A binary overlaps the 21st harmonic of B, while the 5th harmonic of the C binary overlaps the 6th harmonic of binary A.  We have checked what problems this might cause, by removing each of the binaries separately, and we find very similar results to fitting for them all simultaneously.  

\subsection{Iterative Method}
\label{sec:iterative}
We have also used another independent method for disentangling the lightcurves.  The results of the two procedures can be used to check each other.  

In the iterative approach, we follow the schematic steps outlined in Table \ref{tbl:iterative}.  We start with the original lightcurve time series denoted as TABC where the ``T'' stands for time series, and the `A', `B', and `C' signify that all three binaries are contained in T.  We then start by producing a phase-folded, binned and averaged orbital lightcurve (hereafter, denoted as `F') for one of the binaries (e.g., A), by first removing from the time series the intervals when eclipses from the other two binaries (e.g., B and C) occur.  The residual time series is then phase folded, binned and averaged to produce FA$_{a,I}$.  The subscripts `a' and `I' signify that we are starting the cleaning process by working along track `a' (see Table \ref{tbl:iterative} for the track definitions) and this will be a first-level product (`I').

The next step is to subtract fold FA$_{a,I}$ from TABC, using a three-point local Lagrangian-interpolation to calculate the flux to be subtracted at each observed photometric phase of the binary. The result is a time series comprised of the blended lightcurves of only two of the three binaries, and we denote this product as, e.g., TBC.  This completes the first-level products.  In all there are three preliminary folds, one for each binary, and three time series\footnote{Note, for practical reasons, we added a constant flux to these time series in such a way that the flux of the very first data point retained the same value as in the original time series. In this manner, we replaced the varying light of the extracted binary with a constant extra light.}, each containing two of the binaries.  See the second row of Table \ref{tbl:iterative}.

To produce the second-level (`II') products, we take each of the time series from level I, comprised of two binaries, e.g., TBC, remove the eclipses of either one of the binaries, and produce a phase-folded, binned, and averaged orbital lightcurve of the other binary, e.g., FB$_{a,II}$.  Then, as in level I, we subtract off the folded lightcurve of that binary to produce a time series containing only a single binary.  Schematically, TBC - FB$_{a,II}$ = TC, and TBC - FC$_{a,II}$ = TB.  The net result of the level II products are two semi-independent folded orbital lightcurves for the A, B, and C binaries (six folds in all), and two semi-independent time series for binaries A, B, and C (six in all).  See the third row of Table \ref{tbl:iterative} for the full set of second-level products.

The final step is to take all six time series and fold them about the orbital period of the single binary remaining in each one.  This yields two semi-independent pairs of phase folded orbital lightcurves, e.g., FC$_{a,III}$ and FC$_{b,III}$ for each binary.  Refer to the fourth row of Table \ref{tbl:iterative} for the set of final folded lightcurves.

\begin{table*}
\centering
\caption{Logical Tree for Iterative Disentanglement}
\begin{tabular}{lccc}
\hline
\hline
Level & Track a & Track b & Track c \\
\hline
0  &  TABC  &   TABC &  TABC   \\ 
I &  FA$_{a,I}$, TBC  &   FB$_{b,I}$, TAC & FC$_{c,I}$, TAB  \\ 
II & FB$_{a,II}$, TC; FC$_{a,II}$, TB & FA$_{b,II}$,  TC; FC$_{b,II}$, TA & FA$_{c,II}$, TB; FB$_{c,II}$, TA \\ 
III & FC$_{a,III}$ : FB$_{a,III}$ & FC$_{b,III}$ : FA$_{b,III}$ & FB$_{c,III}$ : FA$_{c,III}$    \\ 
\hline 
\label{tbl:iterative} 
\end{tabular}
\end{table*} 

We applied the complete iterative disentangling method to two different initial time-series. The first was for the original time-series obtained from the {\em TESS} data with the use of the {\tt Fitsh} pipeline of Andr\'as P\'al \citep{2012MNRAS.421.1825P}. Second, in order to reduce the non-physical scatter of the extracted lightcurves, we removed a 6-day-long section of the lightcurve between BJD 2458418.4 and 2458424.7 due to its large slope and, furthermore, we carried out a minor detrending operation with the software package of {\tt W\=otan} \citep{2019AJ....158..143H} to remove some additional, slight flux-level variations on a time-scale of 10-15 days. In this manner, the noise-level of the disentangled lightcurves was reduced significantly, without any changes in the structure. Therefore, for our analysis, we used the data series obtained from this slightly detrended second time-series.

\section{Archival data}
\label{sec:archival}
\begin{figure*}
    \centering
    \includegraphics[width=0.99\linewidth]{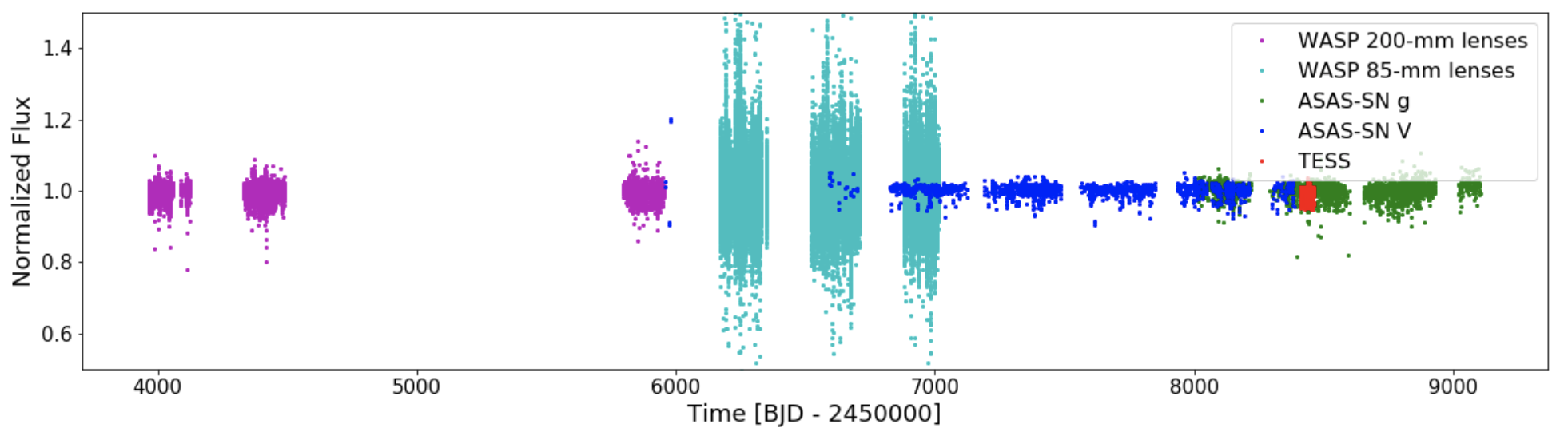}
    \caption{Archival data from ASAS-SN (blue and green symbols), WASP 200-mm lenses (magenta symbols), WASP 85-mm lenses (turquoise symbols), and {\em TESS} (red symbols) highlighting the baseline covered by  the photometry.}
    \label{fig:archival_ALL}
\end{figure*}

A search for archival data on TIC 168789840 reveals that there are a couple of rich sources of historical photometry.  Figure~\ref{fig:archival_ALL} highlights the baseline covered by the available archival observations of the target from ASAS-SN, WASP and {\em TESS}; the corresponding ephemerides for the three EBs are listed in Table \ref{tab:Fourier_ephemeris}. 

\subsection{All-Sky Automated Survey for Supernovae (ASAS-SN)}
\label{sec:asassn}
TIC 168789840 was observed by the ASAS-SN (\citealt{2014ApJ...788...48S}, \citealt{2017PASP..129j4502K}) Sky Patrol from BJD 2456000 to 2459100, with excellent coverage over the last 7 years.  In all, there are 4746 archival photometric data points available.  After renormalizing the green-band to the visual-band observations, we carried out a BLS \citep{bls} transform of the data to see which of the three binary EBs we could recover.  The top two highest peaks in the BLS transform were of the 1.570 day and 8.217 day periods (from binaries A and B, respectively).  We then used the Fourier cleaning tool described in Section \ref{sec:FT} to remove these two periods from the data.  The BLS transform of the cleaned ASAS-SN lightcurve then reveals the 1.306 day primary eclipses of the C binary.  In the top panels of Figure~\ref{fig:a_w_folds} we show the ASAS-SN data folded about the three periods determined from these data.  

\subsection{Wide-Angle Search for Planets (WASP)}
\label{sec:wasp}
The field of TIC 168789840 was observed by the WASP-South transit search between 2006 and 2014. WASP-South was an array of 8 cameras located in Sutherland, South Africa \citep{2006PASP..118.1407P}.  Between 2006 and 2011 the cameras used 200-mm, f/1.8 lenses, observing with a 400--700-nm filter and using a 48$\arcsec$ photometric extraction aperture. Between 2012 and 2014 the cameras had 85-mm, f/1.2 lenses with an SDSS-$r$ filter and a 112$\arcsec$ extraction aperture. TIC 168789840 is the brightest star in both-size apertures (the next brightest is 2.5 magnitudes fainter). Observations had a typical 12-min cadence, and where obtained on clear nights spanning 150 days in each of 2006, 2007, 2011, 2012, 2013 and 2014. A total of 126\,000 photometric data points were recorded.  However, we found that the S/N was better using only the 18,000 data points taken with the 200-mm lens.  

\begin{figure*}
  \centering
  \includegraphics[width=0.3\linewidth]{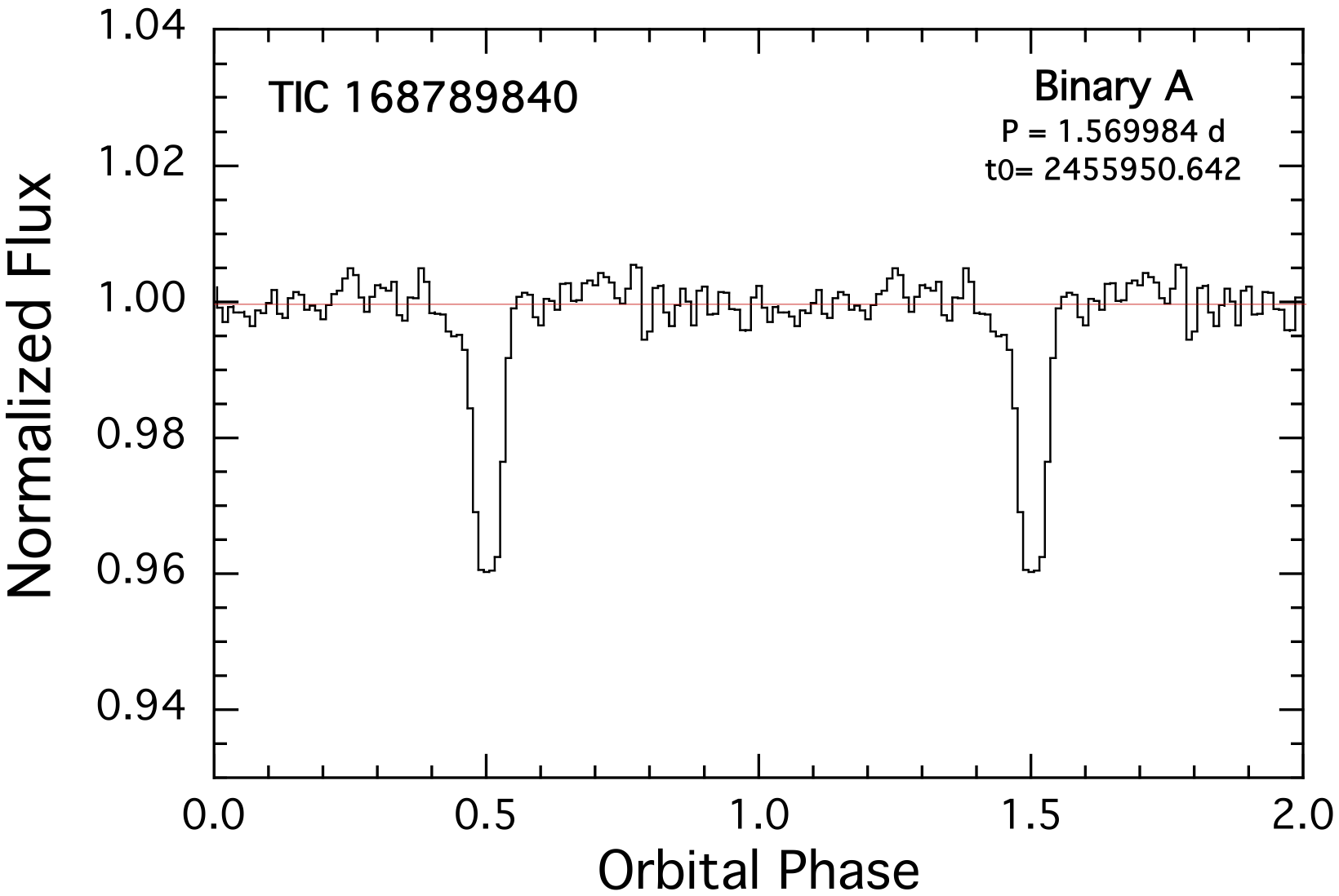}
  \includegraphics[width=0.3\linewidth]{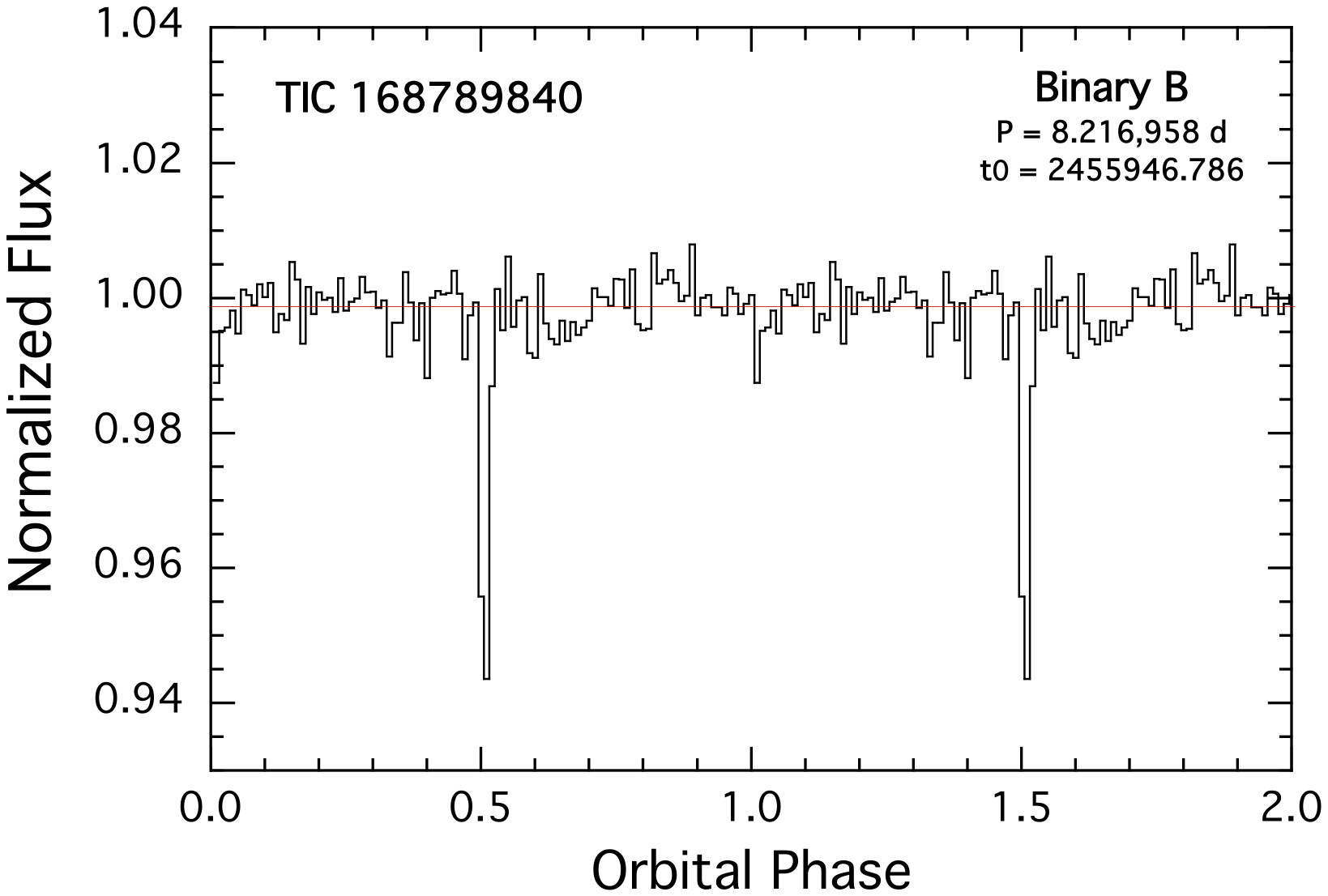}
  \includegraphics[width=0.3\linewidth]{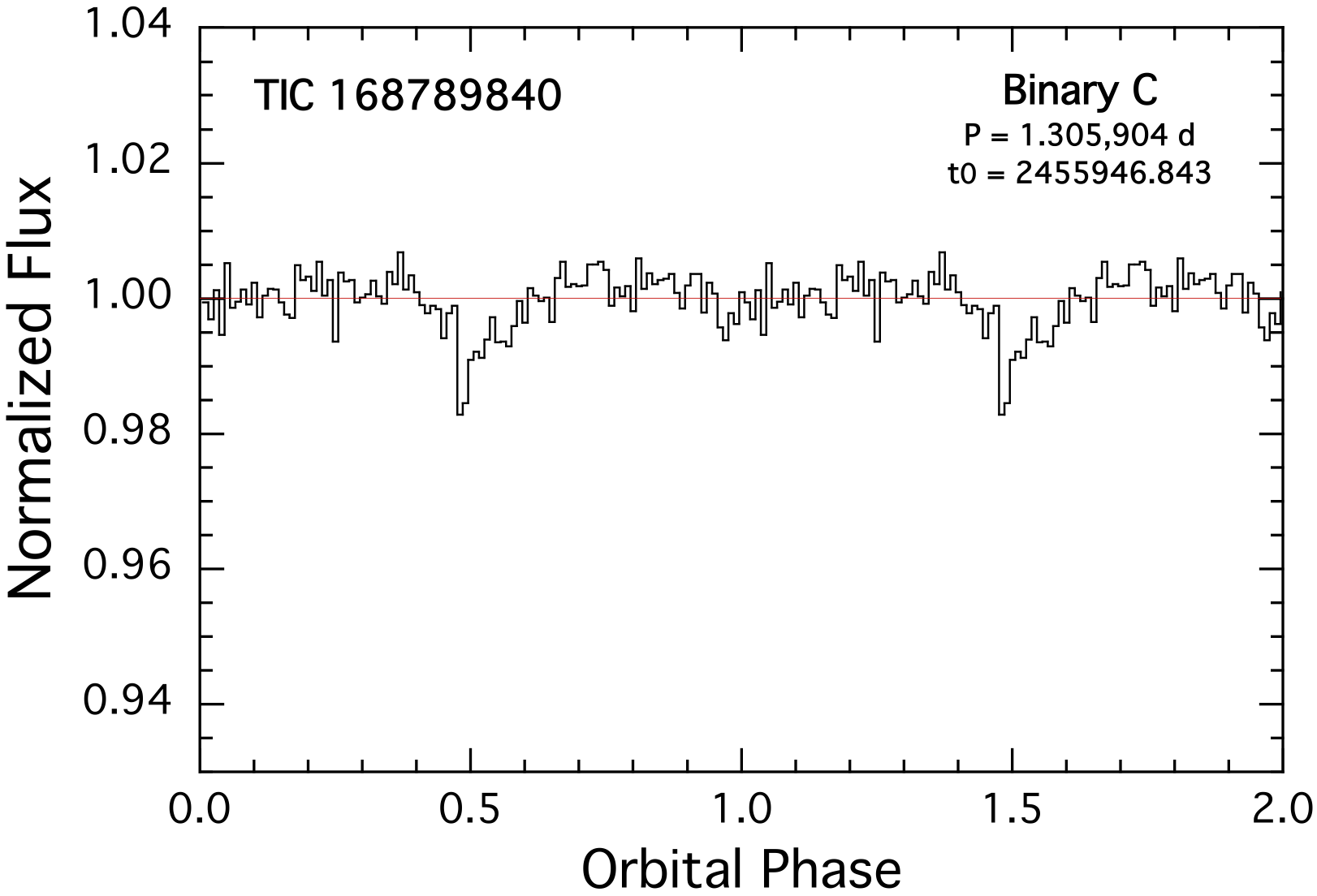}
  \includegraphics[width=0.3\linewidth]{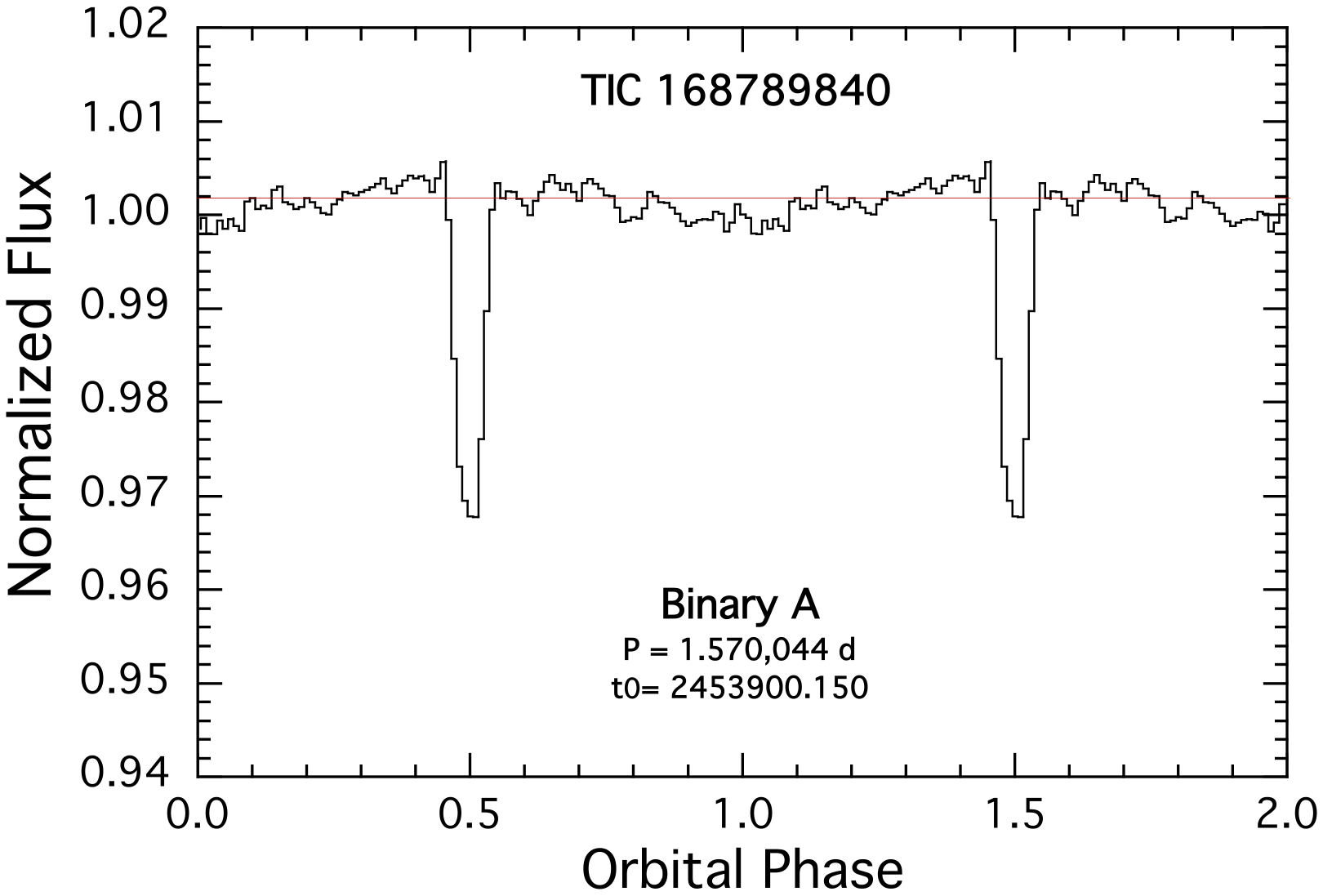}
  \includegraphics[width=0.3\linewidth]{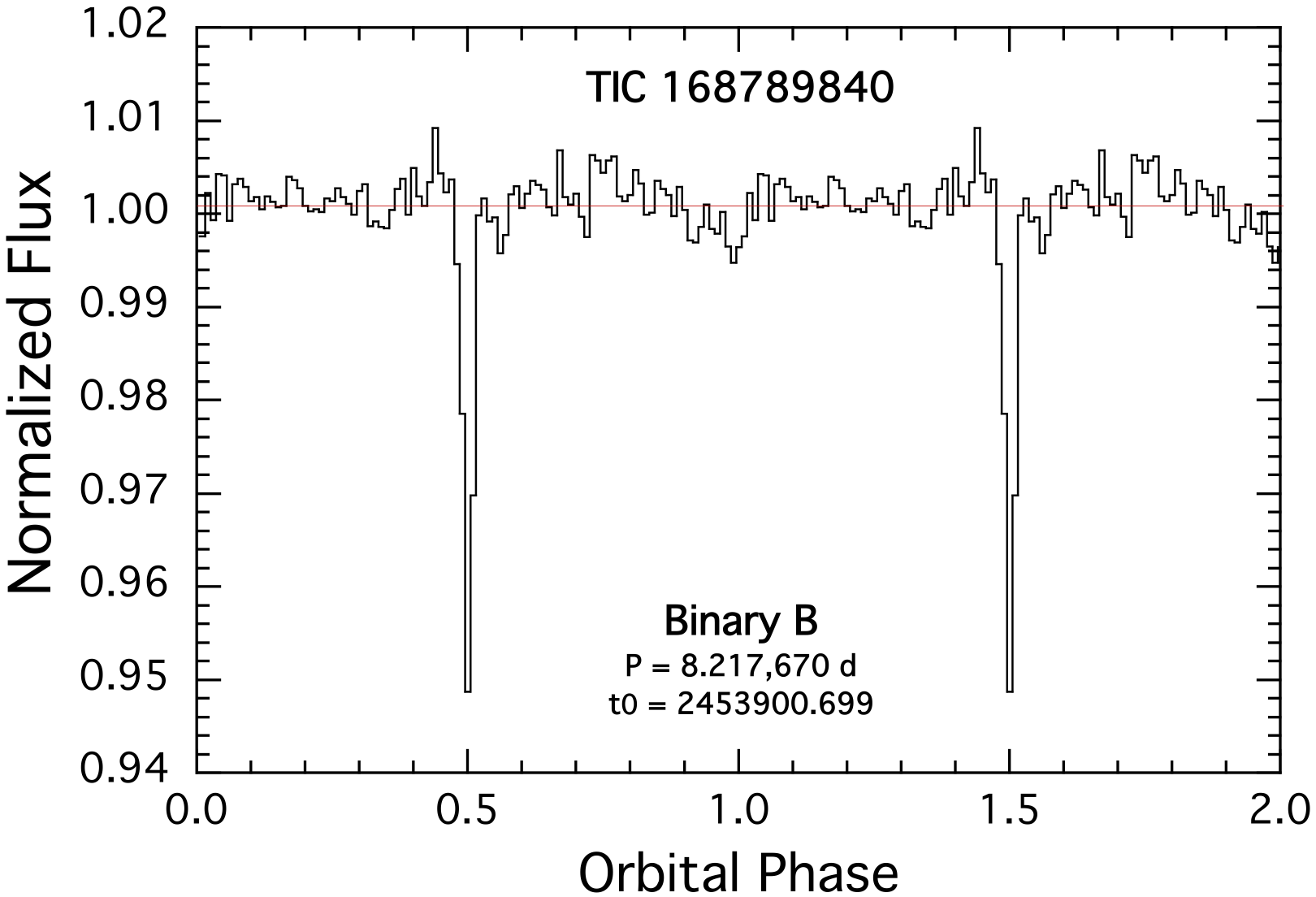}
  \includegraphics[width=0.3\linewidth]{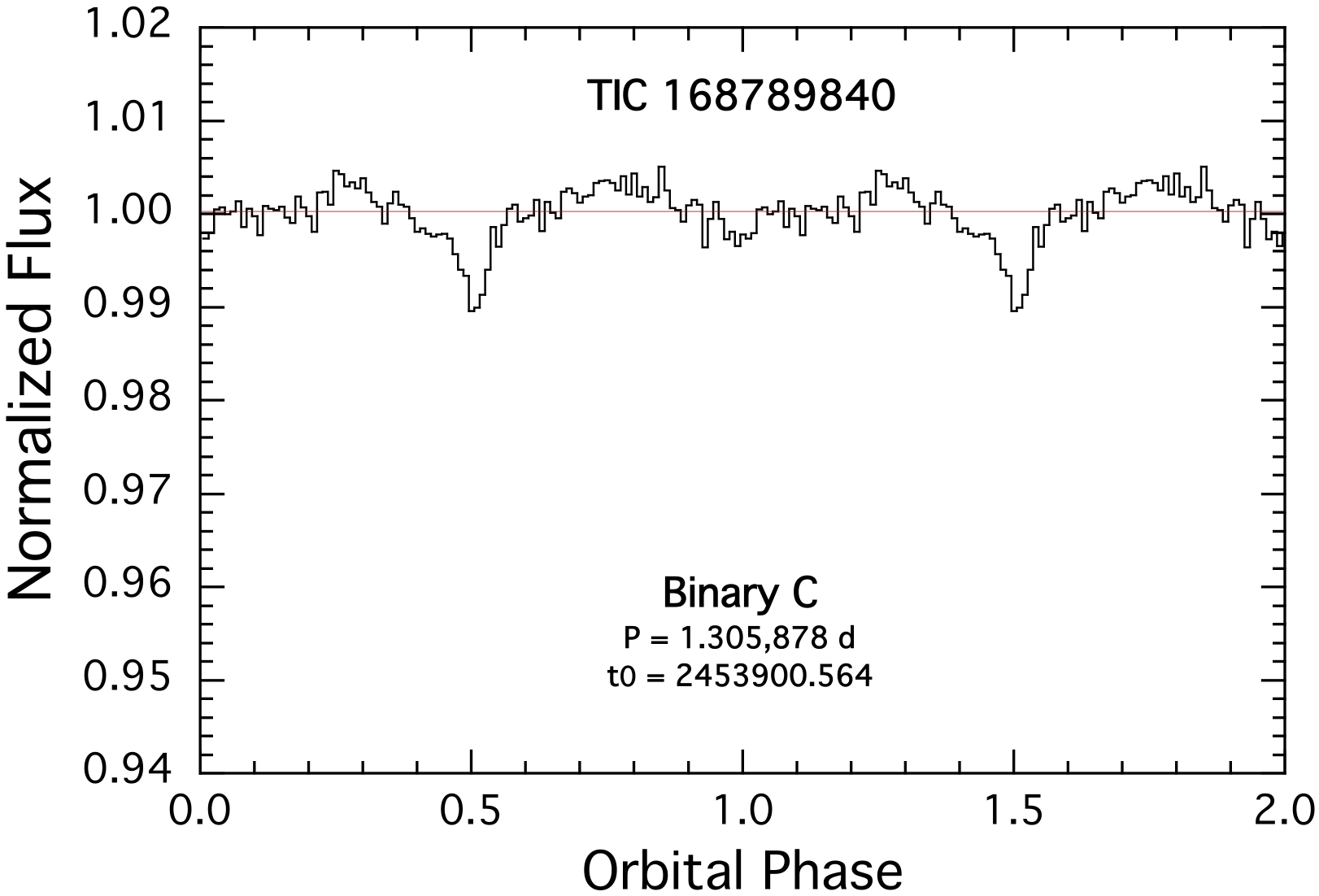}
\caption{Folds of the ASAS-SN (top) and WASP (bottom) archival data for the A, B, and C binaries about their respective orbital periods.  For the A and B folds, the raw data were used.  Before constructing the fold about the C period we removed the orbital profiles of the A and B binaries by Fourier filtering (see section \ref{sec:FT}).  For each binary we have written the fold period and the epoch of the primary eclipse on the plot.  The fold has been shifted by half an orbital period for aesthetic reasons.}
\label{fig:a_w_folds}
\end{figure*}

We analyzed the WASP data in the same manner that we did for the ASAS-SN data.  Again, the eclipses from the A and B binaries were the easiest to find.  We then cleaned the data of these two periods, and easily detected the eclipses of the C binary.  The bottom panels of Figure~\ref{fig:a_w_folds} show the WASP data folded in the same manner as the ASAS-SN data.

\begin{table*}
\centering \caption{Derived ephemerides for the three EBs in TIC 168789840 from ASAS-SN, {\em TESS} and WASP.}
 \label{tab:Fourier_ephemeris}
 \begin{tabular}{c|cc|cc|cc}
 \hline \hline
Binary               & \multicolumn{2}{c|}{A}       & \multicolumn{2}{c|}{B}        & \multicolumn{2}{c}{C}       \\
\hline
${\bf TESS}$ & & & & & & \\
Period [days] & \multicolumn{2}{c|}{1.570101} &  \multicolumn{2}{c|}{8.217173}& \multicolumn{2}{c}{1.305934} \\
T0 [BJD - 2450000] &  \multicolumn{2}{c|}{8412.3855} &  \multicolumn{2}{c|}{8411.9008} &  \multicolumn{2}{c}{ 8413.6822} \\
\hline
${\bf ASAS-SN}$ & & & & & & \\
Period [days] & \multicolumn{2}{c|}{1.569984} &  \multicolumn{2}{c|}{8.216958}& \multicolumn{2}{c}{1.305904} \\
T0 [BJD - 2450000] & \multicolumn{2}{c|}{5950.642} & \multicolumn{2}{c|}{5946.786} & \multicolumn{2}{c}{5946.843} \\
\hline
${\bf WASP}$ & & & & & & \\
Period [days] & \multicolumn{2}{c|}{1.570044} & \multicolumn{2}{c|}{8.217670}& \multicolumn{2}{c}{1.305878} \\
T0 [BJD - 2450000] & \multicolumn{2}{c|}{3900.150} & \multicolumn{2}{c|}{3900.699} &  \multicolumn{2}{c}{3900.564} \\
\hline
{\bf Radial Velocities} & \multicolumn{2}{c|}{fixed} & \multicolumn{2}{c|}{fixed} & \multicolumn{2}{c}{fixed} \\

T0 [BJD - 2450000] & \multicolumn{2}{c|}{9151.868} & \multicolumn{2}{c|}{9151.446} &  \multicolumn{2}{c}{ 9151.193} \\
\hline
\hline
Global Fitted Periods$^a$ & \multicolumn{2}{c|}{ 1.570013(9)} & \multicolumn{2}{c|}{8.217111(30)} &  \multicolumn{2}{c}{ 1.305883(6)} \\
\hline
\hline
\end{tabular}

{Notes. (a) The long-term average period is determined from a linear fit to the four independently determined times of eclipse.  In the case of binary B this assumes no change in its center-of-mass velocity over the past 15 years.  In the case of binaries A and C, which we later show to be in a $\sim$3.7 year quadruple orbit, with speeds of $\sim$7 km s$^{-1}$, this could lead to effects as large as 23 parts per million in the reported period. But, much of the latter is averaged over in the WASP and ASAS-SN measurements which span the $\sim$3.7 year orbit. }

\end{table*} 

\section{Follow-up observations}
\label{sec:follow-up}
Upon identification of the system, we had overwhelming support from follow-up observers providing nearly fifty separate measurements from seven different observatories.  These range from photometric measurements to radial velocity and speckle imaging, each helping us to further unravel the nature of the system. 

\subsection{Photometric measurements}

\subsubsection{{\em TESS} Followup Observing Program}

\begin{figure}
  \centering
  \includegraphics[width=0.8\linewidth]{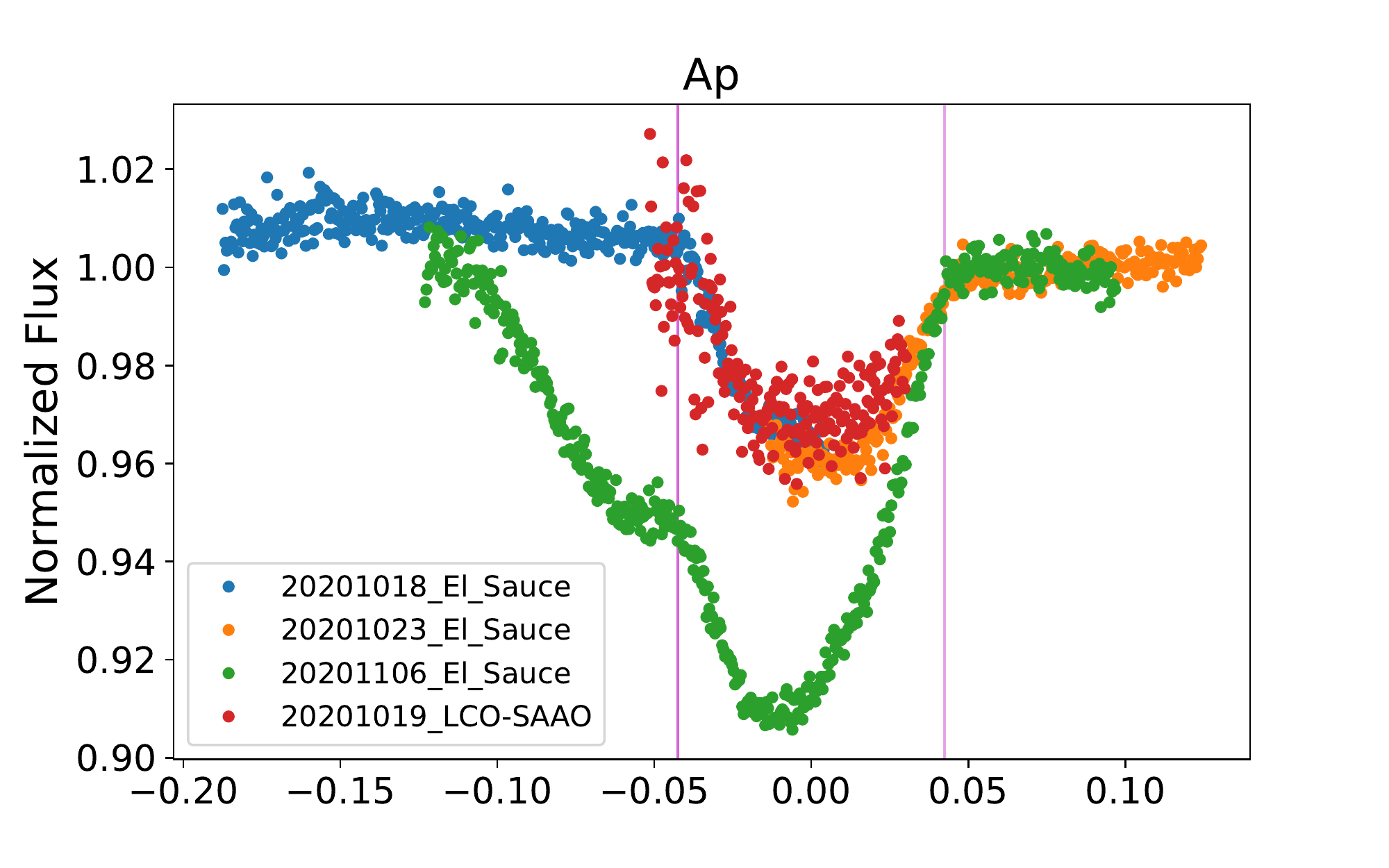}
  \includegraphics[width=0.8\linewidth]{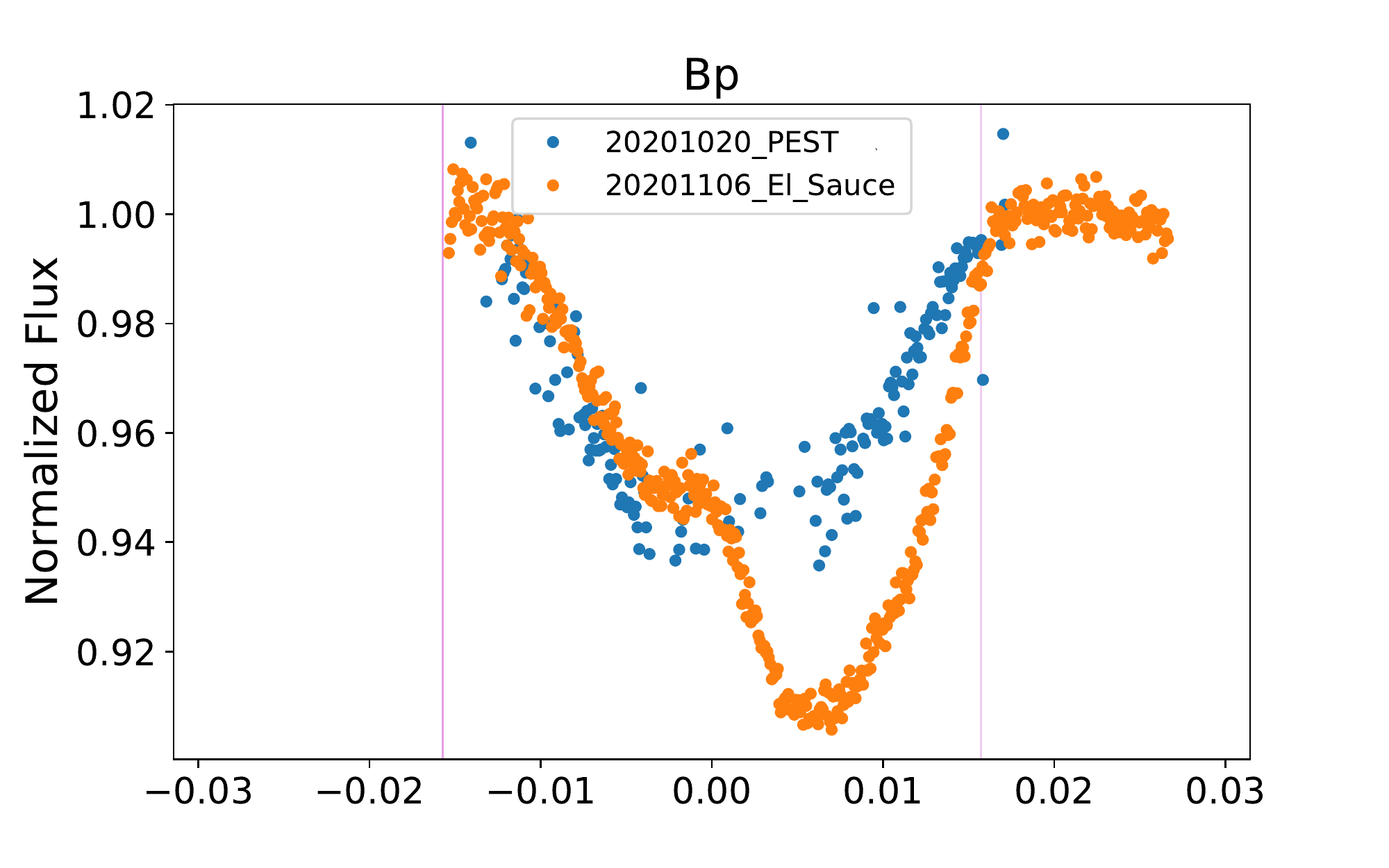}
  \includegraphics[width=0.8\linewidth]{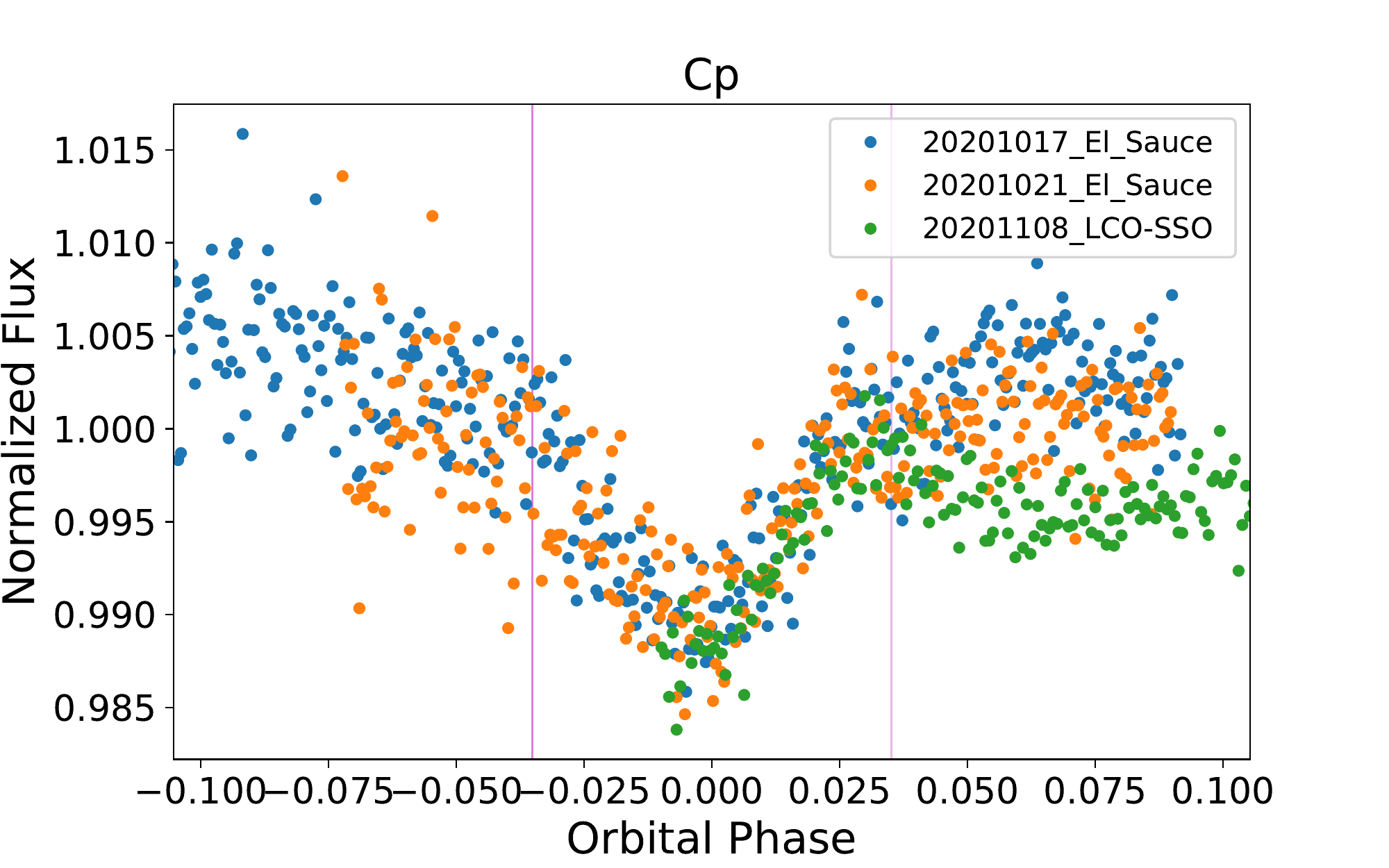}
      \caption{TFOP-led photometric observations of TIC 168789840 confirming that the target is the source of the eclipses detected in {\em TESS} data. Primary eclipses of A, B, and C are shown in the three panels. The vertical red bands represent the corresponding ingress and egress times. The individual measurements are vertically offset for clarity. Some observations cover two eclipses (see text).}
    \label{fig:tfop}
\end{figure}

Photometric follow-up observations were performed through Subgroup 1 of the {\em TESS} Follow Up Observing Program (TFOP) as described in more detail below. We used the {\tt TESS Transit Finder}, which is a customized version of the {\tt Tapir} software package \citep{Jensen:2013}, to schedule our transit observations. These observations, shown in Figure \ref{fig:tfop}, confirm that the target is the source of the different sets of eclipses detected in {\em TESS} data and rule out contamination from nearby sources.  Several of the observations shown in Figure \ref{fig:tfop}, while targeted at one particular eclipse of a given binary, simultaneously observed eclipses from either of the other two binaries.

TIC 168789840 was observed on nine nights with the Evans telescope at El Sauce in Coquimbo Province, Chile.  This system consists of a 0.36-m CDK telescope with a SBIG STT1603-3 CCD, which has an image scale of 1$\farcs$47 pixel$^{-1}$ and $18\farcm8 \times 12\farcm5$ field of view; all observations used the $R_C$ filter.  The observations covered the following UTC dates and eclipses: 2020-10-07 (C primary); 2020-10-18 (A primary);  2020-10-19 (A secondary, C secondary);  2020-10-21 (C primary);  2020-10-22 (A secondary);  2020-10-23 (A primary, C secondary);  2020-11-02 (A secondary);  2020-11-06 (A primary, B primary); and 2020-11-10 (A secondary, B secondary).  The photometric data were extracted using the {\tt AstroImageJ} ({\tt AIJ}) software package \citep{Collins:2017}.

The Perth Exoplanet Survey Telescope (PEST) is a 0.3 m telescope in Perth, Australia, with  an image scale of 1$\farcs$2 and a $31\arcmin \times 21\arcmin$ field of view.  PEST observed in the $R_C$ filter on UTC 2020-10-20, covering the B primary and A secondary eclipses.  A custom pipeline based on {\tt C-Munipack} \citep{c-munipack} was used to calibrate the images and extract the differential photometry.

Two observations made use of the Las Cumbres Observatory Global Telescope (LCOGT) network \citep{Brown:2013}.  Primary eclipses of both the A and C binaries were observed on UTC 2020-10-19 using a 0.4-m telescope in Sutherland, South Africa.  The LCOGT 0.4-m telescopes are equipped with $2048\times3072$ SBIG STX6303 cameras having an image scale of 0$\farcs$57 pixel$^{-1}$ resulting in a $19\arcmin\times29\arcmin$ field of view.  On 2020-11-08, one of the LCOGT 1.0-m telescopes at Siding Spring Observatory observed this system, covering the C primary and A secondary eclipses.   The $4096\times4096$ LCOGT SINISTRO cameras have an image scale of $0\farcs389$ per pixel, resulting in a $26\arcmin\times26\arcmin$ field of view.  The LCOGT images were calibrated by the standard LCOGT BANZAI pipeline \citep{McCully:2018} and the photometric data were extracted using {\tt AIJ}.

\subsubsection{FRAM}
\begin{figure}
\centering
    \includegraphics[width=1.\linewidth]{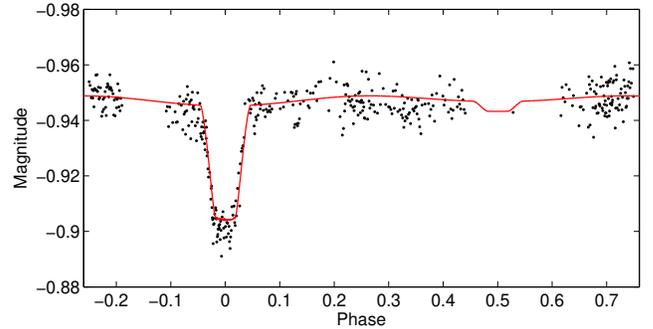}
    \caption{Phase folded lightcurve of ground-based FRAM data showing binary A (filter $R$ was used) plotted against the PHOEBE fit.}
    \label{fig:fram}
\end{figure}

Some follow-up photometric data from ground based observatories were also obtained with a small 30-cm telescope FRAM. It is the Orion ODK 300/2040mm, equipped with the CCD camera MII G4-16000. All observations carried out in standard $R$ filter. The FRAM telescope itself \citep{2019EPJWC.19702008J} is located at the peak of Los Leones, near the town of Malarg\"ue, at the Pierre Auger Observatory, Argentina.  The phase fold of ten separate observations is shown in Figure~\ref{fig:fram}.

\subsection{Spectroscopy}
\label{sec:spec}

\subsubsection{CHIRON}
\label{sec:chiron}

Eight high-resolution optical spectra of TIC 168789840 were taken with the CHIRON fiber echelle spectrometer \citep{chiron} at the CTIO 1.5 m telescope operated by the Small \& Moderate Aperture Research Telescope System (SMARTS) consortium between 2020 November 6 and 21. The spectral resolution is 80,000, exposure time 15 min., and the typical S/N is $\sim$20 per pixel (pixel width $\sim$1.2 km s$^{-1}$). The wavelength calibration is determined from the ThAr spectra taken immediately after the stellar spectra. 

\begin{figure}
  \hglue-0.2cm  \includegraphics[scale=.55]{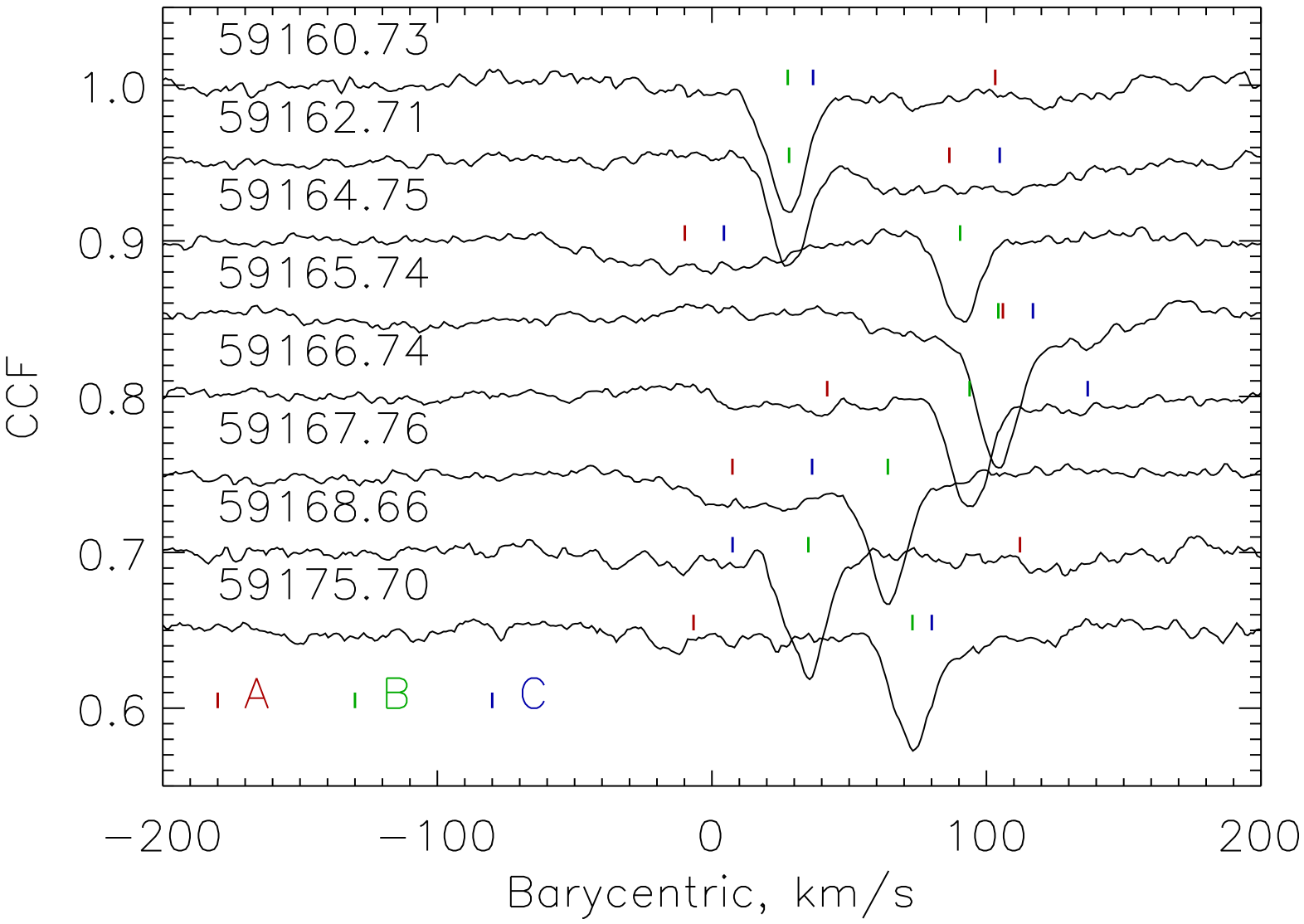}
 \hglue-0.22cm  \includegraphics[scale=.55]{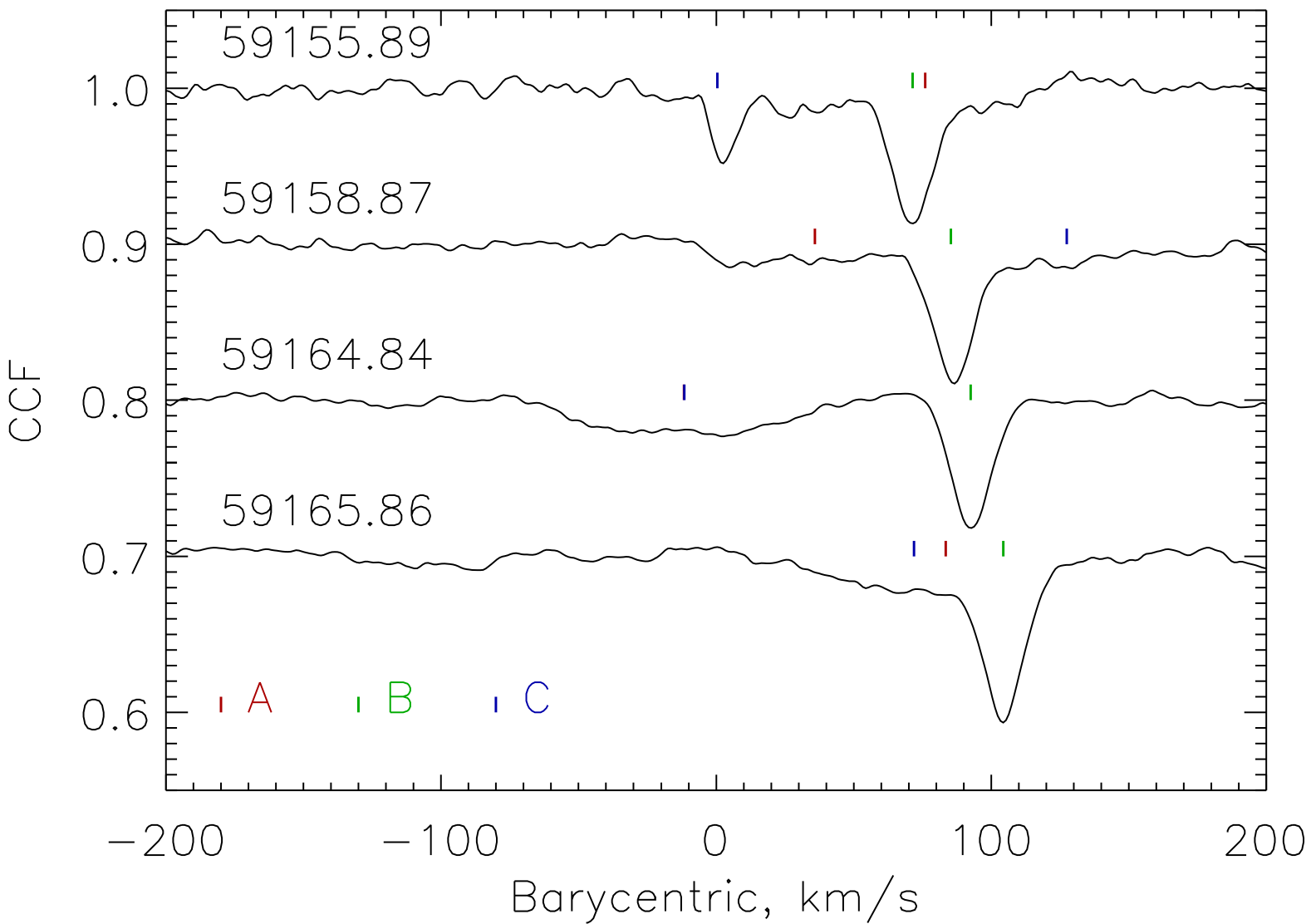}
    \caption{Top panel: Cross-correlation functions (CCFs) of the 8 CHIRON spectra with a binary mask. Bottom panel: Same for the 4 TRES spectra.  The Julian dates of observations are indicated. The RVs of the components corresponding to their orbits are marked by the colored ticks above each CCF. The dip at zero RV in the first TRES spectrum is produced by Moon contamination.  
    \label{fig:chironccf} }
\end{figure}

To find the radial velocities (RVs), the spectra were cross-correlated with a binary mask based on the solar spectrum. Details of this procedure are provided in \citep{2016AJ....152...11T}. Only wavelengths from 480\,nm to 650\,nm are used.  The cross-correlation functions (CCFs) show a narrow dip with an amplitude of 0.07-0.08 and an rms width of 7 km s$^{-1}$ that corresponds to the projected rotation speed of 10.2 km s$^{-1}$,  see Figure~\ref{fig:chironccf}. Moreover, there are broad features resulting from other components with fast rotation. Analysis of  CHIRON spectra shows convincingly that the narrow dip belongs to the primary component of the 8-day pair, B1. A circular spectroscopic orbit  fits both CHIRON and TRES RVs. The RVs of the rapid rotators A1 and C1 are derived by modeling the spectrum (Section \ref{sec:spectrum}).

\subsubsection{Tillinghast Reflector Echelle Spectrograph (TRES)}
\label{sec:TRES}

Four additional spectroscopic observations were made with the Tillinghast Reflector Echelle Spectrograph (TRES; \citealt{2007RMxAC..28..129S}; \citealt{Furesz:2008}) attached to the 1.5 m telescope at the Fred Lawrence Whipple Observatory on Mount Hopkins.  The  wavelength range 3900--9100~\AA\ is covered in 51 orders at a resolving power of 44,000.  The observations were made on November 2, 5, 11, and 12, 2020.  Each spectrum is a combination of three 15-min.~exposures. The flux for each exposure is about 400 photons per  pixel. Small size of the input apertures (fibers) in TRES and CHIRON rules out potential contamination from unresolved sources within a single {\em TESS} pixel. 

\subsection{Spectral Analysis}
\label{sec:spectrum}

We have used the 12 spectra taken with CHIRON and TRES to extract the RVs of the primary stars in the three EBs of the system.  The RVs of the sharp-lined primary of pair B are derived by the standard method, i.e., by cross-correlation of the spectra with a binary mask (CHIRON) or with a template (TRES) and fitting the resulting CCF. However, owing to the blending and rapid axial rotation, the CCFs are not suitable for measuring the RVs of the primaries in binaries A and C; instead, a different approach is needed based on modeling the observed spectra. The light-curve analysis indicates that the secondary components in all three eclipsing pairs are much fainter than the primaries. Therefore, we assume that the contribution  of all secondaries to the spectrum is negligible and model it as a sum of three spectra of the primaries. The light-curve analysis indicates that the fluxes of all primaries in the {\em TESS} band are comparable. 

\begin{figure*}
\centerline{
\includegraphics[width=8.5cm]{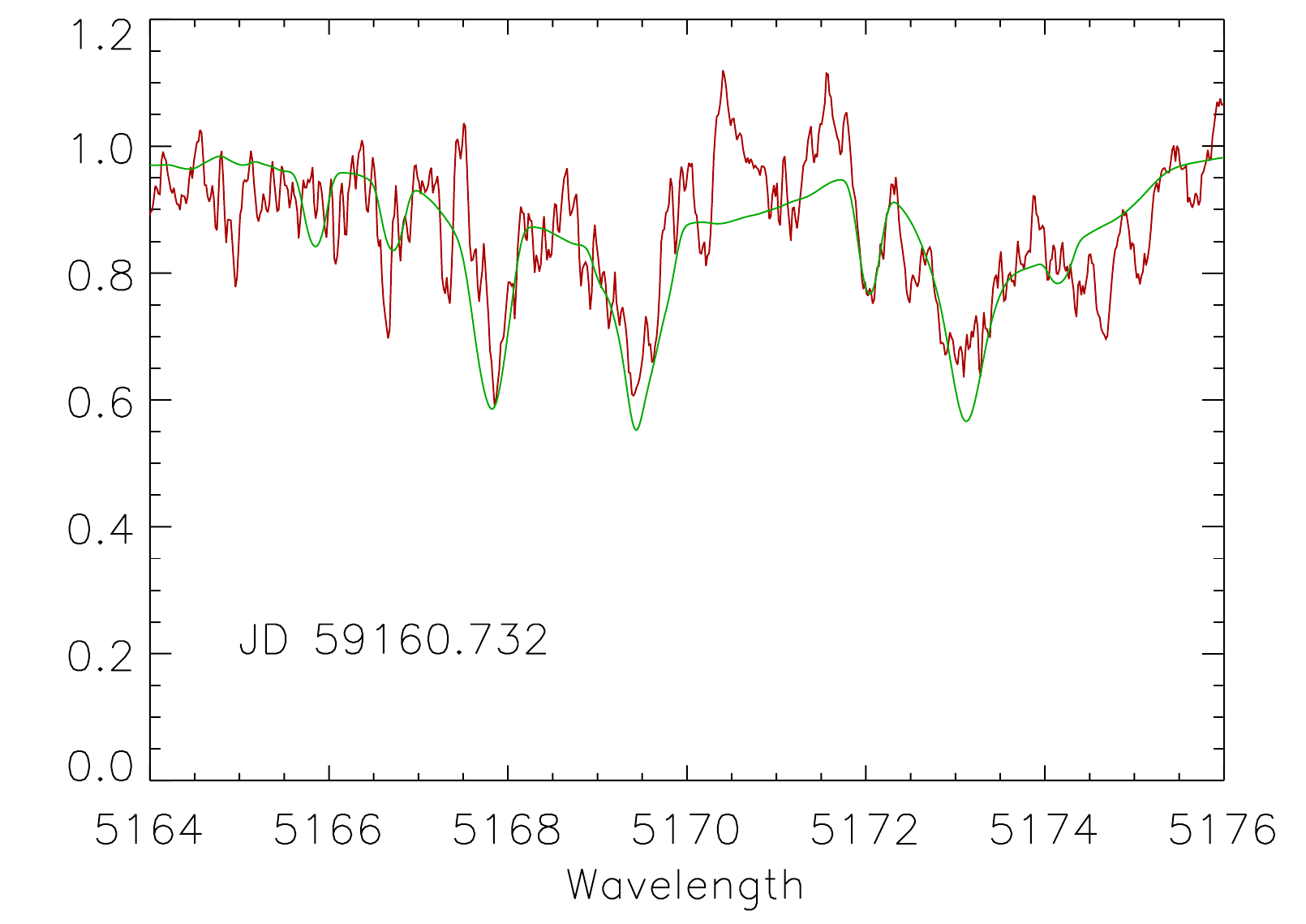} 
\includegraphics[width=8.5cm]{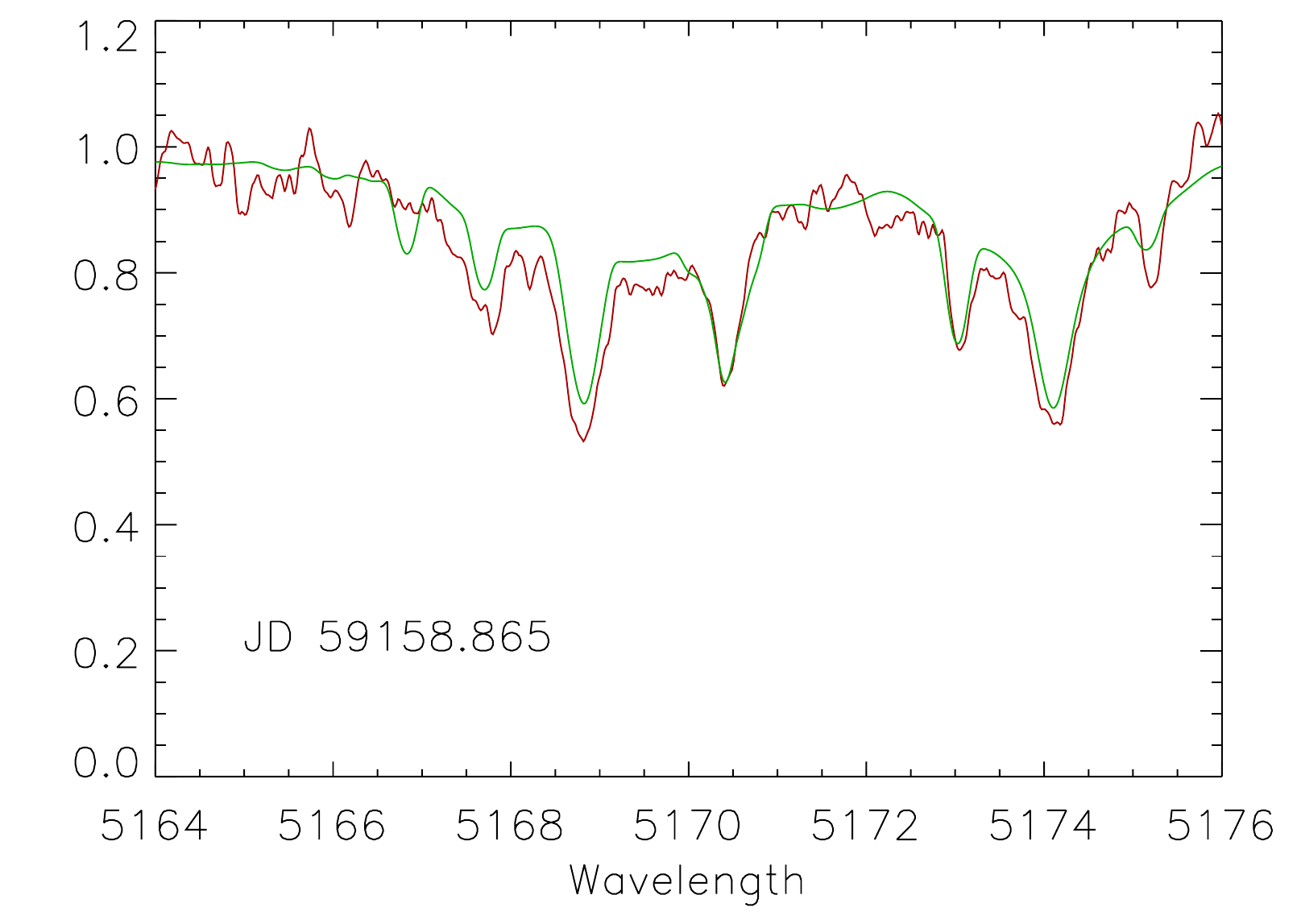} 
}
\caption{The observed spectra with a 10-pixel smoothing (red) are compared to their models (green) in the region around the Mg Ib triplet. Left: CHIRON spectrum, right: TRES spectrum.  \label{fig:modsp} }
\end{figure*}  

 The RVs  of the  narrow B1 dip are  determined by the standard procedure (approximation by  a Gaussian function).  These RVs (as well as the 4 RVs from TRES) correspond to the  circular orbit of B presented below. 

For modeling the spectra, we use the stellar parameters determined from the MCMC analysis  (see Section \ref{sec:sysparms}).  Assuming synchronous stellar rotation with their respective orbits and zero obliquity, we compute equatorial velocities of 48, 10.2, and 58 km s$^{-1}$, which approximate the projected velocities because these binaries are eclipsing. 

The orders of the echelle  spectra were merged together and normalized by  the continuum.  The merging  procedure is  not perfect and leaves residual  waves in the continuum in  the  area where the orders overlap. This minor defect is neglected here. The  merged spectrum is constructed on a logarithmic wavelength grid with a  step of 1 km s$^{-1}$ ranging  from 480\,nm  to 650\,nm.  The normalized  spectrum  was also correlated  with the same solar mask in the wider $\pm$400 km s$^{-1}$ range. These ``wide'' CCFs are slightly sub-optimal in comparison with the standard order-by-order CCFs in terms of the photon noise.  However, the order-merged and normalized spectrum is needed for modeling. The TRES  spectra were transformed to the same order-merged format using the wavelength calibration provided in the headers.

\begin{figure*}
\centerline{
\includegraphics[width=8.5cm]{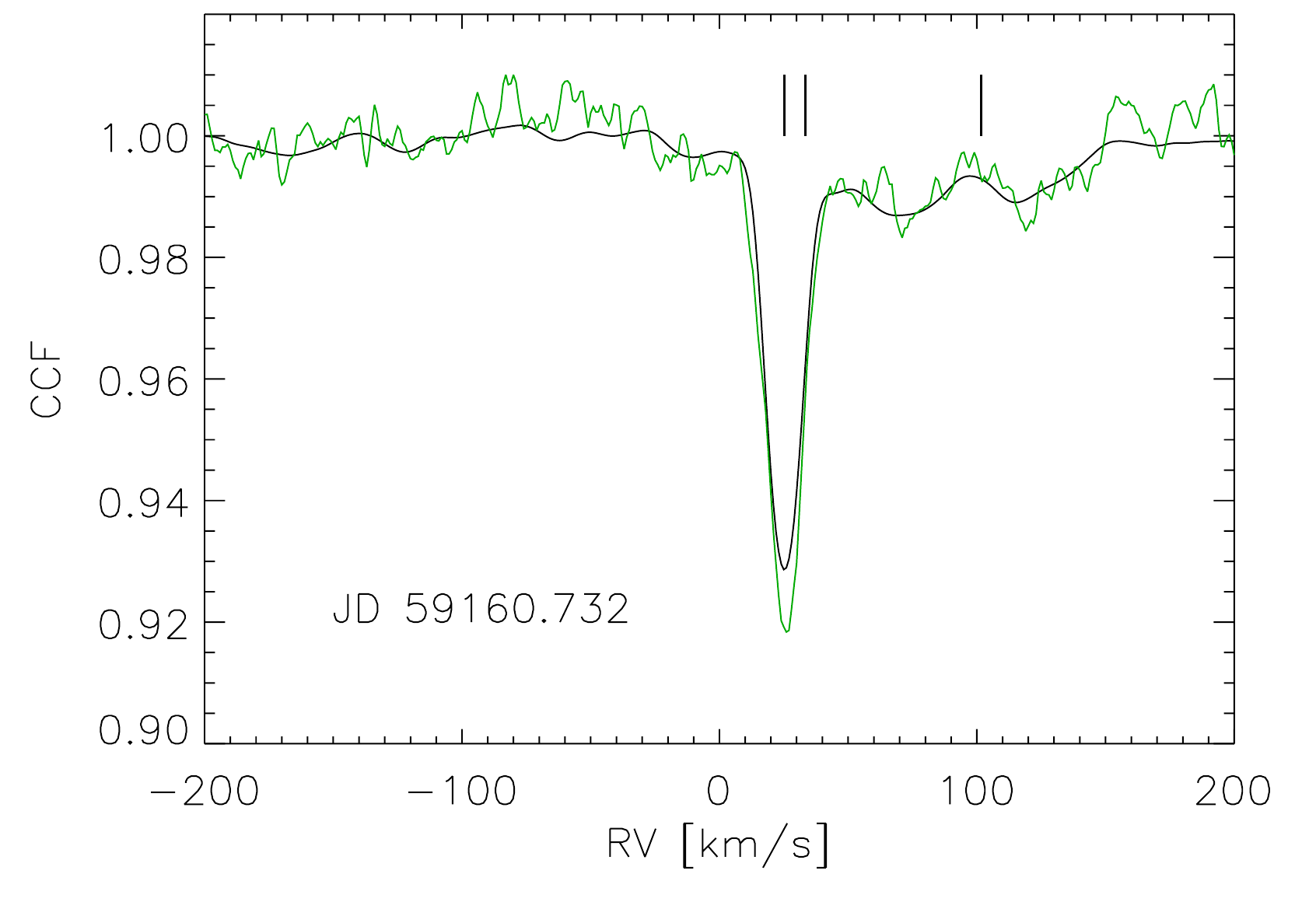} 
\includegraphics[width=8.5cm]{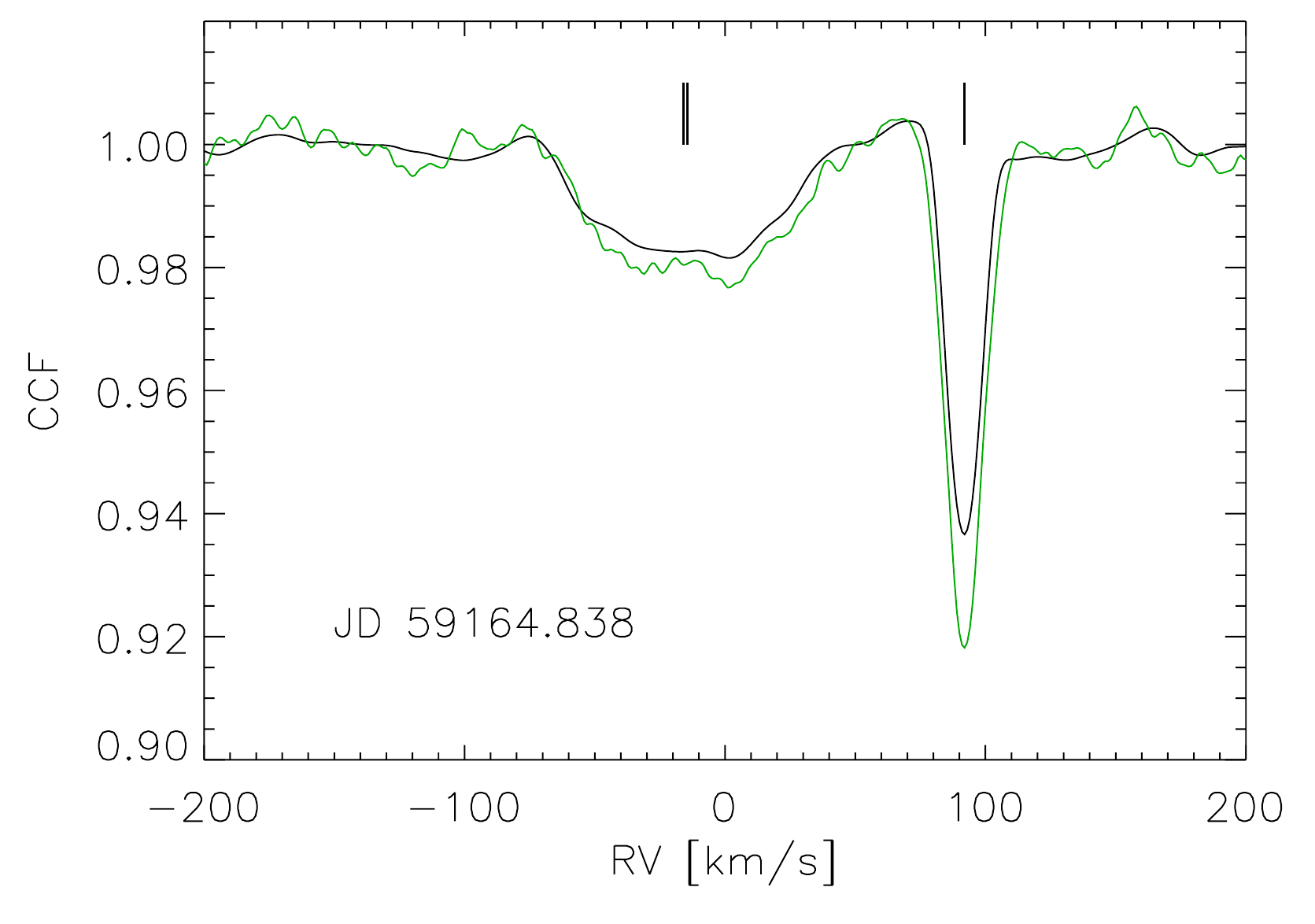} 
}
\caption{Comparison between the CCFs computed for the real (green) and modeled (black) CHIRON  spectra. Both  CCFs are  computed  independently, not fitted. The  vertical lines  mark  the RVs  of the  components  assumed  in the  model.  \label{fig:modccf} }
\end{figure*}  

We use as a template the synthetic spectrum from the POLLUX library \citep{2012ASInC...6...63P}  with  $T_{\rm  eff} =  6200$\,K,  $\log g=4.0$, and [Fe/H]  = $-0.5$.  The solar-metallicity template, chosen initially, has lines deeper than observed. The template is rotationally broadened using  the calculated equatorial velocities. The instrumental broadening is also included, but in the context of the present study it is negligible. Rotational broadening assumes a linear limb-darkening coefficient of 0.68 (solar value).

The orbital parameters of the primary stars in A and C were initially determined using the masses of the components from the MCMC system analysis (Section \ref{sec:sysparms}) and then iteratively improved; the orbit of  B is well defined, and its RVs are assumed to  be accurately  known. Initially, we  also assumed that the relative contributions of all stars to the  spectrum are equal, but then refined the relative fluxes to A:B:C=0.3:0.4:0.3, making B the slightly brighter star.  As we see below (Section \ref{sec:soar}), the two components resolved by  speckle interferometry contain binaries AC and B, respectively.  The magnitude difference in the $I$ band of 0.27 mag, implies that the flux ratio AC:B=0.56:0.44, so B could be even a little brighter than assumed here.

The fitting program  selects one of the observed spectra and compares it to the model. The templates of the three stars are shifted by their respective RVs (known from the orbital elements) and by the barycentric correction and summed in proportion to the assumed relative fluxes. Figure~\ref{fig:modsp} shows two examples comparing models to the observed spectrum. They match qualitatively. Despite the smoothing, the observed spectra are noisy. A better  assessment of the model  is obtained by  correlating it with the same solar mask and  comparing the observed and modeled CCFs. This is  illustrated  in Figure~\ref{fig:modccf}  for  two  dates. The  first spectrum, taken on JD 2459160, did  not have a well-defined broad dip in the CCF because A1 and C1 had  different RVs. On JD 2459164, the dips of A1 and C1 overlapped, producing a  clear signature in the CCF. Note that the dip amplitude of B1 apparently differs from the model by a variable factor.  This could be caused by the fact that the star is a 0\farcs4 visual binary,  so the components  can be mixed in  slightly different proportions, depending on the guiding. The CCF outside the dips is not constant;  it varies  owing  to random  coincidences between  spectral lines  and  mask. To  some  extent,  this variation  is captured by the modeled CCF.

So far, the model uses the pre-computed RVs, without any fitting. Taking these RVs as the initial guess, we fit the RVs of A1 and C1 to minimize the sum of squares between the observed spectrum and its model. Fitting of the two parameters is done using the {\tt amoeba} minimizer \citep{press}. The RVs of B1 and other parameters (flux ratios, rotation speeds) are assumed known. A  version of  the code fitting all 3 RVs gives for B  the same results as fitting the CCF dips. We also tried to fit 4 parameters, including the  relative flux of B. Its  best-fit value of 0.36 is  found consistently (the rms  scatter is 0.01) on all  dates except 59164, when {\tt amoeba} converged slowly and  the best ratio returned by the code is 0.30.

The RVs of A and C are used to determine their orbital parameters which, in turn, are used as the initial guess in further work. Depending on the details (initial guess, fitting tolerance), the resulting RVs of A and C may differ by $\sim$1 km s$^{-1}$, except the dates where the dips of A and C strongly overlap and the differences may be larger.

\begin{figure*}
\includegraphics[width=0.318\linewidth]{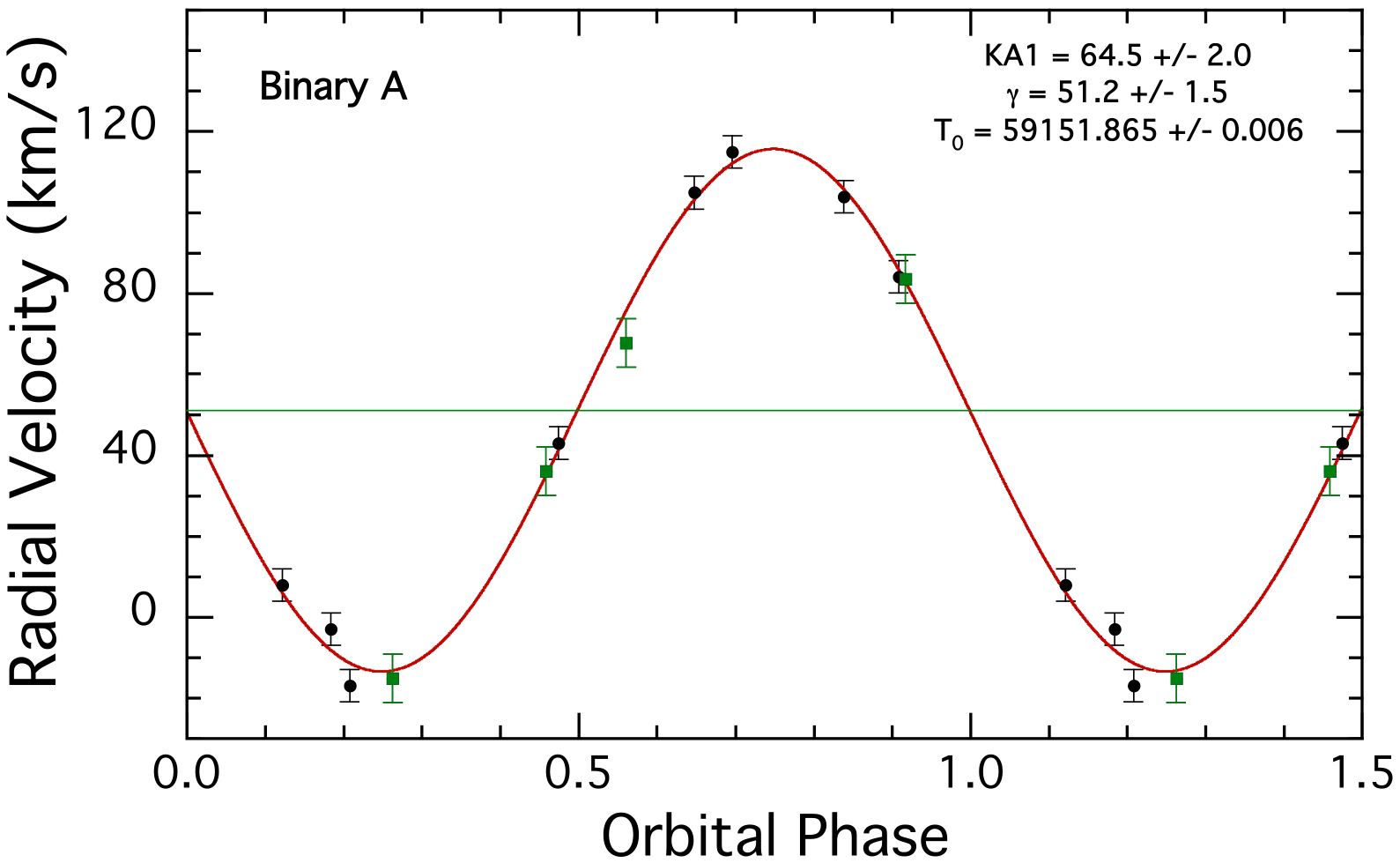} 
\includegraphics[width=0.3\linewidth]{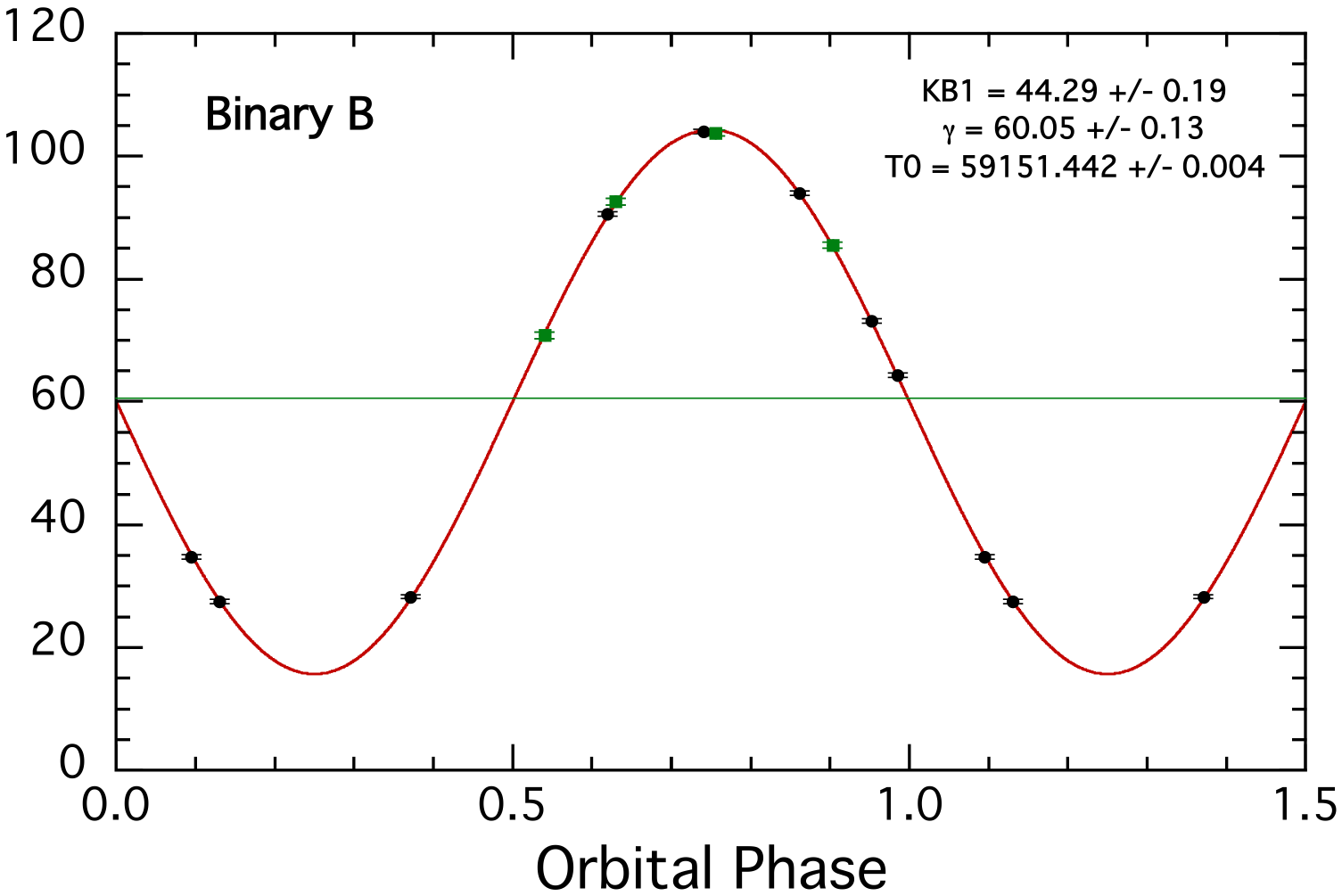} 
\includegraphics[width=0.29\linewidth]{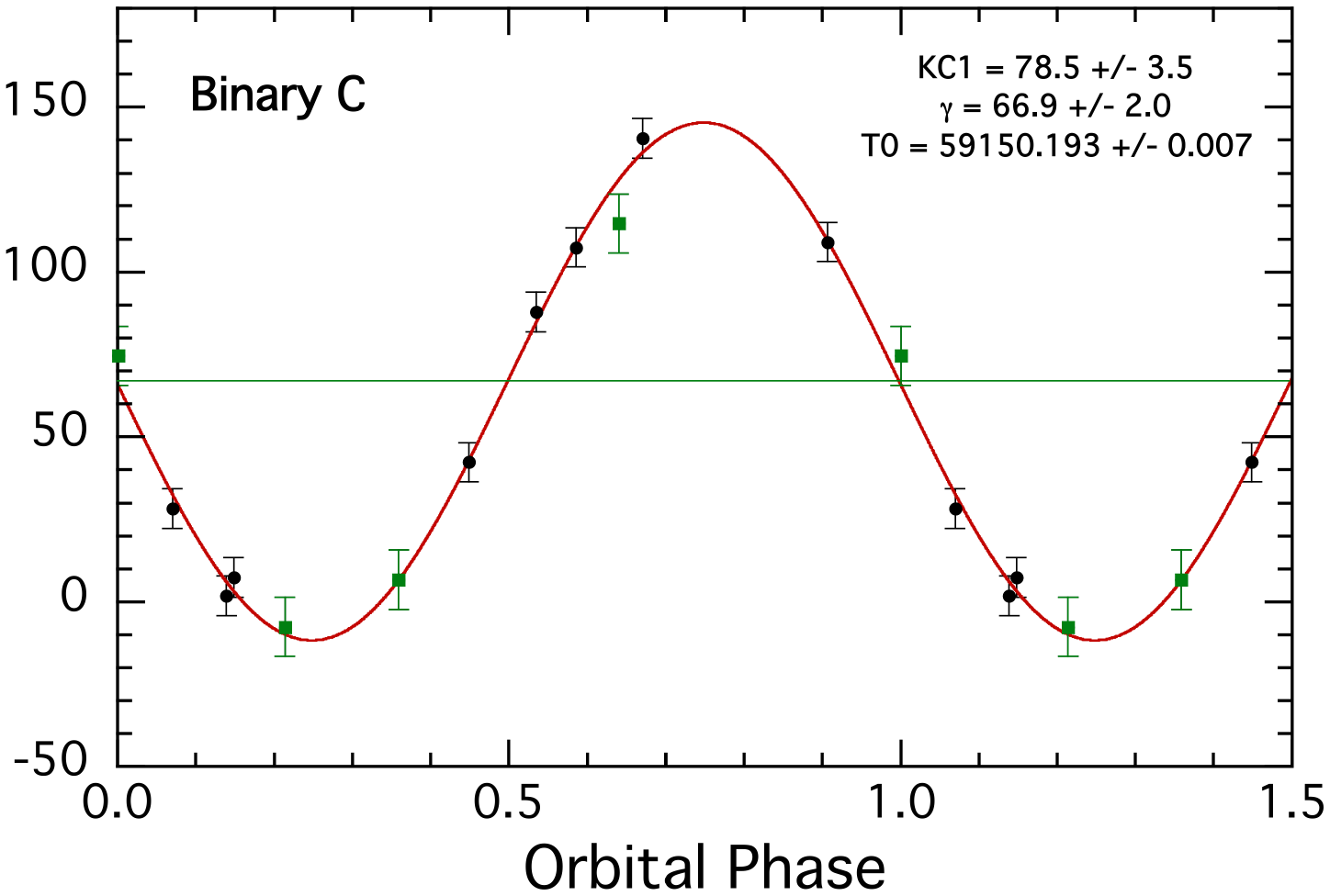} 
\caption{Spectroscopic orbits. CHIRON RVs are plotted as black circles while the TRES RVs are indicated with green squares.}
 \label{fig:orb} 
\end{figure*}  

Minimization of  the quadratic distance  between the spectrum  and its model  is  mathematically  equivalent  to  maximizing  their  product, i.e., the cross-correlation. Owing to the artefacts of the spectrum and the  intrinsic  mismatch  between  real  spectra  and  templates,  the residuals are  much larger than  the statistical errors. For  the same reason,  estimation   of  the  errors  of  the   derived  RVs  appears problematic.


Table~\ref{tab:rv} lists the RVs of  all 3 components derived from the CHIRON and TRES spectra. The RVs of B1 (middle column) come  from direct fitting of the  CCF dip (CHIRON) or the CCF  with a non-rotating  template (TRES).  The RVs of  A1 and  C1 are determined by the spectrum modeling  described above. The RVs of A1 and C1 on BJD 59164, when their dips overlap, are less certain.

\begin{table}
\center
\caption{Radial Velocities}
\label{tab:rv}
\begin{tabular}{l   c c c} 
\hline
\hline
BJD        &     A1    &   B1   & C1 \\
$-2400000$ & \multicolumn{3}{c}{ (km ~s$^{-1}$) } \\ 
\hline
\multicolumn{4}{c}{CHIRON} \\
59160.7361  &    104.9&  27.51 &  28.3 \\
59162.7152  &     84.1&  28.30 & 107.5 \\
59164.7560  &   -16.8:&  90.58 &   7.4: \\
59165.7460  &    103.9& 104.01 & 109.1 \\
59166.7443  &     43.0&  94.01 & 140.5 \\
59167.7603  &      8.0&  64.30 &  42.3 \\
59168.6616  &    114.9&  34.78 &   1.9 \\
59175.7079  &     -2.8&  73.15 &  87.9 \\
\multicolumn{4}{c}{TRES} \\
59155.8896  &   67.7 &  70.84  &   6.7 \\
59158.8696  &  36.2  &  85.52  & 114.7 \\
59164.8419  &  -15.0 &  92.62  & -7.6  \\
59165.8692  &  83.6  & 103.74  & 74.6 \\
\hline
Orbital Element & & & \\
\hline
$K_1$  [km s$^{-1}$] & $63.7 \pm 1.5$ &  $44.28 \pm 0.14$ &  $78.9 \pm 2.6$ \\
 $V_0$  [km s$^{-1}$] &  $51.5 \pm 1.1$ &  $60.03 \pm 0.10$ &  $66.7 \pm 1.6$ \\
 $T_0$$^a$  & 9151.868(6) & 9151.446(4) & 9150.193(7) \\
$\sigma$$^b$  [km s$^{-1}$] & 4 & 0.34 & 6 \\
\hline
\end{tabular} 

{Notes. (a) $T_0$ is the epoch of the descending node of the RV curve (i.e., the eclipse time) in BJD - 2450000. (b) $\sigma$ is the rms residuals of the RV points from the fit.  The rough estimated relative error bars on the individual RV points were scaled until $\chi^2$ per degree of freedom was unity.}

\end{table}

The elements of circular spectroscopic orbits fitted to the RVs are also listed in  Table~\ref{tab:rv}, and the RV plots are shown in Figure~\ref{fig:orb}. Each orbit is  based on 12 RVs, 8 from CHIRON and 4 from TRES. The TRES RVs are given a lower weight in the fits (with the latter relative error bars taken to be 1.5 times larger than for the CHIRON points). The epoch $T_0$ corresponds to the primary eclipse, so the argument of periastron is fixed to $\omega = 90^\circ$. An attempt to fit an eccentric orbit of B gives $e=0.005 \pm 0.005$, so we assumed the orbit to be circular in subsequent analysis. 

Given the estimated masses of the primary-star components (see Section \ref{sec:sysparms}), the spectroscopic orbits constrain the mass ratios.  The final estimates of the components' masses are given in Section \ref{sec:sysparms} using all available information, including the system SED and the analyses of the photometric lightcurves. 

Interestingly, the  systemic velocities of A and C deviate from the velocity of B in the opposite sense, and their mean, 59 km s$^{-1}$, is close to the velocity of B.  This tells us that binaries A and C orbit each other with a period of the order of several years, while binary B belongs to the visual secondary component (see Section \ref{sec:soar}).

\subsection{Speckle Imaging}
\label{sec:soar}

TIC 168789840 was observed with the speckle camera at the Southern Astrophysical Research Telescope (SOAR) on October 27, 2020 (JY 2020.8236). The instrument and data processing are described by \cite{2018PASP..130c5002T}. Several series of 400 images with exposure time of 24.4\,ms per frame were taken in the $I$ band (824/170 nm) using the iXon-888 electron-multiplication CCD camera. The image cubes are processed by the standard speckle method. Orientation on the sky and pixel scale (15.81 mas) are determined from calibration binaries with well-known positions. The latest results from this instrument and references to other publications can be found in \cite{2020AJ....160....7T}. 

\begin{table}
\centering
\caption{Measurements of AC,B at SOAR}
\begin{tabular}{ccc c c}
\hline
\hline
Date & P.A. & Sep. & $\Delta m$ & Filt.    \\  
(JY)  & (deg) & (arcsec) & (mag) & \\  
\hline
 2020.8236 & 257.74 & 0.4230  &  0.27 &  I \\
 2020.8368 & 257.61 & 0.4233  &  0.29 &  I \\
 2020.9243 & 257.62 & 0.4235  &  0.28 &  I \\
 2020.9243 & 257.69 & 0.4243  &  0.31 &  V \\
\hline 
\label{tbl:speckle} 
\end{tabular}
\end{table} 

\begin{figure}
    \centerline{
    \includegraphics[width=0.98\columnwidth]{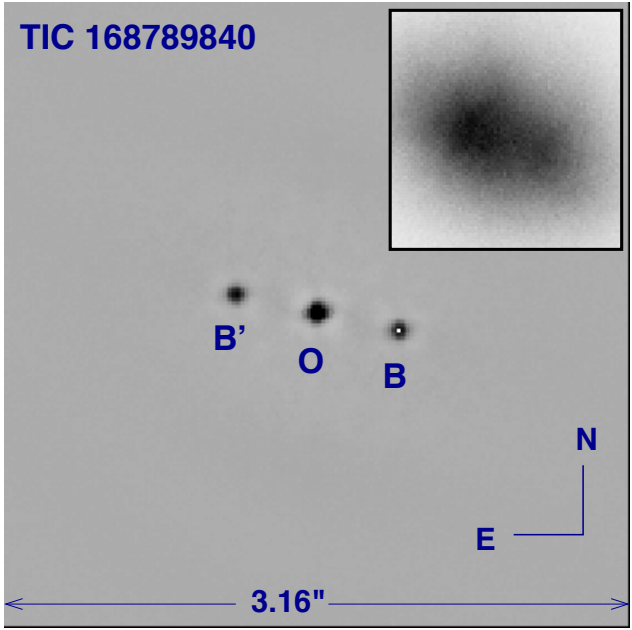}}
    \caption{Speckle auto-correlation function of TIC 168789840 (in negative rendering) recorded on 2020 October 27 at SOAR. Two  peaks B and B' on both sides of the center O indicate that it is a resolved pair; the true peak corresponding to the secondary component  is marked by the white dot. The field size is 3\farcs16, binary separation 0\farcs423; data taken in a wider field (up to 16\farcs2) show absence of other fainter sources. The insert shows a  long-exposure image produced from the same data cube where the pair is partially resolved owing to the good 0\farcs6 seeing.
    \label{fig:soar} }
\end{figure}

The object was clearly resolved into a 0\farcs42 pair at position angle of 257\fdg7 with a magnitude difference $\Delta I = 0.27$ mag (Figure ~\ref{fig:soar}). The true quadrant was determined from the shift-and-add images. Owing to the excellent 0\farcs6 seeing on that night, the pair is partially resolved even in the classical sense in the re-centered and co-added images produced from the data cubes. Approximation of the semi-resolved classical image by two Moffat functions provides independent confirmation of the magnitude difference derived from speckle processing. Observation was repeated on November 1 and December 3, 2020, and practically the same results were obtained (Table~\ref{tbl:speckle}). 
 Data over a wider field were also taken to ascertain the absence of other faint sources at larger separations, up to 8\arcsec. The contrast limit for detection of other companions is about 4.0 mag at 1\arcsec\,separation and 5.5 mag at 3\arcsec and further out.

One of the resolved components (the primary) is a close pair consisting of binaries A and C. However, separation of this inner pair should be less than $\sim$30 mas, otherwise it would be detectable by the asymmetry in the speckle power spectrum; no such asymmetry is found. 

\section{Analysis of System Parameters} 
\label{sec:sysparms}

The results of follow-up observations, along with the original {\em TESS} lightcurve and archival data, allowed for an extensive analysis of the system parameters.  First, we determined a number of dimensionless ratios for each of the binaries, e.g., $R/a$ and $T_{\rm eff}$ ratios using {\tt PHOEBE} \citep{2011ascl.soft06002P} and {\tt Lightcurvefactory} \citep{2013MNRAS.428.1656B,2017MNRAS.467.2160R,2018MNRAS.478.5135B} from analysis of the disentangled photometric lightcurves. In a second step, we combined these ratios with the measured system spectral energy distribution (`SED') to determine the stellar parameters for all six stars using an MCMC analysis. These analyses are described in detail in the following sections.

\subsection{{\tt PHOEBE} and {\tt Lightcurvefactory} Analysis}
\label{sec:phoebe}

We start the analysis of the system parameters by first fitting the disentangled photometric lightcurves with two different binary lightcurve emulators:  {\tt PHOEBE} \citep{2011ascl.soft06002P} and {\tt Lightcurvefactory} \citep{2013MNRAS.428.1656B,2017MNRAS.467.2160R,2018MNRAS.478.5135B}.  To further produce two independent sets of results, we use  {\tt PHOEBE} with the Fourier disentangled lightcurves (see Figure~\ref{fig:recons}), and {\tt Lightcurvefactory} with the iteratively disentangled lightcurves (see discussion in Section \ref{sec:FT}).  In both of these analyses we used two simplifying assumptions: (i) circular orbits for all three pairs, and (ii) fixed effective temperatures for the primary stars in all three binaries. A logarithmic limb-darkening law was also applied for both analyses. 

\begin{figure}
    \centering
    \includegraphics[width=0.98\columnwidth]{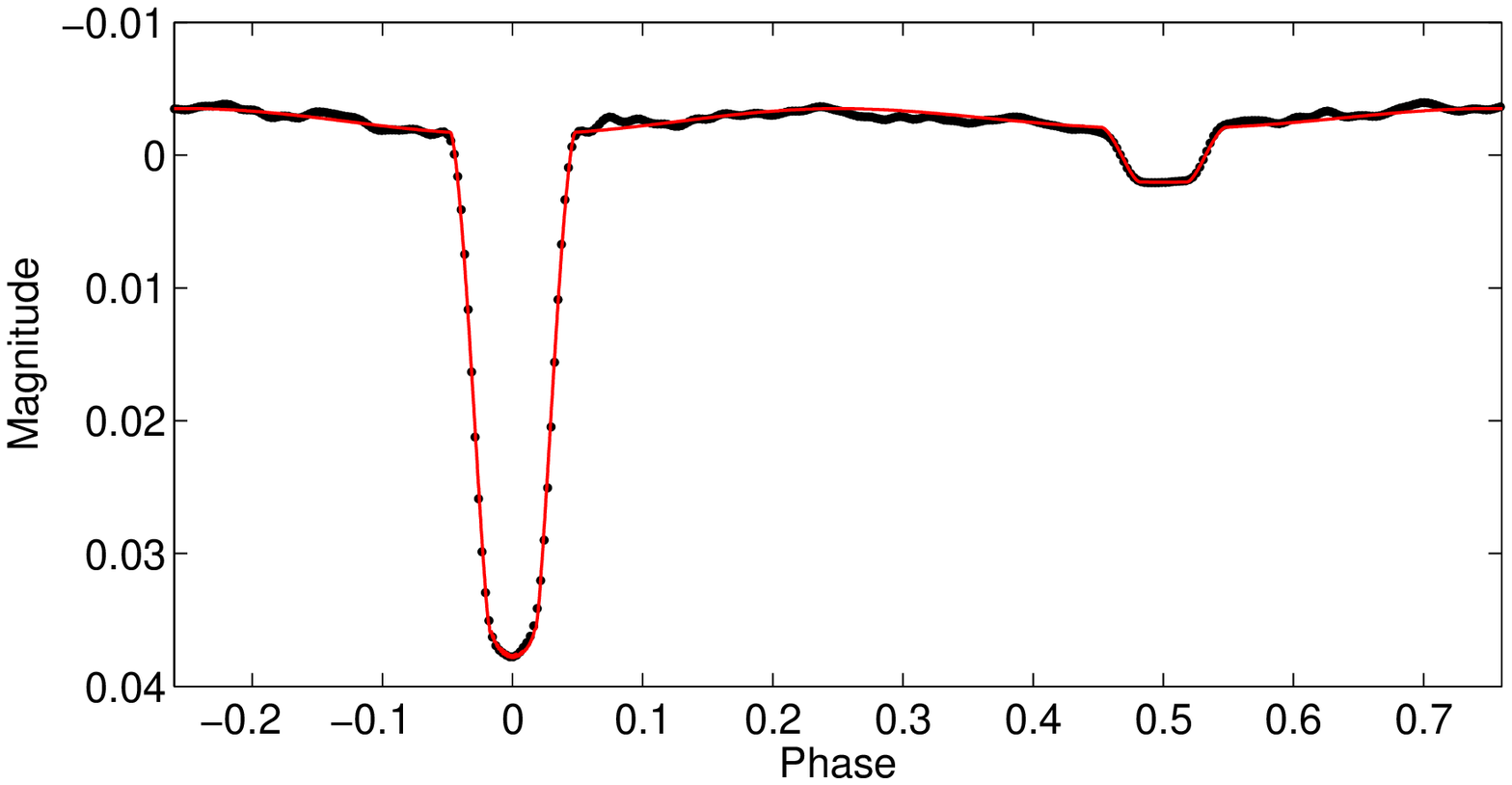}
    \includegraphics[width=0.98\columnwidth]{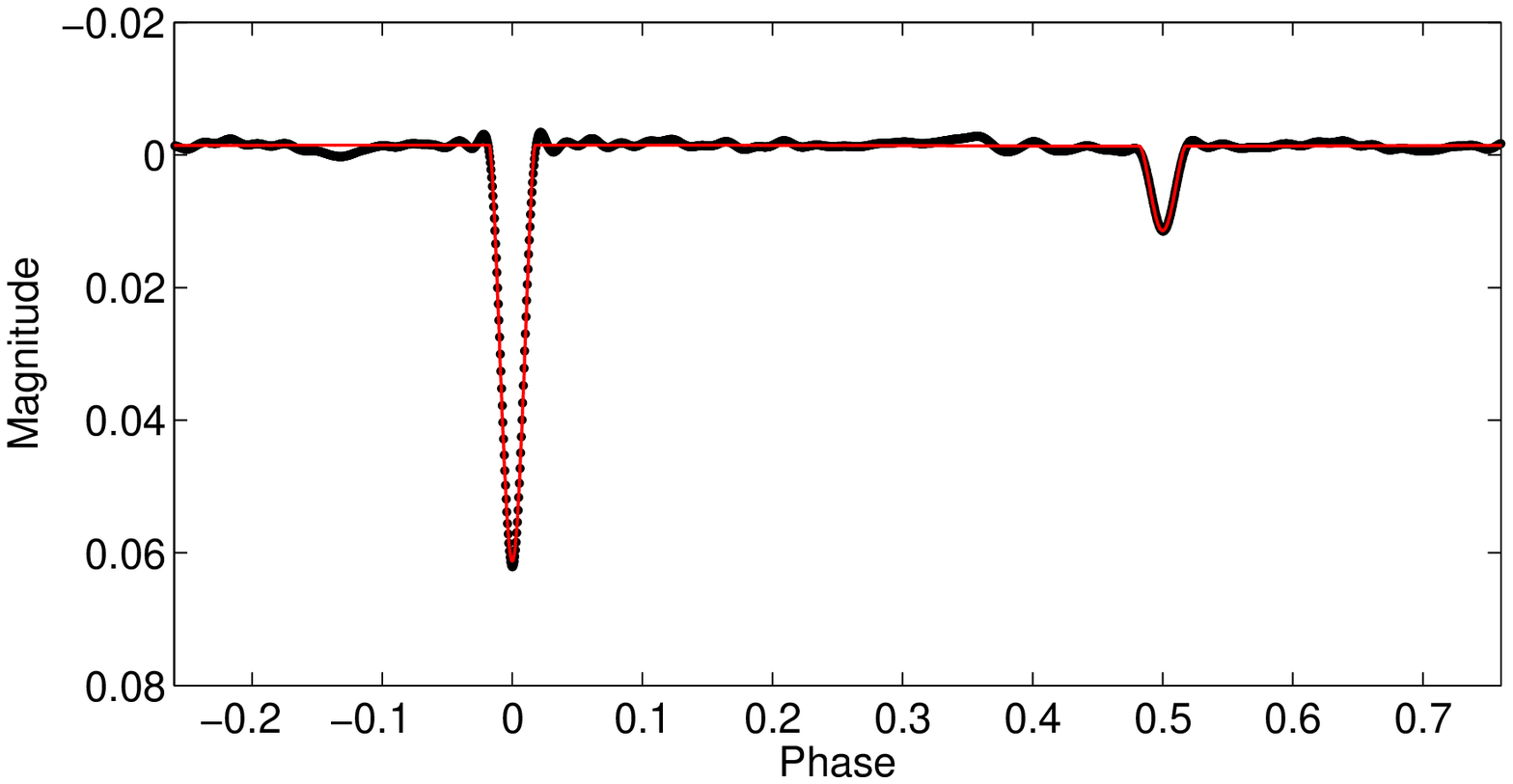}
    \includegraphics[width=0.98\columnwidth]{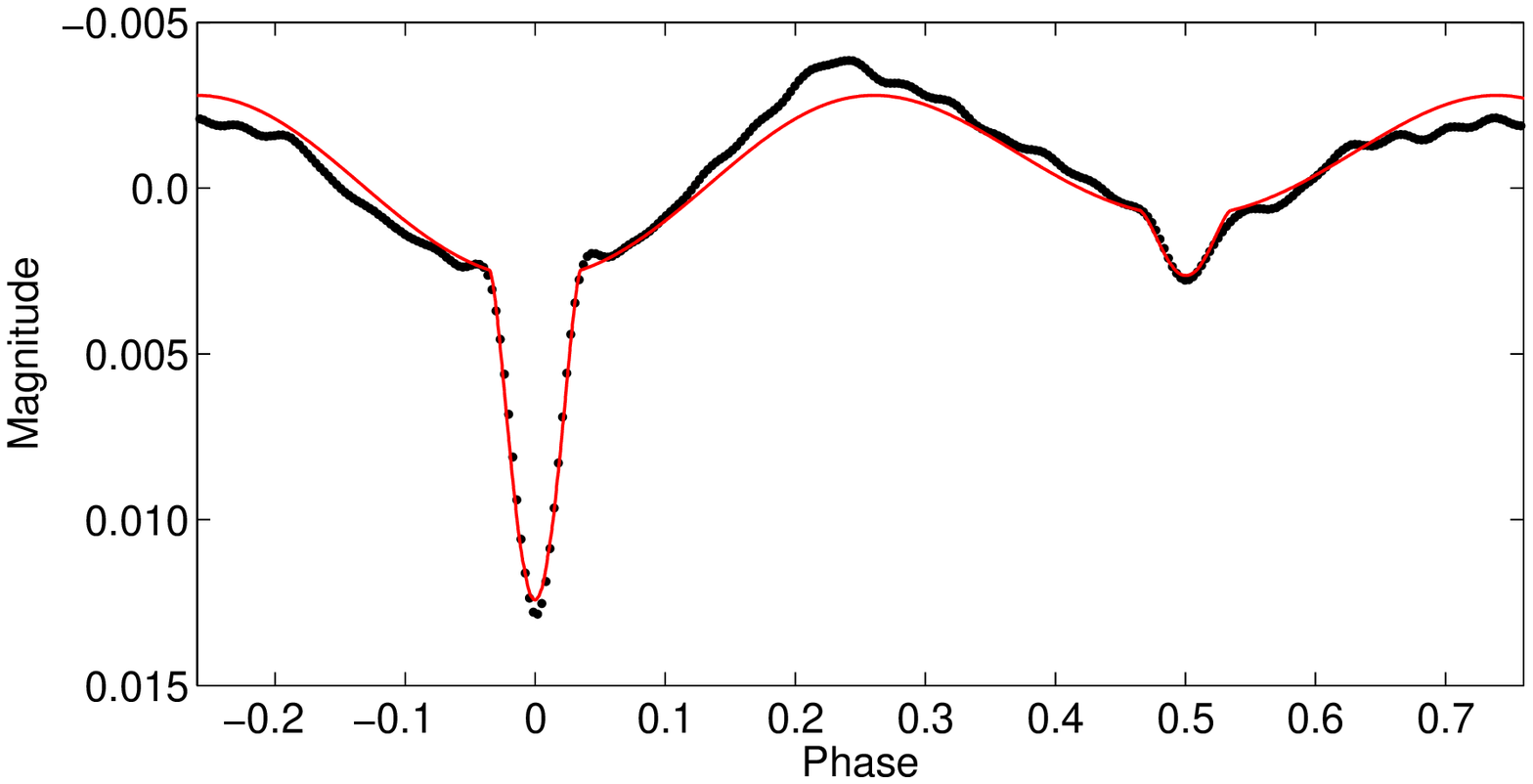}
\caption{{\tt PHOEBE} fits of the fold of binary A (top), B (middle), and C (bottom).}
\label{fig:phoebe_fits}
\end{figure}

\begin{figure}
    \centering
    \includegraphics[width=0.98\columnwidth]{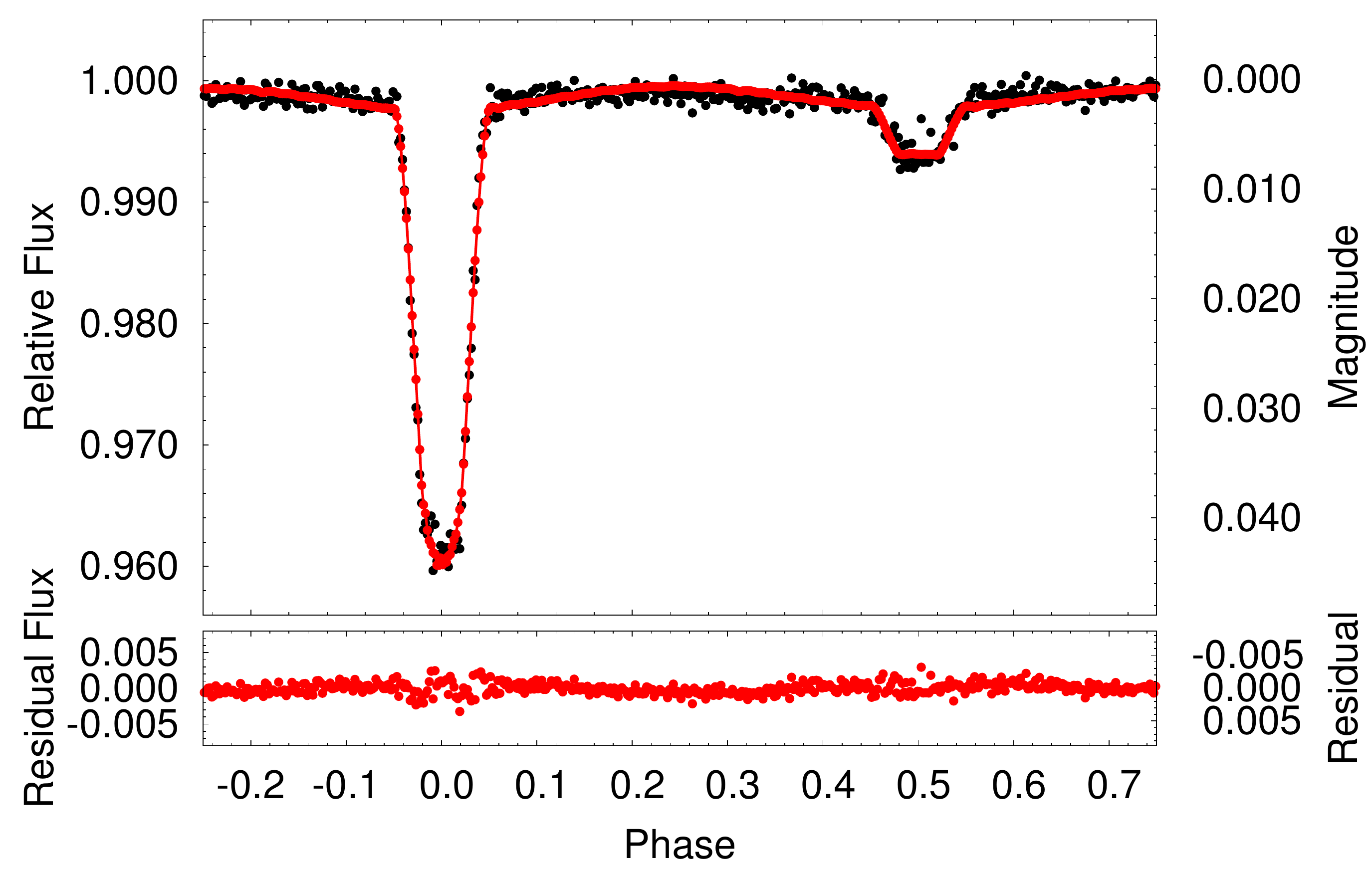}
    \includegraphics[width=0.98\columnwidth]{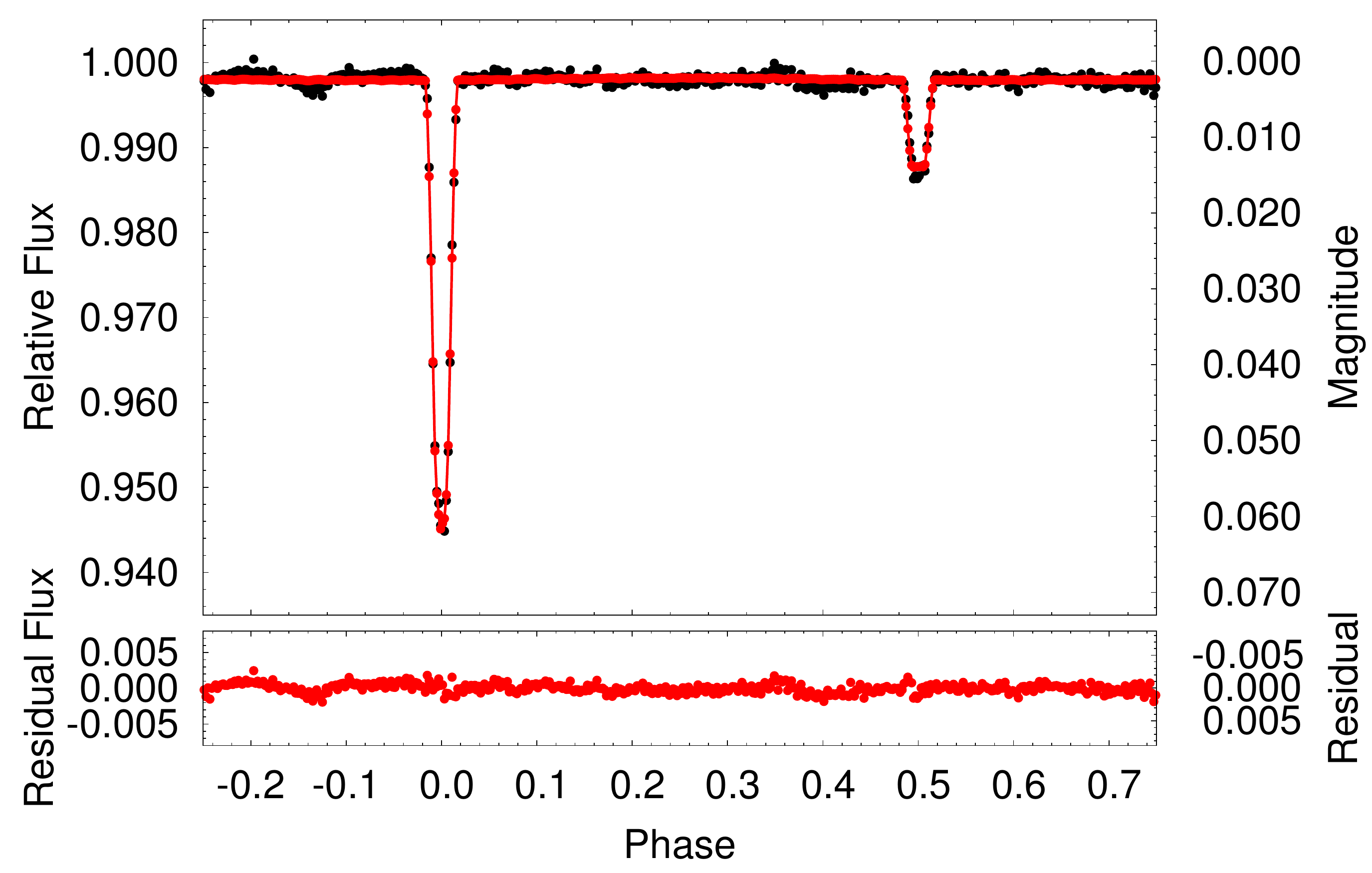}
    \includegraphics[width=0.98\columnwidth]{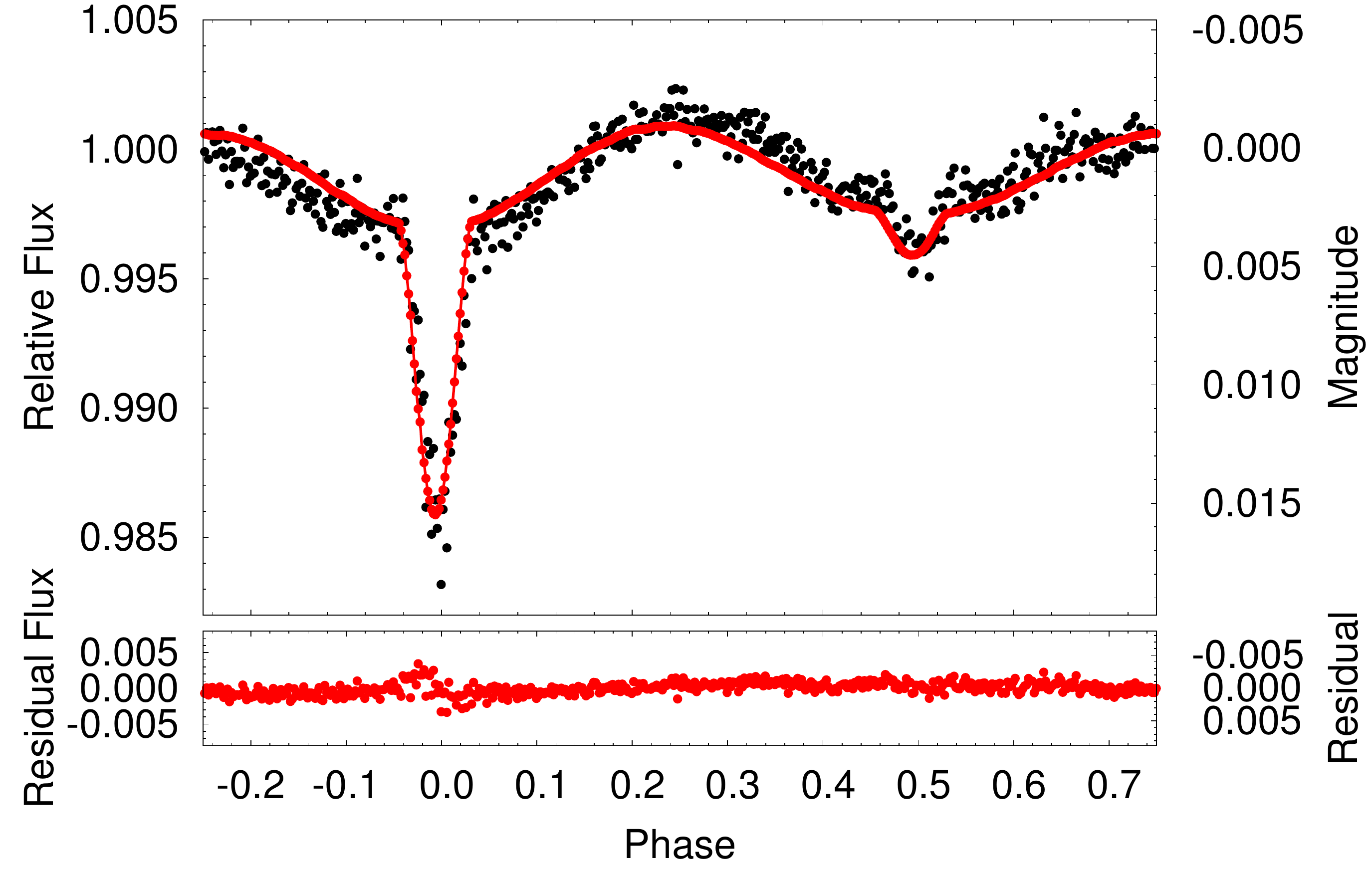}
\caption{{\tt Lightcurvefactory} fits of the fold of binary A (top), B (middle), and C (bottom).}
\label{fig:lcf_fits}
\end{figure}

The results of the {\tt PHOEBE} fit to the Fourier disentangled lightcurves are shown in Figure~\ref{fig:phoebe_fits}, while the fits to the iteratively disentangled lightcurves using {\tt Lightcurvefactory} are shown in Figure~\ref{fig:lcf_fits}.  

The resulting dimensionless parameters derived from these two fits are given in Table \ref{tbl:fitpar}.  These fits allowed for the determination of six values of scaled stellar radii, $R/a$ (where $a$ is the semi-major axis of the orbit), three values of primary to secondary $T_{\rm eff}$ ratios, as well as the three orbital inclination angles. Furthermore, in order to get temperature ratios of the primaries of the different binaries with respect to each other, we made additional runs with {\tt Lightcurvefactory}, simultaneously fitting the blends of any two of the three pairs (i.e. the time-series TAB, TAC, TBC -- see  Table~\ref{tbl:iterative}). We regard the consistency between the two independent analyses of the dimensionless system parameters seen in Table \ref{tbl:fitpar} to be quite encouraging. 

In addition, we find the `third light' parameters for each binary, i.~e., the amount of the extra flux contribution over the flux of the eclipsing binary being considered.  Both {\tt PHOEBE} and {\tt Lightcurvefactory} have built-in functionality to solve for the third light parameter in any given lightcurve.  We note, however, that the third light values are not particularly accurate at this stage of the analysis.  This is remedied by the fact that the analysis described here, as well as the more complete MCMC analysis described in Section \ref{sec:mcmc}, are done iteratively, and the results become progressively more accurate upon iteration.

\begin{table*}
\centering
\caption{Fitted Parameters Based on the {\em TESS} Photometric Lightcurves}
\begin{tabular}{p{4cm} p{5cm} p{4cm}}
\hline
\hline
Fitted Parameter&{\tt Lightcurvefactory}&{\tt PHOEBE}\\
&RVs\slash Iterative Disentanglement&Fourier Disentanglement\\\hline
$R_{\rm A1}/a_{A}$ &  $0.215 \pm 0.002$  &   $0.217 \pm 0.002$    \\
$R_{\rm A2}/a_{A}$ &  $0.093 \pm 0.004$  &    $0.087 \pm 0.003$  \\
$R_{\rm B1}/a_{B}$ & $0.077 \pm 0.003$  &     $0.066 \pm 0.002$  \\
$R_{\rm B2}/a_{B}$ & $0.031 \pm 0.002^{*}$  &    $0.056 \pm 0.002^{*}$  \\
$R_{\rm C1}/a_{C}$ &  $0.233 \pm 0.009$  &   $0.246 \pm 0.007$     \\
$R_{\rm C2}/a_{C}$ &  $0.094 \pm 0.011$  &   $0.107 \pm 0.005$     \\
\hline
$T_{\rm eff, A2}/T_{\rm eff,A1}$ & $0.590 \pm 0.014$  & $0.554 \pm 0.005$ \\
$T_{\rm eff, B2}/T_{\rm eff,B1}$ & $0.692 \pm 0.009$  & $0.681 \pm 0.004$ \\
$T_{\rm eff, C2}/T_{\rm eff,C1}$ & $0.590 \pm 0.018$  & $0.626 \pm 0.007$   \\
$T_{\rm eff,A1}/T_{\rm eff, B1}$ & $1.034 \pm 0.038$  & ... \\
$T_{\rm eff,A1}/T_{\rm eff, C1}$ & $1.024 \pm 0.047$ &  ...  \\
$T_{\rm eff, C1}/T_{\rm eff,B1}$ & $0.953 \pm 0.037$ &  ...  \\
\hline
Inclination A [deg] & $89.6 \pm 0.5$ & $89.6 \pm 0.4$  \\
Inclination B [deg] & $88.5 \pm 0.5$ & $88.1 \pm 0.3$ \\
Inclination C [deg] & $75.9 \pm 1$   & $74.7 \pm 0.5$  \\
\hline
Third Lights after Iteration$^\dag$ & & \\
\hline
Third light to A & \multicolumn{2}{c}{$0.707 \pm 0.038$~~~~~~~~~~~~~~~~~~}  \\
Third light to B & \multicolumn{2}{c}{$0.604 \pm 0.047$~~~~~~~~~~~~~~~~~~}  \\
Third light to C & \multicolumn{2}{c}{$0.688 \pm 0.057$~~~~~~~~~~~~~~~~~~}  \\
\hline 
\label{tbl:fitpar} 
\end{tabular}

{Notes. (a) The dimensionless quantities in this table for the three EBs were derived from the {\em TESS} lightcurves that were disentangled using two independent methods (see text for details), and two different binary lightcurve emulators ({\tt Lightcurvefactory} and {\tt Phoebe}). ($^*$) The disparity in the radius of B2 from the two different approaches is due to the differences in the eclipse width and depth from their respective disentanglement methods. ($\dag$) The third light results are arrived at after iterating the lightcurve analysis as described in Section \ref{sec:phoebe} with the MCMC analysis of the system parameters as described in \ref{sec:mcmc}} 

\end{table*} 

\subsection{MCMC Analysis of the Stellar Parameters}
\label{sec:mcmc}

We now combine the results of the dimensionless system parameters with several other pieces of information and constraints to solve for all of the stellar parameters for the six stars.  
Our approach is to fit for the six stellar masses and a common age, while making the explicit assumption that all the stars in the sextuple are coeval and that there has been no mass transfer among the constituent stars.  We also employ as constraints (i) the measured SED for the system, (ii) {\em MIST} stellar evolution tracks  \citep{dotter16,choi16,paxton11,paxton15,paxton19}, and (iii) \citet{castelli03} model atmospheres.  When this analysis was carried out, there was no Gaia distance information for this object in DR2 \citep{2018A&A...616A...2L}.  Therefore, we also fit for the distance to the source as well as the unknown interstellar extinction.  

In Table \ref{tbl:mcmc} we summarize exactly what the MCMC fitted parameters, the constraints, and the output parameters are.  In all, we are fitting 9 free parameters.  On the other side of the ledger there are 12 easily identified constraints ($R/a$ and $T_{\rm eff}$ ratios, and RVs; see middle column of Table \ref{tbl:mcmc}).  In addition there are 26 SED points, {\em MIST} evolution tracks\footnote{The {\em MIST} tracks were for an assumed solar chemical composition.}, \citet{castelli03} model atmospheres\footnote{The \citet{castelli03} model atmospheres were also for an assumed solar chemical composition and for a fixed $\log g = 4.0$, which well matches the primary stars in the problem (see Table \ref{tbl:parms}) that contribute 97\% of the system light.}, and the assumption of a coeval evolution of the system without mass transfer.   These latter items are hard to quantify in terms of a `number' of constraints; whether they are adequate will be determined by the uncertainties in the results.  In the end, we hope to determine 21 independent parameters of the system, as listed in the 3rd column of Table \ref{tbl:mcmc}.

To carry out this fit for the 9 free parameters we used a Markov Chain Monte Carlo (`MCMC', see, e.g., \citealt{2005AJ....129.1706F}) code modeled after the one used in \citet{2020MNRAS.494.5118K} and Rappaport, Kurtz, Handler, et al.~(2020; submitted to MNRAS), but modified to handle six stars.  For the initial MCMC runs, the priors on the six stellar masses, the age and distance of the system, and the interstellar extinction were taken to be uniform over sufficiently large ranges so as to include all plausible values.  For the final runs, the ranges of the priors were somewhat narrowed, but the priors remained uniform over their respective ranges.  In all cases, and for all parameters, the range of priors was wider than $\pm 4 \sigma$ of the finally determined parameter error bars.

For each link in the MCMC chain we know the trial masses and the system age.  From the evolution tracks we then also know all the corresponding radii and effective temperatures.  The masses, combined with the known orbital periods, yield the semi-major axis of each of the three binaries.  With this information we can check how well the $R/a$ values and temperature ratios match the input values (see Section \ref{sec:phoebe}).  The stellar radii and effective temperatures are then used in conjunction with the trial distance and $A_V$ value, along with the atmosphere models, to compute the model composite SED. These all contribute to $\chi^2$ in assessing whether to retry the step or make another jump from that point.  For each addition to $\chi^2$ we assume Gaussian distributed uncertainties in the RV values, the $R/a$ values, temperature ratios, and the uncertainties in the SED points.

\begin{figure*}
  \centering
  \includegraphics[width=0.4\linewidth]{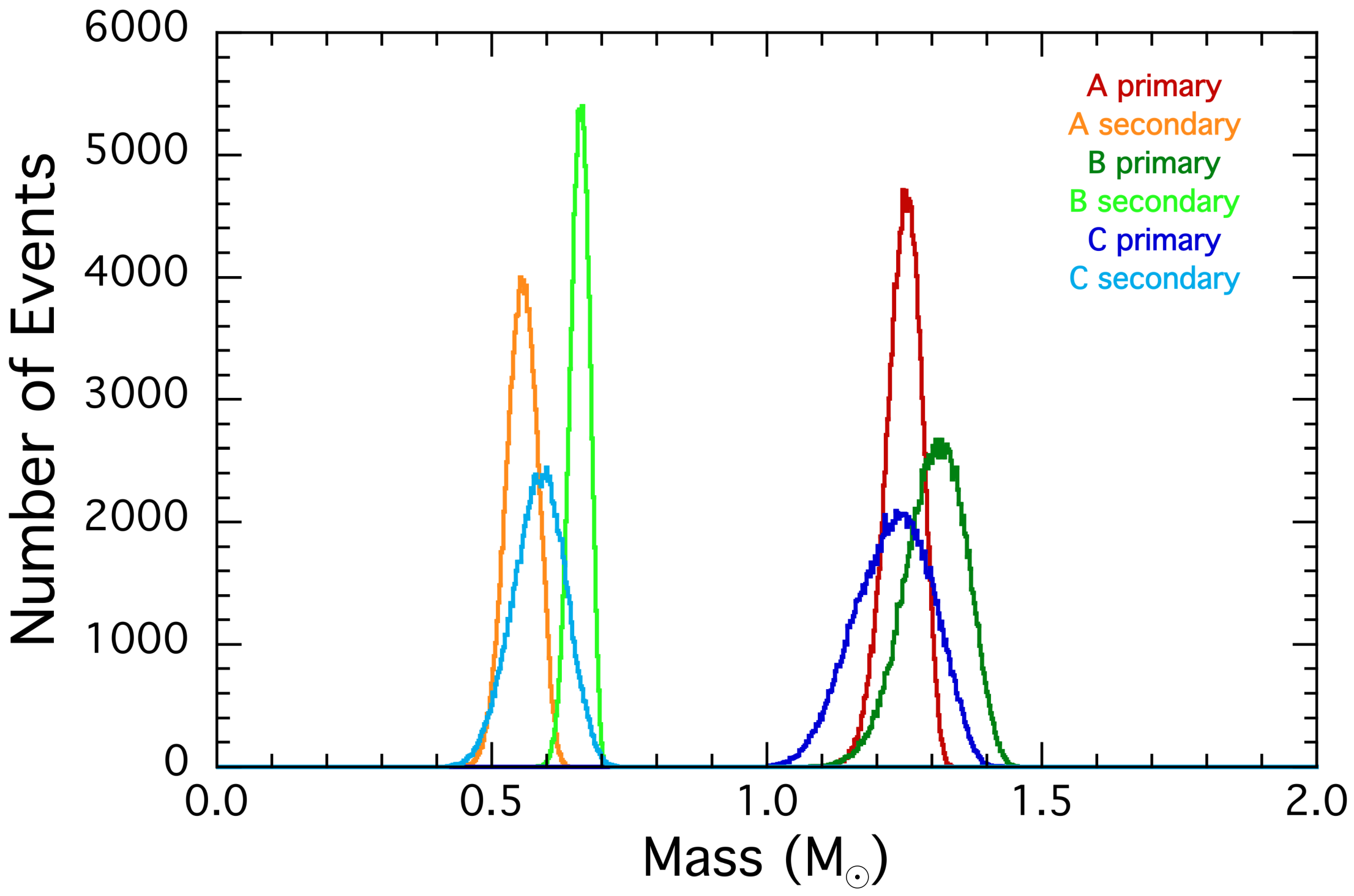}
  \includegraphics[width=0.4\linewidth]{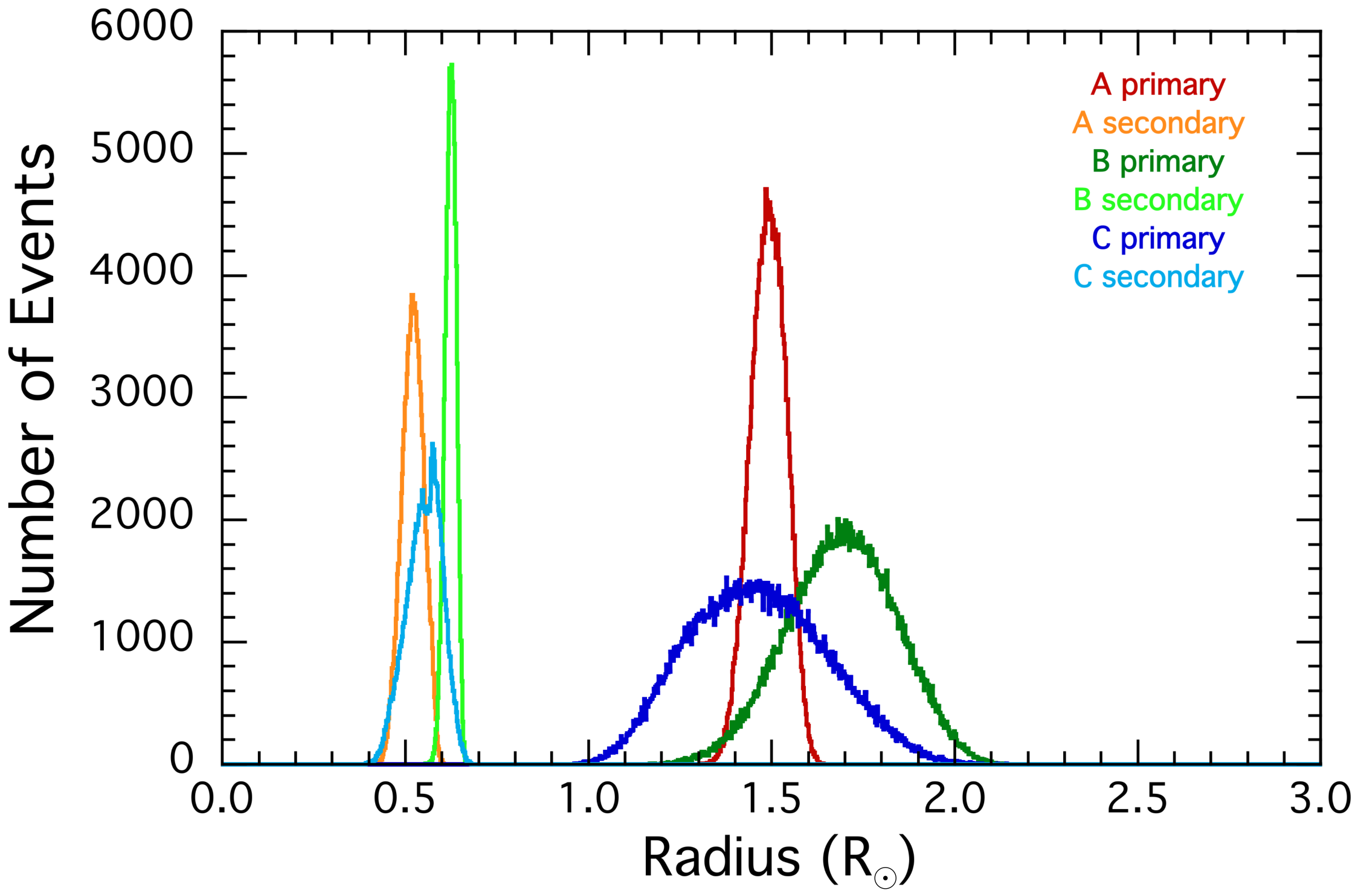}
  \includegraphics[width=0.4\linewidth]{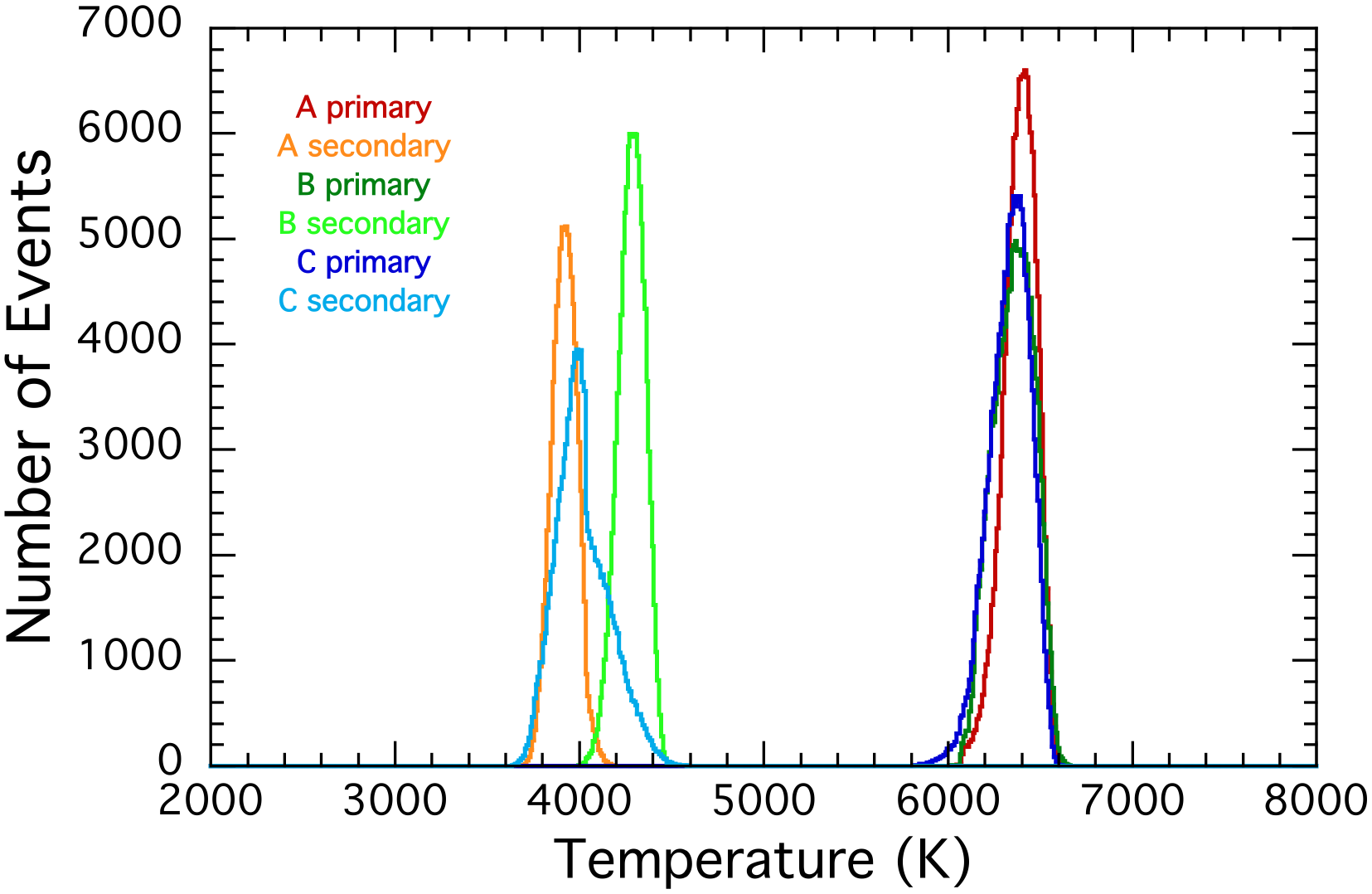} 
  \includegraphics[width=0.4\linewidth]{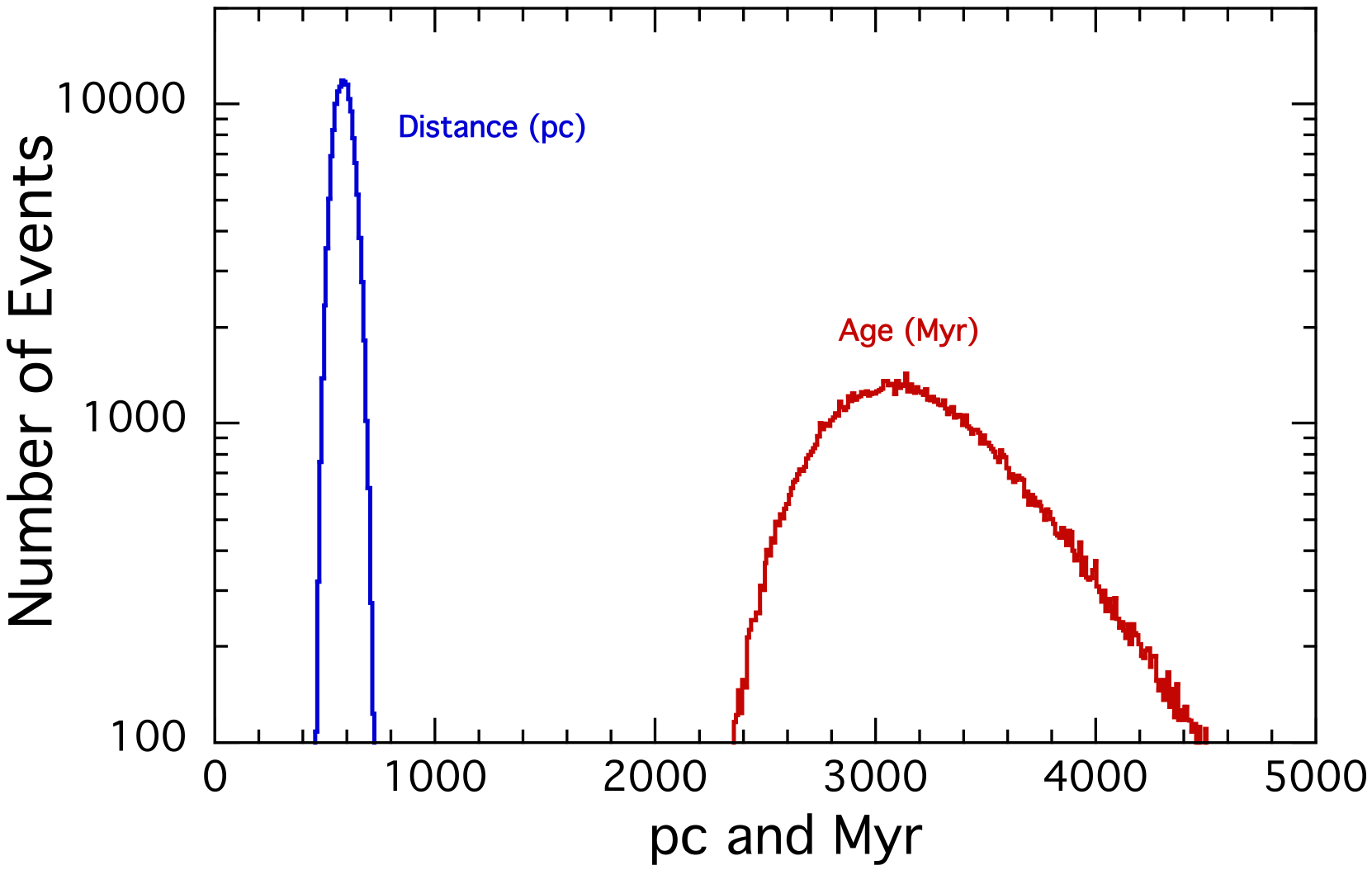}
\caption{MCMC outputs showing the distributions of system parameters.  Note the similarity in each of the three primaries and secondaries, leading to our discussion of the binaries as ``triplets''.}
\label{fig:mcmcdist}
\end{figure*} 

We ran a dozen independent MCMC chains of 20 million links each to arrive at our results. Table \ref{tbl:parms} lists our fits to the system parameters, with uncertainties for the masses, radii, and $T_{\rm eff}$s.  We also list in the Table several other parameters for each star that may be helpful in making sense of future RV or imaging observations of this system, e.g., the expected orbital velocities.  Figure~\ref{fig:mcmcdist} shows the posterior distributions for the six stellar masses, the six radii, and the six $T_{\rm eff}$ values. The fourth panel in that figure gives the distributions of distance to the sextuple as well as of its age.

The three short-period binaries would seem to be very similar `triplets', each with a more massive primary (of $\sim$1.2 $M_\odot$) that is slightly evolved off the MS, and a secondary that is sub-solar and unevolved.  The main difference among these three binaries is that one of them has an orbital period which is $\sim$5 times longer than the other two.  

\begin{figure*}
\centering
\includegraphics[width=0.65\textwidth]{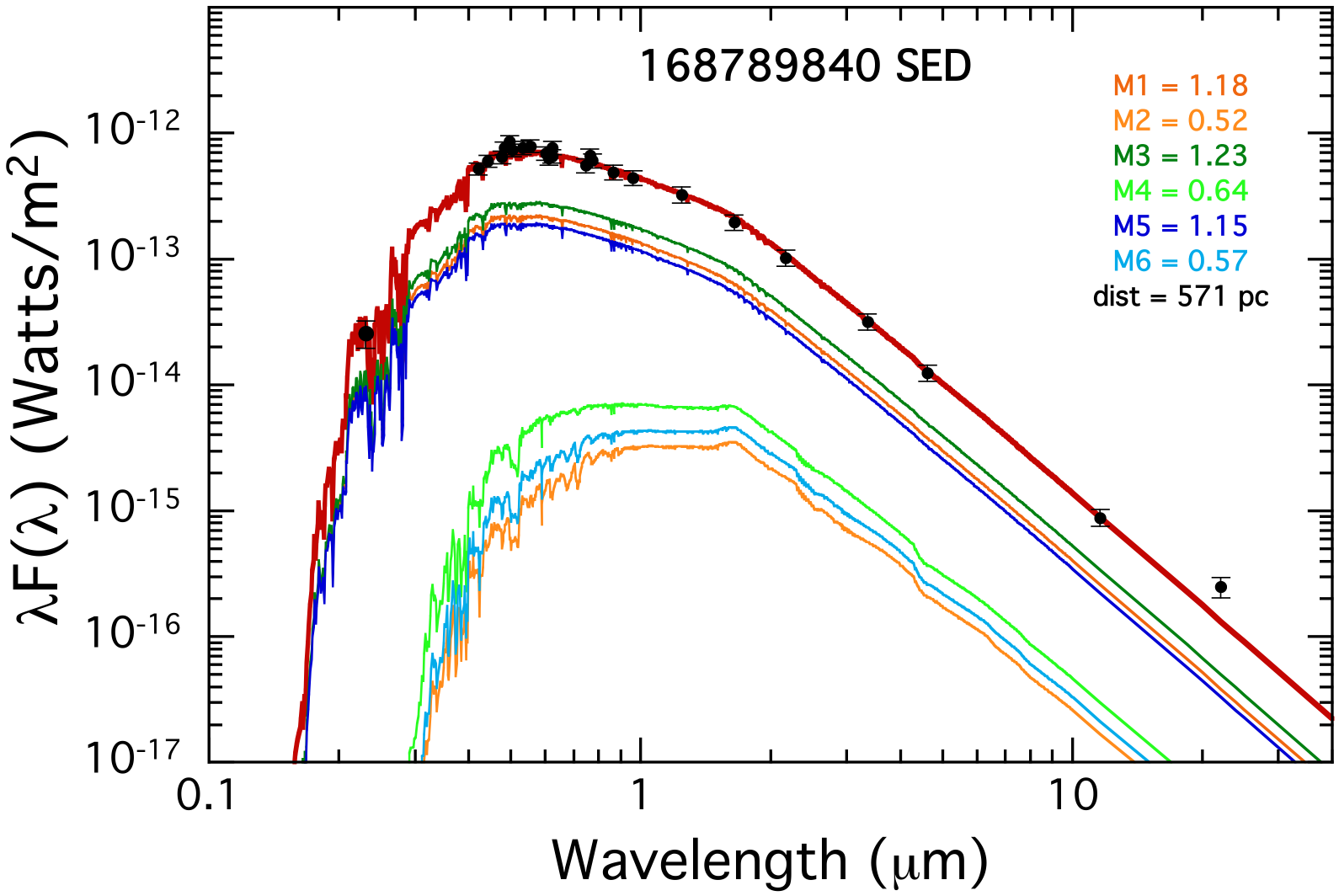}
\caption{SED diagram for TIC 168789840. The six curves are the model contributions to the SED from the individual stars, while the heavy red curve is the sum of the contributions.  The black points with error bars are the measured points (from VizieR; \citealt{vizier}).}
\label{fig:sed}
\end{figure*}

In Figure~\ref{fig:sed}  we show the best fit to the SED data (from VizieR; \citealt{vizier}).  The six thin curves are the contributions to the SED from the individual stars.  The heavy red curve is the sum of the contributions.  The black points with error bars are the measured points.  Note that the Galex \citep{galex07} NUV point is right on the model curve (though hard to notice).  The best fit is for a distance of 571 pc in this figure ($584 \pm 70$ pc in Table \ref{tbl:parms}) and an $A_V$ value of 0.28.  Now that the Gaia EDR3 \citep{gaia3} are available, we have checked our fitted photometric distance with the parallax-determined value of $593 \pm 150$ pc, and find strong agreement.

\begin{figure*}
\centering
\includegraphics[width=0.65\textwidth]{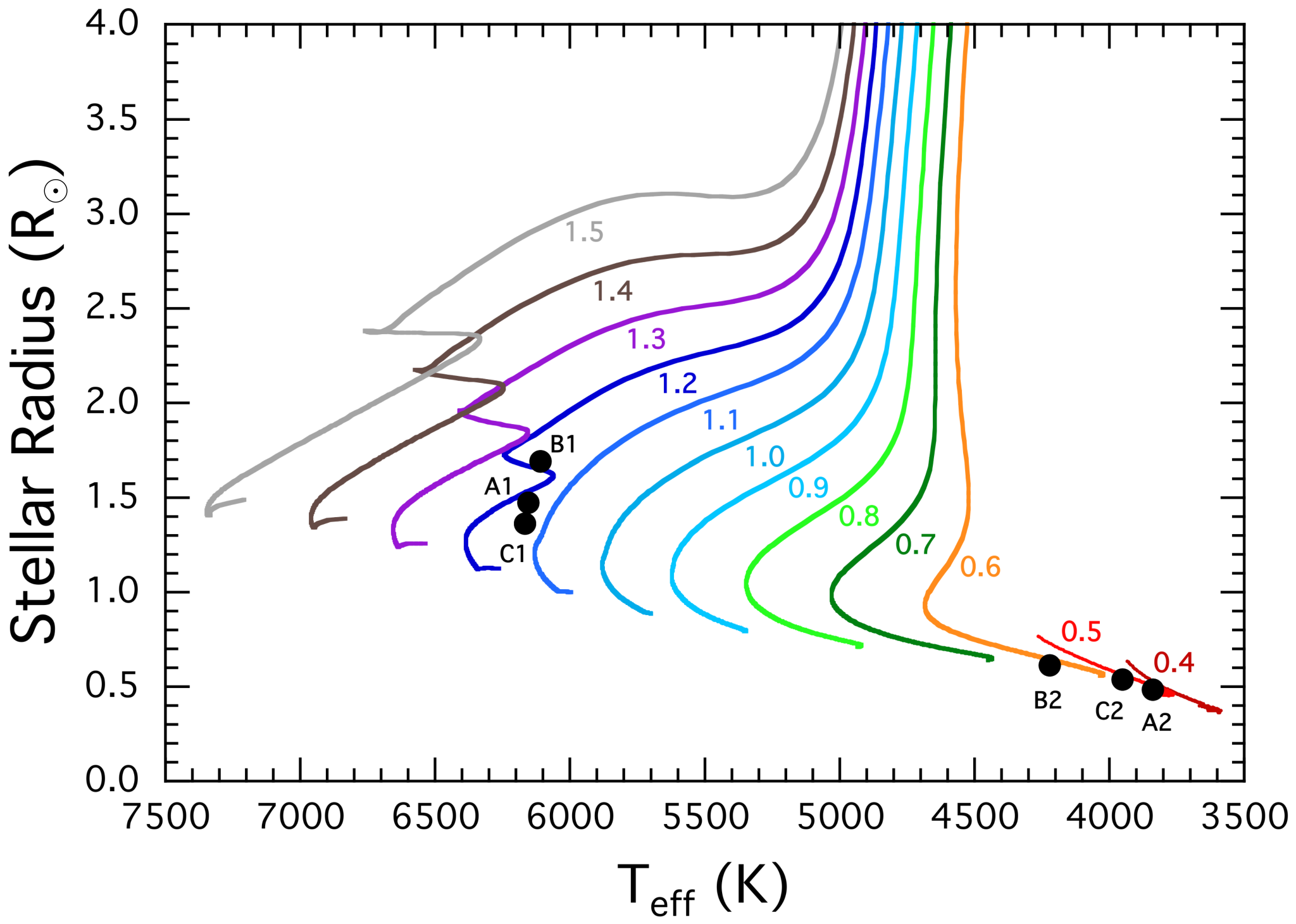}
\caption{The location of the six stars in TIC 168789840 in the plane of stellar radius and effective temperature.  The tracks are taken from the {\em MIST} library (see Table  \ref{tbl:mcmc} for references).  The number next to each track is the corresponding stellar mass in units of $M_\odot$.}
\label{fig:hrdiagram}
\end{figure*}

In Figure~\ref{fig:hrdiagram} we show the location of the six stars of TIC 168789840 in the plane of stellar radius and effective temperature, with superposed stellar evolution tracks.  All three of the primary stars lie close to the evolution track for a 1.2 $M_\odot$ star and have distinctly evolved away from the main sequence (between the TAMS and sub-giant phase).  The three secondary stars are clearly sub-solar and near the main sequence.

\begin{figure*}
\centering
\includegraphics[width=0.65\textwidth]{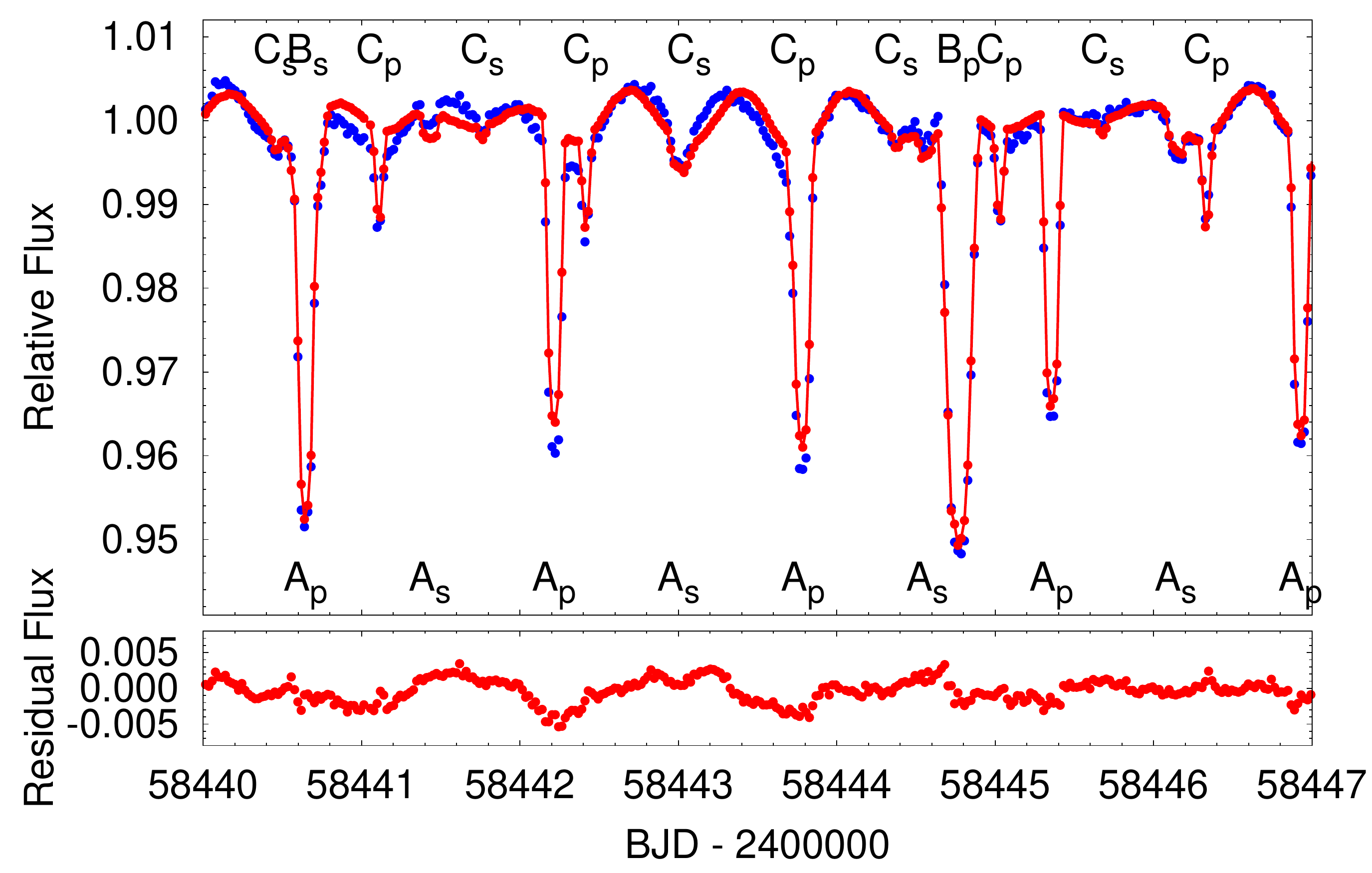}
\caption{{\tt Lightcurvefactory} photodynamical model for the overlapping set of eclipses superposed on a 7-day segment of the {\em TESS} data. Blue points are the data in 30-min cadence while the red curve is the model.  The different EB eclipses are marked (e.g. A$_\mathrm{p}$ marks the primary eclipse of binary A).}
\label{fig:LCF_fit}
\end{figure*}

To our knowledge, this is the first time that a fit for the properties of six stars, based largely on a set of overlapping photometric lightcurves and composite SED information, has been attempted.  As a final demonstration of how the {\tt Lightcurvefactory} photodynamical model using all these parameters fits the original {\em TESS} lightcurve we show in Figure~\ref{fig:LCF_fit} the two curves superposed over a 7-day segment of the {\em TESS} lightcurve.  The correspondence with the actual data is quite gratifying.

\begin{table}
\centering
\caption{System Parameters and Constraints in the MCMC Analysis}
\begin{tabular}{lll}
\hline
\hline
    Fitted Parameters   &     Constraints$^a$   &    Output     \\    
\hline
$M_{\rm A1}$ &   $R_{\rm A1}/a_{A}$ &     $M_{\rm A1}$   \\ 
$M_{\rm A2}$ &   $R_{\rm A2}/a_{A}$ &     $M_{\rm A2}$   \\ 
$M_{\rm B1}$ &  $R_{\rm B1}/a_{B}$ &      $M_{\rm B1}$   \\ 
$M_{\rm B2}$ &   $R_{\rm B2}/a_{B}$ &     $M_{\rm B2}$   \\ 
$M_{\rm C1}$ &   $R_{\rm C1}/a_{C}$ &     $M_{\rm C1}$   \\ 
$M_{\rm C2}$ &   $R_{\rm C2}/a_{C}$ &     $M_{\rm C2}$   \\ 
system age & $T_{\rm eff, A2}/T_{\rm eff,A1}$  & system age \\
distance & $T_{\rm eff, B2}/T_{\rm eff,B1}$ &  distance  \\
extinction $A_V$ & $T_{\rm eff, C2}/T_{\rm eff,C1}$ &  extinction $A_V$ \\
  & $K_{\rm A1} $ &  $R_{\rm A1}$ \\
  & $K_{\rm B1}$ &  $R_{\rm A2}$ \\
  & $K_{\rm C1}$  &  $R_{\rm B1}$ \\
  & 26 SED points &  $R_{\rm B2}$ \\
  & coeval assumption & $R_{\rm C1}$ \\
  & {\em MIST} evolution tracks & $R_{\rm C2}$ \\
  & Kurucz model spectra & $T_{\rm eff, A1}$ \\
    &  &  $T_{\rm eff, A2}$ \\
    &  & $T_{\rm eff, B1}$ \\ 
    &  & $T_{\rm eff, B2}$ \\
    &  & $T_{\rm eff, C1}$ \\
    &  & $T_{\rm eff, C2}$ \\
\hline 
\label{tbl:mcmc} 
\end{tabular}

{Notes. (a) The $R/a$ and temperature ratio constraints come from the light-curve emulator analysis of the disentangled {\em TESS} lightcurves for the three binaries. $K_{\rm A1}$ is based on the RV analysis of the CHIRON and TRES spectra (see text).  `Coeval' assumption means that all six stars in the system are assumed to have been born at the same time, and that no mass transfer has occurred among them.  The {\em MIST} stellar evolution models are from \citet{dotter16,choi16,paxton11,paxton15,paxton19}, while the `Kurucz' model atmospheres are from \citet{castelli03}.}  

\end{table} 

\begin{table*}
\centering
\caption{Computed Parameters for the Six Stars in TIC 168789840}
\begin{tabular}{lcccccccc}
\hline
\hline
    star   &     Mass   &    Radius  &   $T_{\rm eff}$  &    Lumin  &  a  &  $K$   &  $v \sin i$ &  $\log g$ \\    
             &     ($M_\odot$) & ($R_\odot$) & K & ($L_\odot$) & ($R_\odot$) & km s$^{-1}$ & km s$^{-1}$ & cgs \\
\hline
A1 &   $1.25 \pm 0.05$ &     $1.49 \pm 0.07$  &  $6400 \pm 125$  &   3.39 &    6.9&    70.7 &    48.5 &    4.18 \\ 
A2 &   $0.56 \pm 0.04$ &     $0.52 \pm 0.04$  &  $3923 \pm 100$  &   0.07 &    6.9 &   153.1 &    17.5 &    4.73 \\ 
B1 &   $1.30 \pm 0.08$ &     $1.69 \pm 0.22$  &  $6365 \pm 170$  &   3.95 &   21.4 &    44.4 &    10.1 &    4.12 \\ 
B2 &   $0.66 \pm 0.03$ &     $0.62 \pm 0.02$  &  $4290 \pm 110$  &   0.12 &   21.4 &    87.1 &     3.8 &    4.67 \\ 
C1 &   $1.23 \pm 0.10$ &     $1.45 \pm 0.28$  &  $6350 \pm 160$  &   2.74 &    6.1 &    75.4 &    51.5 &    4.24 \\ 
C2 &   $0.59 \pm 0.07$ &     $0.56 \pm 0.07$  &  $3990 \pm 190$  &   0.07 &    6.1 &   154.3 &    20.9 &    4.72 \\ 
\hline 
System     &  dist  & age & $A_V$ & & & &  & \\
                 & (pc)   & (Myr) & (mag) & & & & &  \\
\hline
      & $584 \pm 70$  &  $3160 \pm 624$  &  $0.28 \pm 0.06$ & & & & & \\
\hline
\label{tbl:parms} 
\end{tabular}

{Notes.  All the parameters result from an MCMC study of the system constraints (see text).  ``a'' is the binary semi-major axis.}  

\end{table*} 

\subsection{Inferences on the quadruple and sextuple orbits}

The SOAR speckle auto-correlation function (Figure~\ref{fig:soar}) shows two images of comparable brightness separated on the sky by 0\farcs423, in agreement with, but more accurate than, the new Gaia EDR3 results \citep{gaia3} of 0\farcs374 $\pm$ 0\farcs021.  We had tentatively argued in Sections \ref{sec:spectrum} and \ref{sec:soar}, that the brighter of the images was comprised of the A and C binaries, while the slightly fainter image (with $\Delta I \simeq 0.27$ mag) was the B binary by itself.  Here we further quantify that argument.

\begin{figure}
  \centering
  \includegraphics[width=0.98\columnwidth]{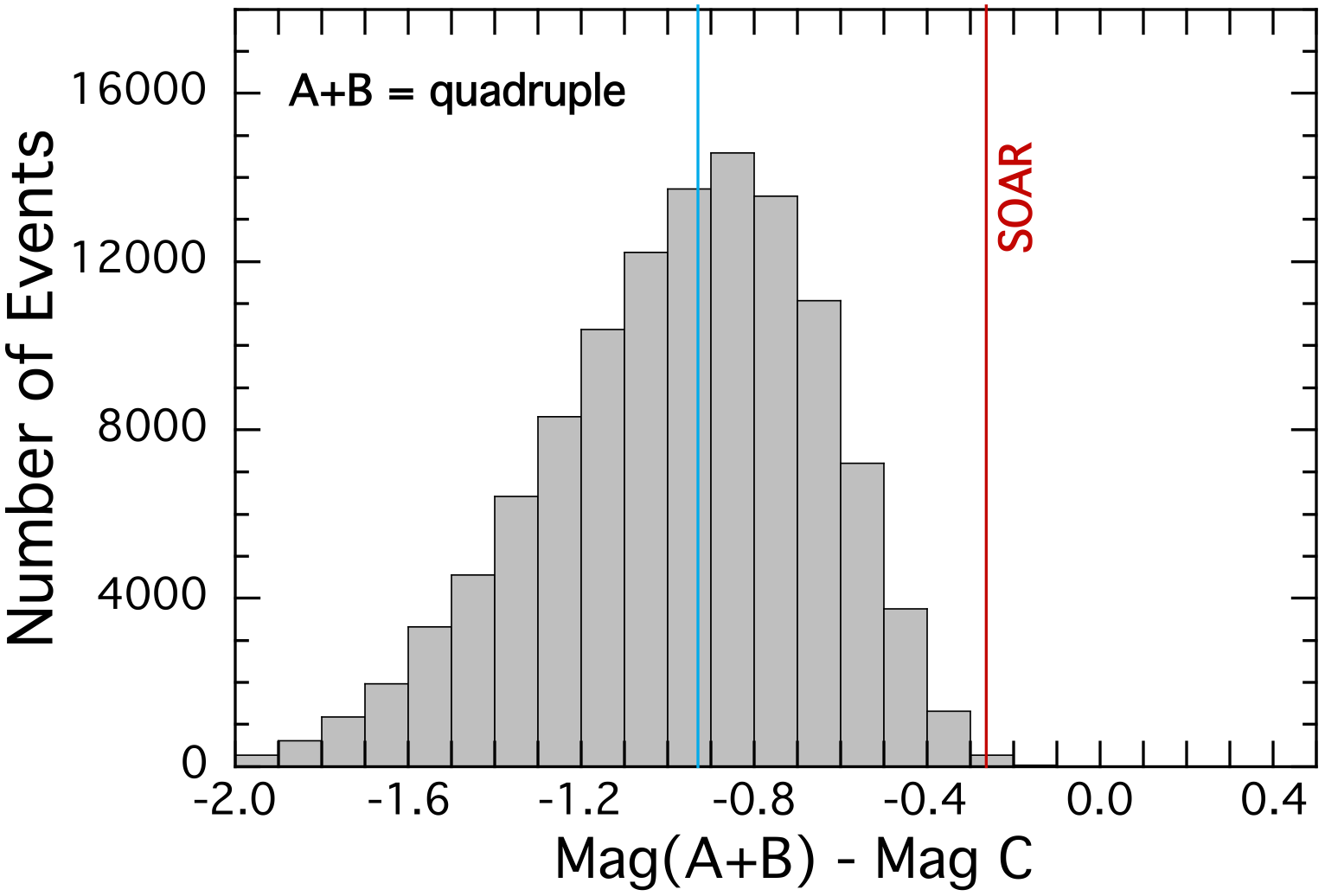}
  \includegraphics[width=0.98\columnwidth]{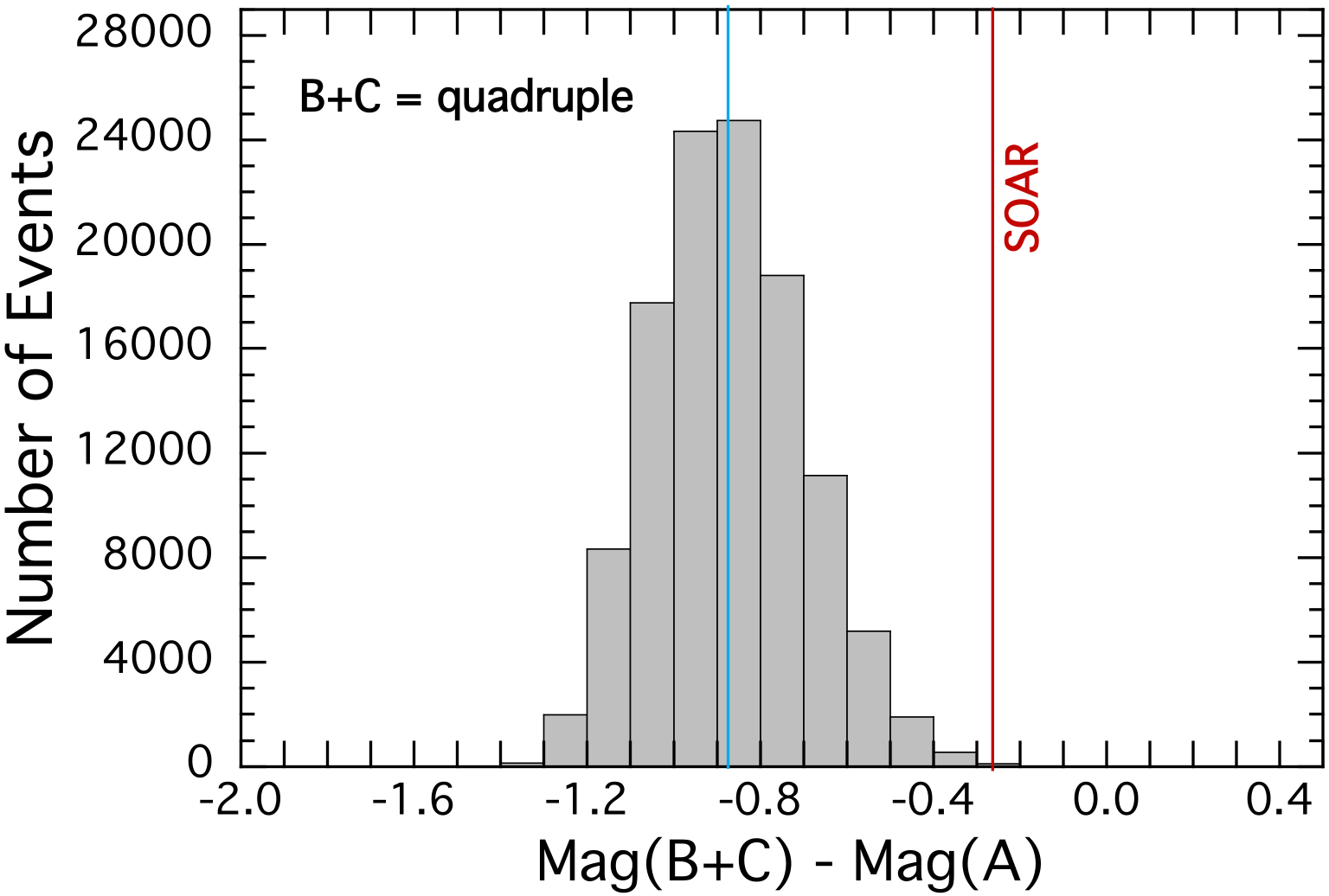}
  \includegraphics[width=0.98\columnwidth]{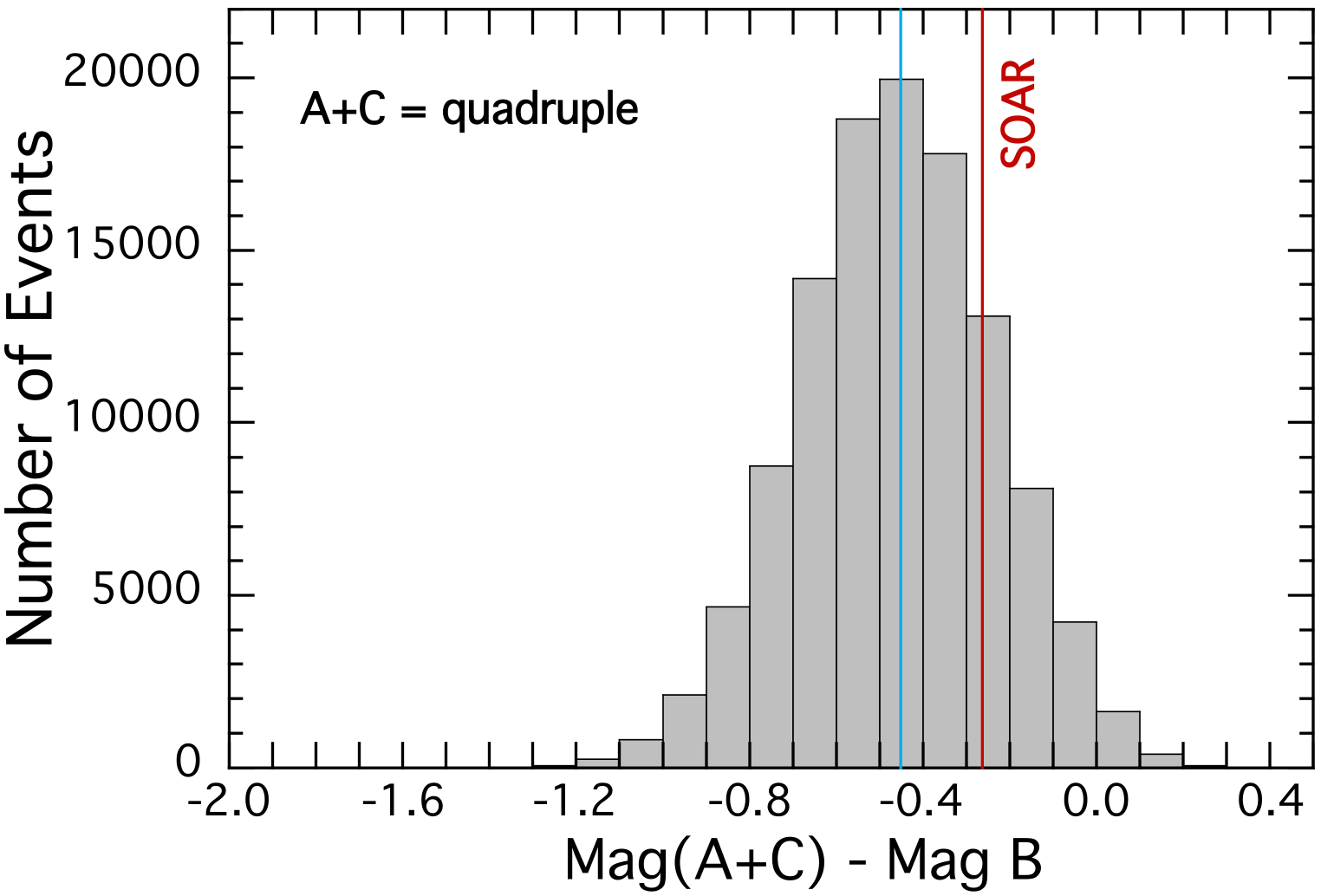}
\caption{Brightness ratio distributions in each of the possible scenarios of the identity of the inner quadruple.  For both the AB and BC quadruple possibilities, the measured value from speckle imaging lies on the extreme of the distributions.  For AC, however, the measured value is well within the expected range.  This provides strong supporting evidence to the RV analysis which concludes that AC is the inner quadruple.}
\label{fig:brightness}
\end{figure}

We utilized the results of our MCMC analysis of the system parameters to predict the brightness of each binary in the $I$ band during each link in the MCMC chain. In Figure~\ref{fig:brightness} we show distributions of the $I$-magnitude difference between the two images under the assumption that the inner quadruple is comprised of A+B, B+C, or A+C, respectively.  In the first two cases, the measured SOAR magnitude difference of 0.27 mag has almost no plausible probability of agreement with model $\Delta I$ values of $-0.94 \pm 0.31$ mag and $-0.88 \pm 0.17$ mag, respectively (see Figure~\ref{fig:brightness}).  By contrast, the hypothesis that A+C form the inner quadruple is consistent with the measured value of $\Delta I$ with a  value of $-0.45 \pm 0.20$ mag\footnote{We have also verified that the $G$ magnitude difference ($G_{\rm AC}-G_{\rm B}$) reported in the new Gaia EDR3 release of -0.34, is in agreement with similar distributions from our MCMC analysis in the $G$ band.} . 

The difference between the center-of-mass RVs of A and C also indicates that they are bound together in an orbit with a period of a few years. Thus, we are confident that the sextuple consists of an inner quadruple comprised of binaries A and C having a sky separation of $\lesssim 30$\,mas, which in turn is orbited by binary B at a current-epoch sky projection of 0\farcs423.  These two angular separations amount to projected physical separations of $\lesssim 18$ AU and 250 AU, respectively. Circular orbits with these separations, coupled with the masses given in Table \ref{tbl:parms}, would correspond to orbital periods of $\lesssim 40$ yr and $\sim$1700 yr, respectively.  

Depending on the exact separation of binaries A and C, there could well be observable Eclipse Timing Variation (ETV) effects. We have therefore attempted an ETV analysis of the eclipse times of the binaries A and C, similar to that used previously, e.g., in \cite{2019A&A...630A.128Z}.

\begin{figure}
\centerline{
\includegraphics[width=9cm]{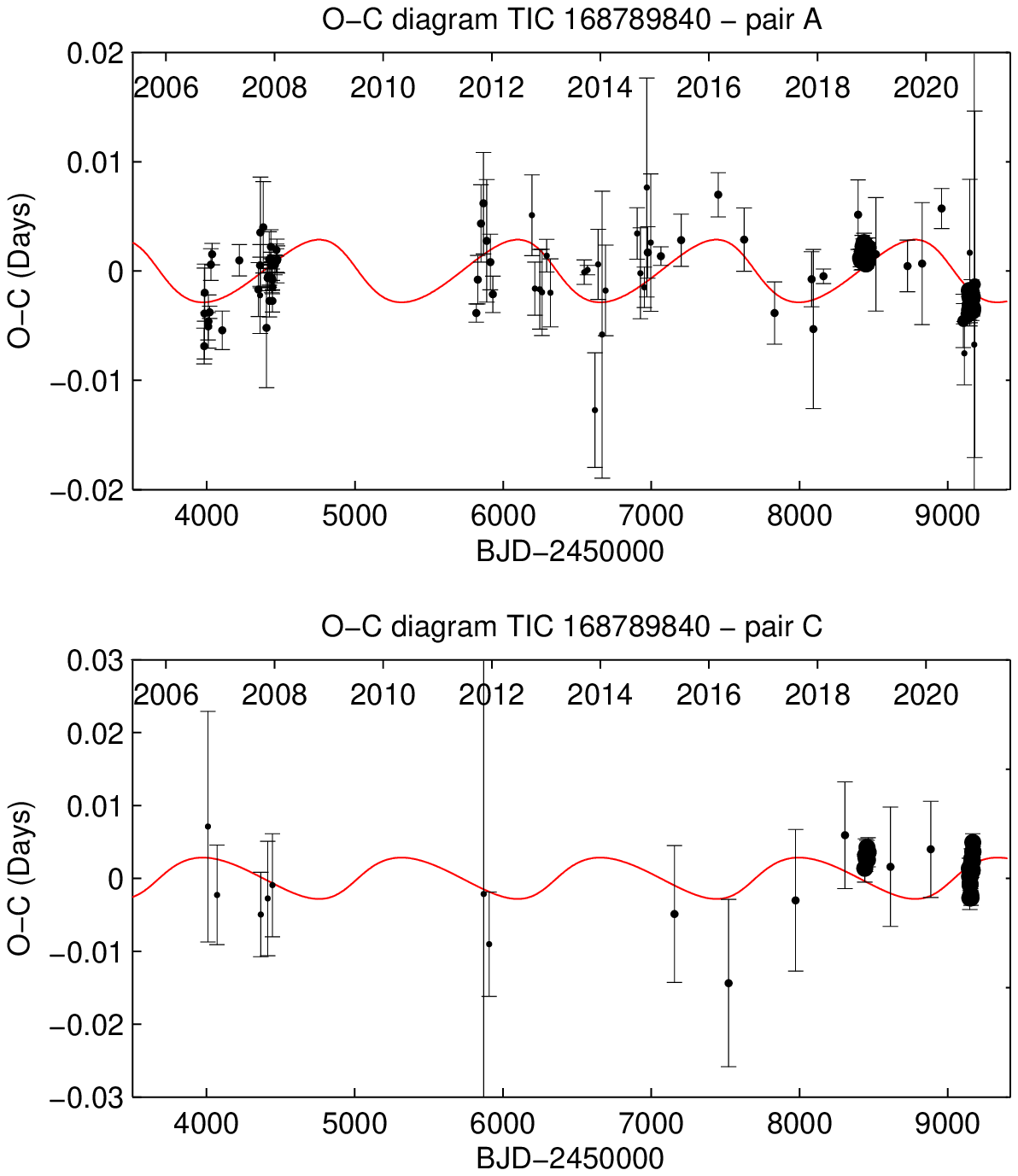} }
\caption{$O-C$ diagram of binary A (top) and binary C (bottom).  There are significantly more eclipse times for binary A than for C, and the former are more accurately measured. The orbits of the A and C binaries were fit jointly with a common set of orbital parameters (e.g., eccentricity, $e$, longitude of periastron, $\omega$, and time of periastron passage, $\tau$). The relative amplitudes of the two orbits were tied with a fixed ratio of $a_A/a_C=M_C/M_A \simeq 1$.
\label{fig:etv} }
\end{figure}

The top panel of Figure~\ref{fig:etv} shows the Observed minus Calculated ($O-C$) diagram for binary A, which has the more readily detectable eclipses in the archival WASP and ASAS-SN data. Here there seems to be a fairly clear periodic variation with a $\sim$3.7 yr period, an amplitude of 0.0029 days, and an orbital eccentricity of 0.28.

The detection of a similar corresponding ETV for binary C is tricky and yields rather uncertain results. The reason is that the eclipses are too shallow and are barely visible in the WASP and ASAS-SN  lightcurves.  The $O-C$ diagram for binary C is shown in the bottom panel of \ref{fig:etv}.  Due to the relatively poor archival coverage of binary C, we adopted the following approach.  Since we know the masses of both pairs A and C (Table \ref{tbl:parms}), their respective amplitudes in the $O-C$ diagram should follow the relation: $a_A/a_C=M_C/M_A \simeq 1 \pm 0.14$. Therefore, our joint analysis of both binaries used this simplification and both fits in Figure \ref{fig:etv} were produced from a joint orbital solution, i.e., a fixed amplitude ratio, and common set of orbital parameters. As one can see, the predicted variation for pair C is difficult to discern. Much more precise times of eclipses, especially for pair C, are needed to confirm this hypothesis. Hopefully, new {\em TESS} data would serve as an ideal data source in this aspect.

\begin{figure}
\centerline{
\includegraphics[width=8.5cm]{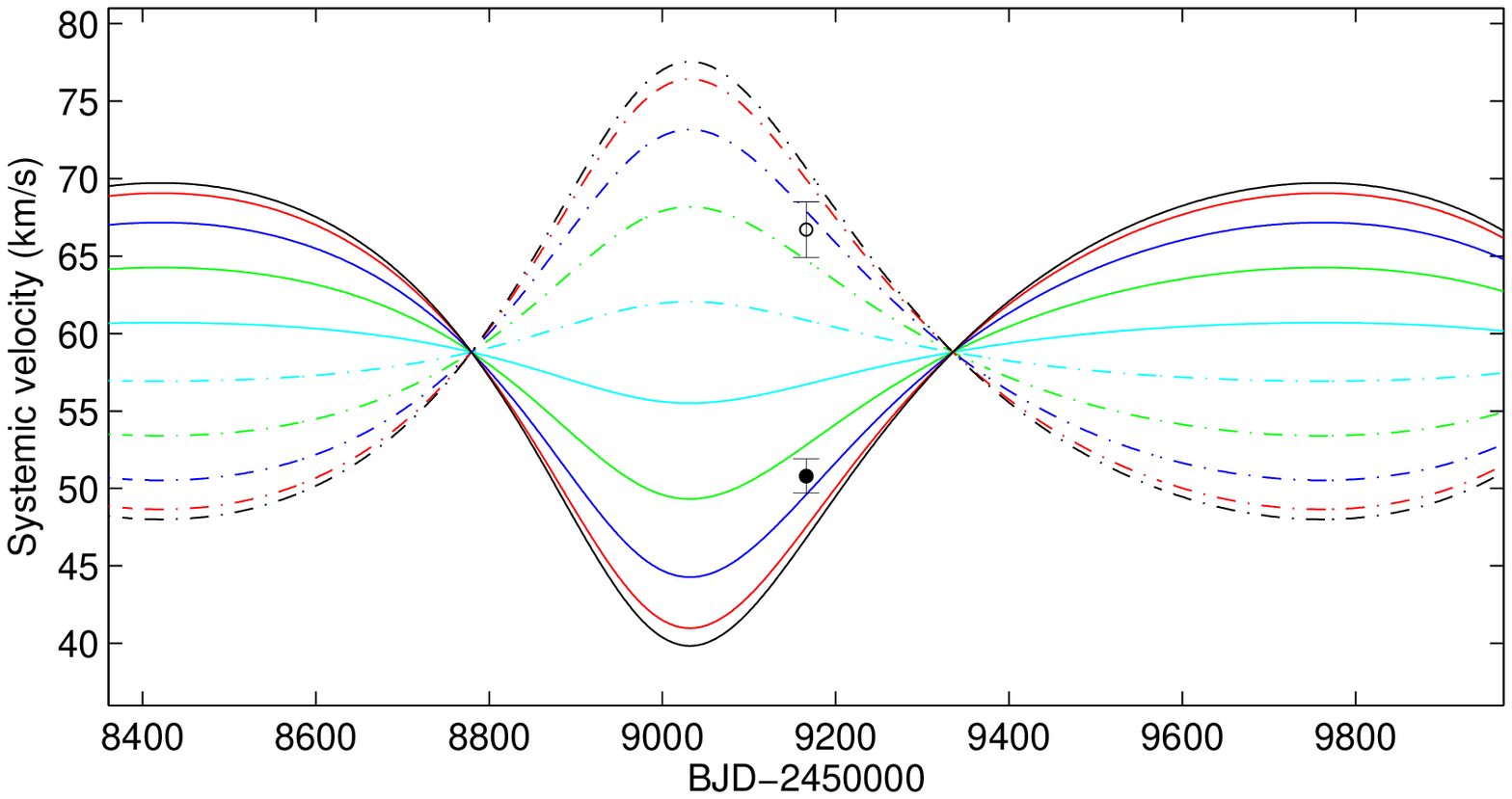}}
\caption{Radial velocity predictions for the binary A and C center of masses based on the ETV solutions presented in Table \ref{tbl:quad} and Figure~\ref{fig:etv}. The colored curves are for inclination angles of the AC quadruple of 10$^\circ$ (cyan), 30$^\circ$ (green), 50$^\circ$ (blue), 70$^\circ$ (red), and 90$^\circ$ (black), where the solid and dashed curves are for the C and A binaries, respectively.  For an inclination of 42$^\circ$, the expected values of $V_0$ for the A and C binaries nicely match what one finds from the RV analyses of the A (solid circle) and C (open circle) binaries (see Table \ref{tab:rv}).
\label{fig:AC_RV}}
\end{figure}

The orbital parameters we have found for the quadruple (AC) are given in Table \ref{tbl:quad}.  However, these should still be taken with caution especially due to poor archival data coverage of binary C.  If we accept the ETV curve as the valid solution for the AC quadruple, then it directly predicts the RVs of A and C as functions of time.  In  Figure~\ref{fig:AC_RV} we show the expected RVs under the assumption that the systemic radial velocity, $V_0$, of binary B $\simeq 59$ km s$^{-1}$ (see Table \ref{tab:rv}) also represents the center of mass velocity of the quadruple AC.  We also plot on the figure the two values of systemic radial velocities, $V_0$, for binaries A and C (see Table \ref{tab:rv}) at the mean time the RVs were taken.  One can readily see that there is a match for the predicted RVs if the orbital inclination of the AC quadruple orbit is of about $42^\circ$, in accord with what the ETV analysis indicates.

\begin{table}
\centering
\caption{{\bf Fitted Parameters for the Inner Quadruple AC}}
\begin{tabular}{lc}
\hline
\hline
    Parameter   &     Value    \\    
\hline
period &   $3.7 \pm 0.6 0$  years \\ 
$a_A \sin i$ & $0.516 \pm 0.110$ au \\
$a_C \sin i$ & $0.510 \pm 0.110$ au \\
eccentricity &   $0.28 \pm 0.05$  \\ 
$\omega_{\rm AC}$ &  $166.0 \pm 25.2$ \\ 
$\tau_{\rm AC}$ &  $2457662 \pm 305$ \\ 
$f(M)$ & $0.011 \pm 0.001$ $M_\odot$ \\
$i$ & $42^\circ$ \\
\hline 
\label{tbl:quad} 
\end{tabular}

{Notes. The fitted projected semimajor axes, $a_A \sin i$ and $a_C \sin i$ were taken to have a fixed ratio in proportion to their measured inverse mass ratio.  $\omega$ and $\tau$ are the argument of periastron and the time of periastron passage, respectively. $f(M)$ is the mass function corresponding to the projected semimajor axes.  The inclination angle, $i$, is inferred from the mass function and the measured masses of the A and C binaries.}  

\end{table} 

For completeness, we note that it is highly unlikely that there are three unrelated EBs that are so precisely aligned along the line-of-sight just by chance. To calculate the probability corresponding to such a coincidence, we first compute the magnitudes of each EB from its flux contribution and find: 11.76, 11.58 and 11.98 Tmag for pairs A, B, and C, respectively. We then compare those with the number of nearby Gaia stars having magnitudes between 11.5 and 12.5 Tmag\footnote{Assuming these are representative of the field of view.}, and with the speckle observation from SOAR. There are 4 such stars in a ${\rm 5\arcmin\times5\arcmin}$ region around the target: Gaia ID 4882948284462670000 (Gmag = 12.11, Tmag = 11.68, using \citealt{2019AJ....158..138S}; 4883001580713720000 (Gmag = 12.19, Tmag = 11.76); 4882947498485530000 (Gmag = 12.77, Tmag = 12.34); and 4883001615073460000 (Gmag = 12.8, Tmag = 12.37). Thus the probability of having one such star within ${\rm 0.03\arcsec}$ of the target (the SOAR limit on the separation of the inner quad)---and unrelated to it---is ${\approx4\times10^{-8}}$ and the probability for a star to be within 4\farcs4 of the target (the outer orbit separation as resolved by SOAR) is ${\approx7\times10^{-6}}$; thus the compound probability that TIC 168789840 is actually three unrelated EBs is ${\approx3\times10^{-14}}$. The equality between the RV of B and the mean RV of A and C is another strong argument that these stars are gravitationally bound in one system.

\section{Summary}
\label{sec:summary}

In this work, we have presented the discovery of the first known sextuply-eclipsing sextuple star system TIC 168789840. Our analysis shows that the orbital periods of the three constituent eclipsing binaries are 1.570 days (binary A), 1.306 days (binary C) and 8.217 days (binary B), such that binaries A and C form an inner quadruple system with a period of about 4 years, and the latter forms the outer subsystem with a period of about 2,000 years. The three eclipsing binaries are practically ``triplets'' with best-fit primary masses and radii of 1.23-1.30 $M_\odot$ and 1.46-1.69 $R_\odot$; secondary masses and radii of 0.56-0.66 $M_\odot$ and 0.52-0.62 $R_\odot$; and primary and secondary effective temperatures of 6350-6400 K and 3923-4290 K, respectively.  

TIC 168789840 is a fascinating system that naturally merits additional observation and analysis. Though quite similar to the famous Castor system, the ``triplet'' nature of TIC 168789840 combined with the presence of three primary and three secondary eclipses enable further investigations into its stellar formation and evolution. Remarkable objects like TIC 168789840 or Castor give us insights on the formation of multiple systems --- a matter of active research and debate. It is well known that components of hierarchical systems have correlated masses \citep{msc}, suggesting accretion from a common source. On the other hand,  disk fragmentation and subsequent migration, driven by accretion, appears to be the dominant mechanism of close binary formation; its crude modeling can explain their statistics \citep{2020MNRAS.491.5158T}.  The above model suggests a tight anti-correlation between the mass ratios of close binaries and the time of companion's formation: low-mass companions are the latest to form. Formation of close binaries  by several mechanisms acting separately or in combination and their migration can be "observed" in numerical simulations of cluster collapse by \citet{2019MNRAS.484.2341B}.

Transient massive disks prone to fragmentation likely result from an accretion burst, caused, e.g.,  by a close approach of two protostars surrounded by gas envelopes. Then one or both stars form secondaries via disk fragmentation and also become bound together in a wide orbit \citep[see][]{2019MNRAS.484.2341B}. Accretion of the remaining gas drives the inner subsystems to closer orbits and shrinks the outer orbit as well. Dissipative dynamics of an accreting triple or quadruple system evolving in a common envelope presumably can align all its orbits in one plane. This scenario could explain formation of tight quadruples like VW LMi with nearly coplanar architecture \citep{2020MNRAS.494..178P} and compact coplanar triples.

With regard to TIC 168789840, we might think that an encounter of the young binary AC with another star B led to its capture on a wide orbit, while strong accretion from the unified envelope, caused by this dynamical event, formed seed secondary companions to all stars by disk fragmentation. The seeds continued to grow and migrate inward, while the intermediate and outer orbits also evolved. The inner quadruple AC indeed resembles the tight coplanar quadruple VW LMi (outer period 1 yr), although in the latter the two inner mass ratios and periods (0.5 and 7.9 days) are not as similar as in AC, and only one inner subsystem is eclipsing. This scenario, still speculative, explains the origin of the doubly-eclipsing inner quadruple AC and predicts that the orbit of AC should be  coplanar with both inner binaries. The outer orbit of B around AC is much wider, and the eclipses of the binary B could be a matter of coincidence. In the Castor system, the outer orbit of $\sim$10 kyr period is not aligned with the 460-yr orbit of the intermediate quadruple, and only one of the three close binaries (the outer one) is eclipsing.

For future measurements, we note that further resolution of the system may be possible with interferometetry.  The axis of the inner quadruple AC is 4 AU or $\sim$ 7 mas, so might be marginally resolved by speckle at 10-m telescope, or certainly at ELT. Consideration should also be given to GRAVITY, if fringe tracking is possible. Future Gaia DR measurements have the potential to detect the $\sim$ 4-yr wobble, with the DR3 catalog being released in December 2020. Additional RV monitoring will also give the AC spectroscopic orbit and upcoming {\em TESS} measurements may detect the ETV securely.

Regarding our ongoing search for multiple star systems, we continue to find more of these systems in the {\em TESS} data through a combination of machine learning (to limit the size of the data set) followed by a visual survey.  {\em TESS} has allowed us to find well over 100 such candidate multi-star systems to date, with the analysis of another sextuple system by Jayaraman, Rappaport, Borkovits, Zasche et al. to follow this in this near future.

\section*{Note Added in Manuscript}
\begin{figure*}
    \centering
    \includegraphics[width=0.8\textwidth]{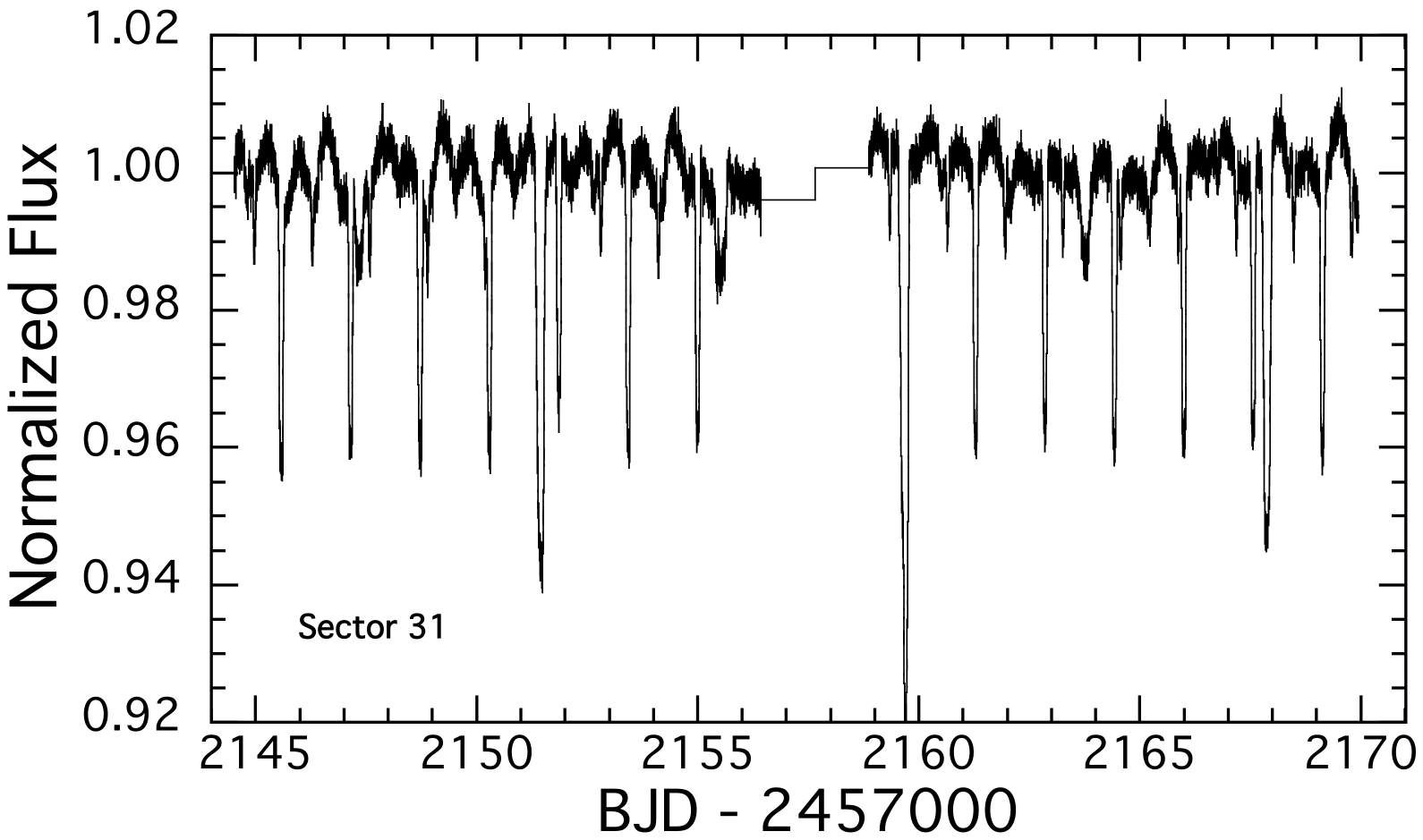}
    \caption{{\em TESS} lightcurve of TIC 168789840 in sector 31.}
    \label{fig:s31}
\end{figure*}

\begin{figure}
    \centering
    \includegraphics[width=0.98\columnwidth]{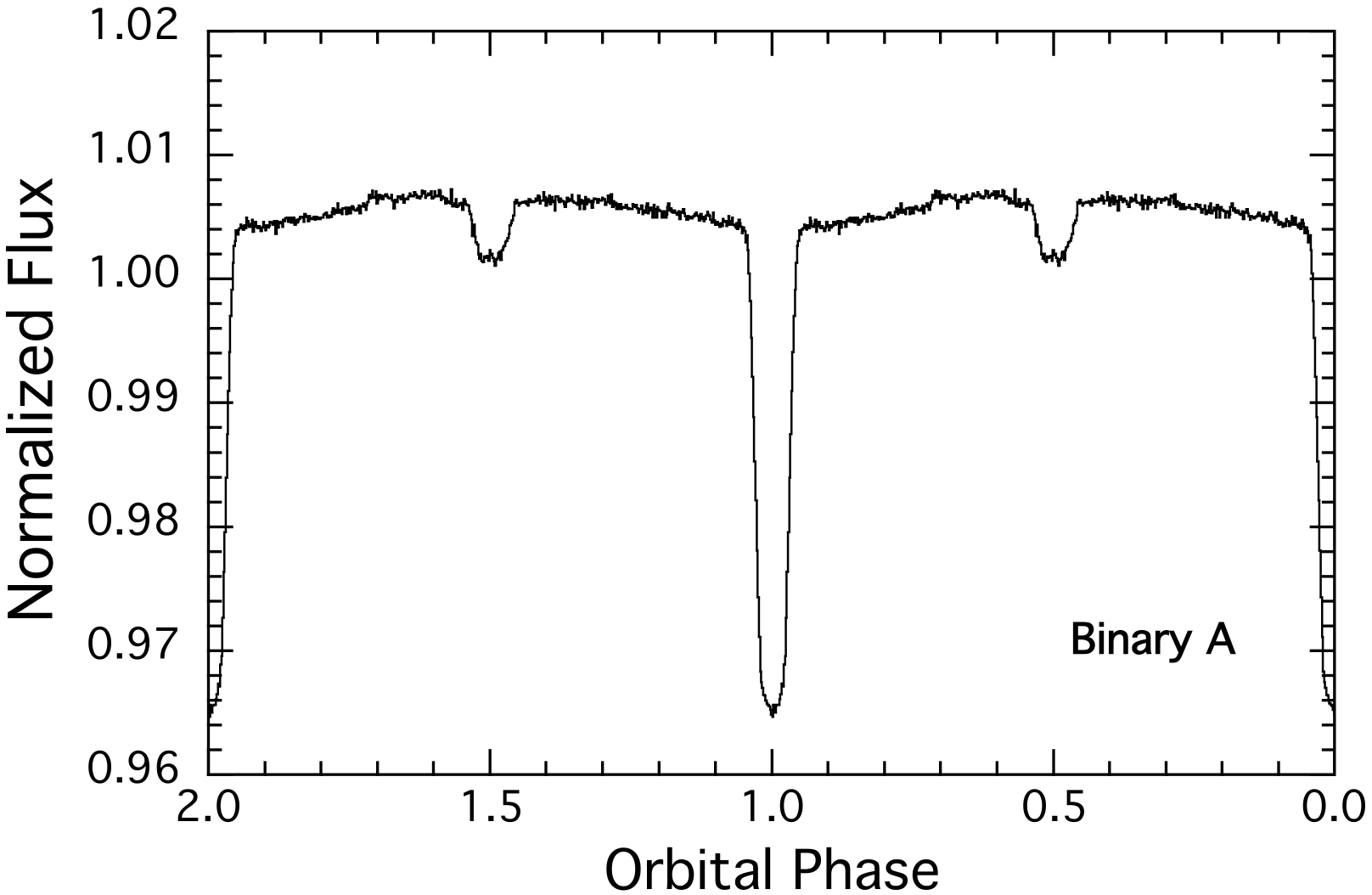}
    \includegraphics[width=0.98\columnwidth]{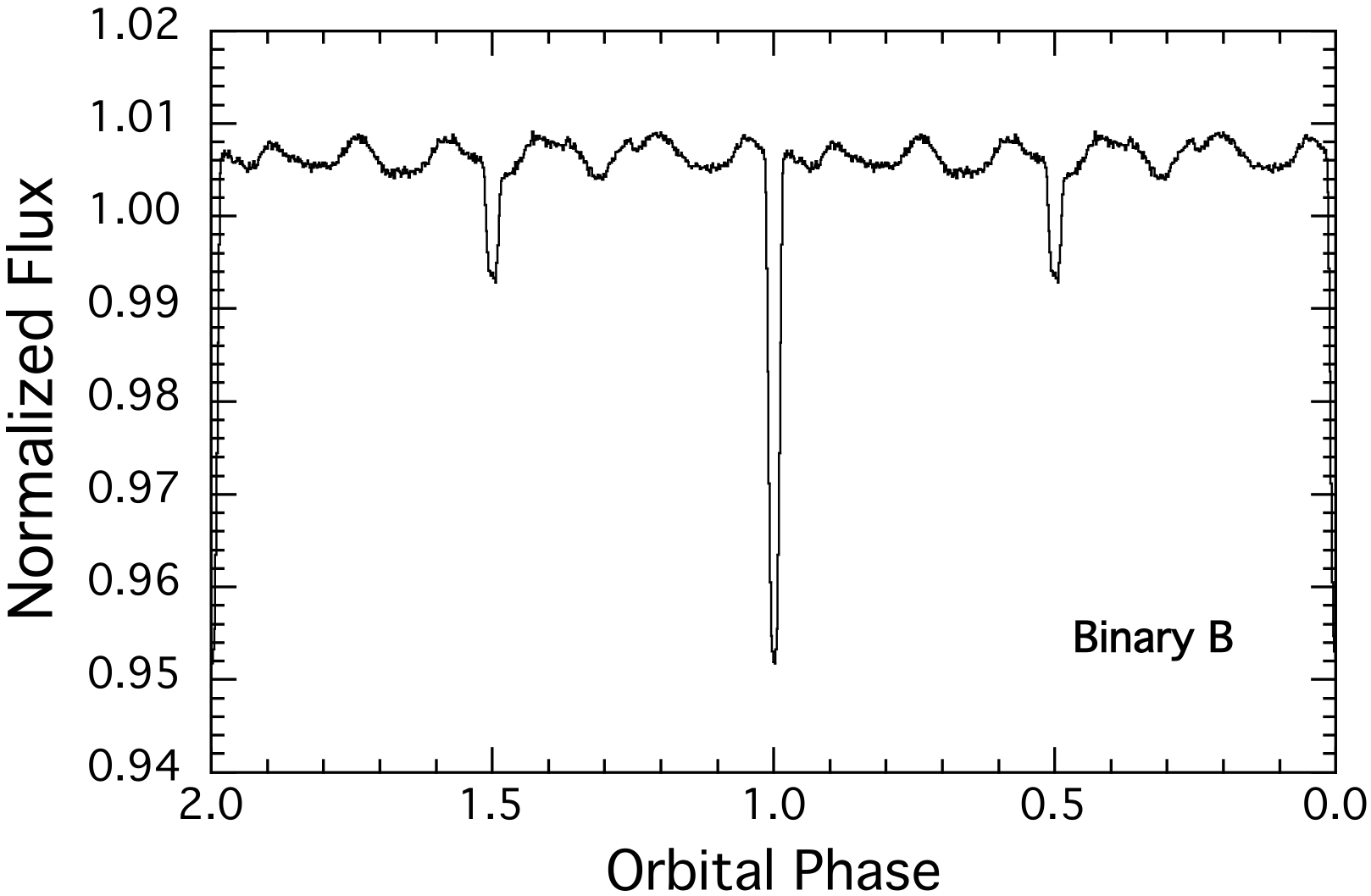}
    \includegraphics[width=0.98\columnwidth]{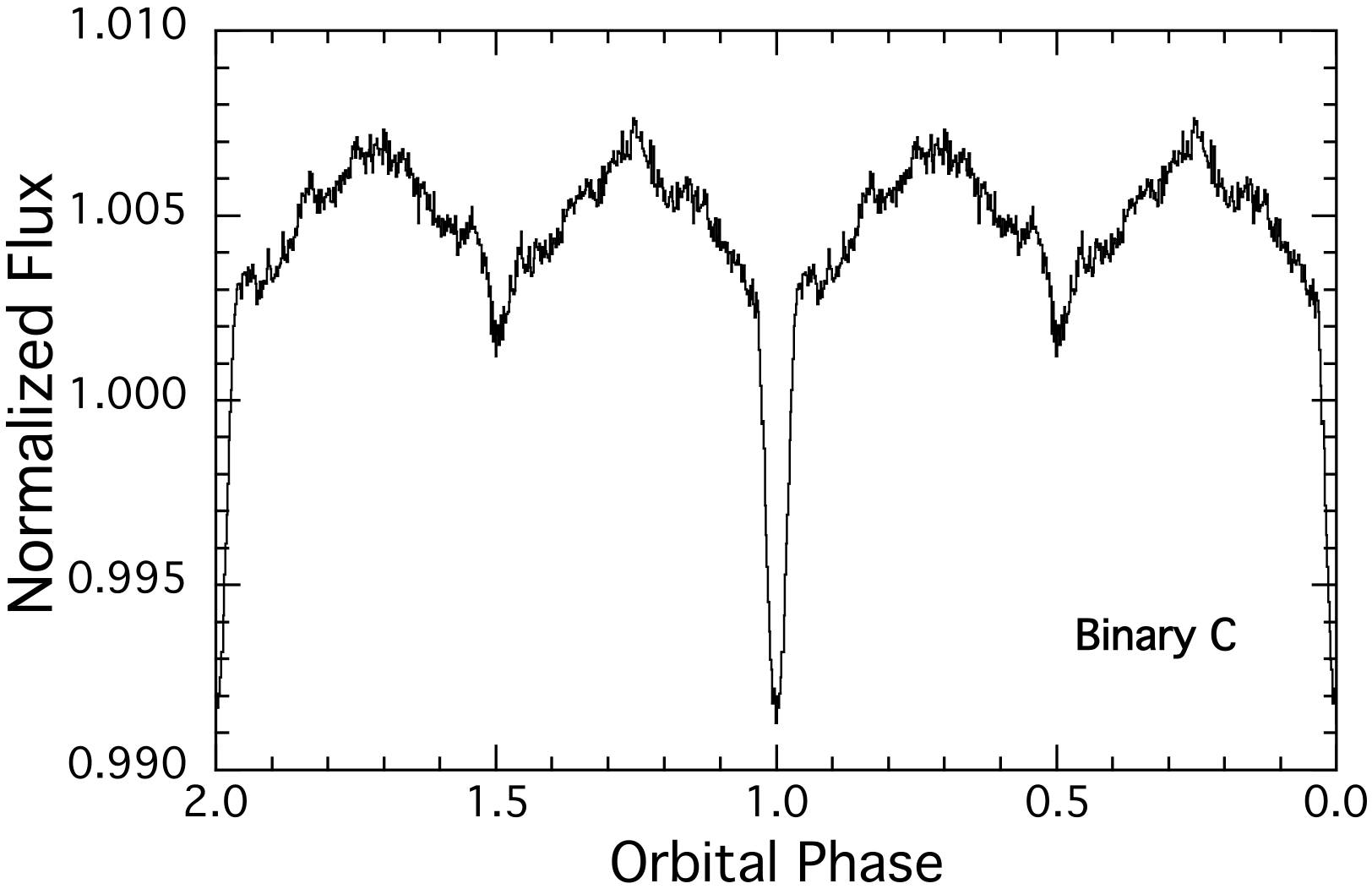}
\caption{Folds of the sector 31, 2-minute cadence, disentangled light curves of binary A (top), B (middle), and C (bottom).}
\label{fig:lcfS31_fits}
\end{figure}

Since this paper was completed, we have received new {\em TESS} data which were taken at 2-minute cadence, thereby greatly improving the temporal resolution.  We show in Figure \ref{fig:s31} the new lightcurve and in Figure \ref{fig:lcfS31_fits} the folded, disentangled lightcurve.  We have redone those parts of the analyses which utilize the {\em TESS} lightcurve and find that the basic answers presented herein do not change significantly.

\acknowledgments

This paper includes data collected by the {\em TESS} mission, which are publicly available from the Mikulski Archive for Space Telescopes (MAST). Funding for the {\em TESS} mission is provided by NASA's Science Mission directorate.

Resources supporting this work were provided by the NASA High-End Computing (HEC) Program through the NASA Center for Climate Simulation (NCCS) at Goddard Space Flight Center.  Personnel directly supporting this effort were Mark L. Carroll, Laura E. Carriere, Ellen M. Salmon, Nicko D. Acks, Matthew J. Stroud, Bruce E. Pfaff, Lyn E. Gerner, Timothy M. Burch, and Savannah L. Strong.

This research has made use of the Exoplanet Follow-up Observation Program website, which is operated by the California Institute of Technology, under contract with the National Aeronautics and Space Administration under the Exoplanet Exploration Program. 

This research is based on observations made with the Galaxy Evolution Explorer, obtained from the MAST data archive at the Space Telescope Science Institute, which is operated by the Association of Universities for Research in Astronomy, Inc., under NASA contract NAS5-26555.

We would like to thank Gil Esquerdo, Mike Calkins, and Amber Medina for obtaining TRES measurements of the system.

We would also like to thank the Pierre Auger Collaboration for the use of its facilities. The operation of the robotic telescope FRAM is supported by the grant of the Ministry of Education of the Czech Republic LM2018102. The data calibration and analysis related to the FRAM telescope is supported by the Ministry of Education of the Czech Republic MSMT-CR LTT18004 and MSMT/EU funds CZ.02.1.01/0.0/0.0/16$\_$013/0001402.

TB acknowledges the financial support of the Hungarian National Research, Development and Innovation Office -- NKFIH Grant KH-130372.

TD acknowledges support from MIT's Kavli Institute as a Kavli postdoctoral fellow.

This work has made use of data from the European Space Agency (ESA) mission {\it Gaia} (\url{https://www.cosmos.esa.int/gaia}), processed by the {\it Gaia} Data Processing and Analysis Consortium (DPAC, \url{https://www.cosmos.esa.int/web/gaia/dpac/consortium}). Funding for the DPAC has been provided by national institutions, in particular the institutions participating in the {\it Gaia} Multilateral Agreement.

Resources supporting this work were provided by the NASA High-End Computing (HEC) Program through the NASA Advanced Supercomputing (NAS) Division at Ames Research Center for the production of the SPOC data products.

This work makes use of observations from the LCOGT network.

\facilities{
\emph{Gaia},
MAST,
TESS,
WASP,
ASAS-SN,
NCCS,
FRAM,
PEST,
CHIRON,
TRES,
SOAR,
LCOGT}

\software{
{\tt Astrocut} \citep{astrocut},
{\tt AstroImageJ} \citep{Collins:2017},
{\tt Astropy} \citep{astropy2013,astropy2018}, 
{\tt Eleanor} \citep{eleanor},
{\tt IPython} \citep{ipython},
{\tt Keras} \citep{keras},
{\tt Keras-vis} \citep{kerasvis}
{\tt Lightcurvefactory} \citep{2013MNRAS.428.1656B,2017MNRAS.467.2160R,2018MNRAS.478.5135B},
{\tt Lightkurve} \citep{lightkurve},
{\tt Matplotlib} \citep{matplotlib},
{\tt Mpi4py} \citep{mpi4py2008},
{\tt NumPy} \citep{numpy}, 
{\tt Pandas} \citep{pandas},
{\tt PHOEBE} \citep{2011ascl.soft06002P},
{\tt Scikit-learn} \citep{scikit-learn},
{\tt SciPy} \citep{scipy},
{\tt Tapir} \citep{Jensen:2013},
{\tt Tensorflow} \citep{tensorflow},
{\tt Tess-point} \citep{tess-point}
}

\bibliography{refs}{}
\bibliographystyle{aasjournal}

\end{document}